\documentclass{article}
\usepackage{arxiv}
\usepackage[utf8]{inputenc} 
\usepackage[english]{babel}
\usepackage[T1]{fontenc}    
\usepackage{hyperref,url,doi}
\usepackage{float,multirow,booktabs}
\usepackage{amsfonts,amsmath,amssymb,amsthm,bbm,bm}
\usepackage{nicefrac, microtype, enumerate,soul}
\usepackage[normalem]{ulem}
\usepackage{graphicx,psfrag,epsf}
\usepackage{subcaption,lscape}
\usepackage[font=small,labelfont=bf]{caption}
\usepackage[table,xcdraw]{xcolor}
\usepackage{algorithm,tabularx}
\usepackage[noend]{algpseudocode}
\usepackage{rotating,titlesec,enumitem}
\usepackage{natbib}
\usepackage{textalpha}

\usepackage{MathCommands}

\newcounter{algsubstate}

\makeatletter
\newcommand{\multiline}[1]{%
	\begin{tabularx}{\dimexpr\linewidth-\ALG@thistlm}[t]{@{}X@{}}
		#1
	\end{tabularx}
}
\makeatother

\newcommand{\transp}{{\sf T}}
\newtheorem{prop}{Proposition}
\newcommand{\Rx}{\textrm{R$_\textrm{x}$}}

\newcommand{\syn}{\textrm{syn}}

\newcommand{\pr}{\textrm{pr}}

\title{Probabilistic Learning of Treatment Trees in Cancer}

\author{{Tsung-Hung Yao}\thanks{Corresponding author.} \\
	Department of Biostatistics\\
	University of Michigan at Ann Arbor\\
	Ann Arbor, MI 48109 \\
	\texttt{yaots@umich.edu} \\
	\And
	{Zhenke Wu} \\
	Department of Biostatistics\\
	University of Michigan at Ann Arbor\\
	Ann Arbor, MI 48109 \\
	\texttt{zhenkewu@umich.edu} \\
	\And
	{Karthik Bharath} \\
	School of Mathematical Sciences\\
	University of Nottingham\\
	Nottingham, UK\\
	\texttt{Karthik.Bharath@nottingham.ac.uk} \\
	\And
	{Jinju Li} \\
	Department of Biostatistics\\
	University of Michigan at Ann Arbor\\
	Ann Arbor, MI 48109 \\
	\texttt{lijinju@umich.edu} \\
	\And
	{Veerabhadran Baladandayuthapan} \\
	Department of Biostatistics\\
	University of Michigan at Ann Arbor\\
	Ann Arbor, MI 48109 \\
	\texttt{veerab@umich.edu} \\
}

\renewcommand{\shorttitle}{Probabilistic Learning of Treatment Trees in Cancer}

\hypersetup{
pdftitle={Probabilistic Learning of Treatment Trees in Cancer},
pdfsubject={Statistics},
pdfauthor={Tsung-Hung Yao, Zhenke Wu, Karthik Bharath, Jinju Li, Veerabhadran Baladandayuthapan},
pdfkeywords={Approximate Bayesian Computation, Dirichlet Diffusion Trees, Patient Derived Xenograft, Precision Medicine, Tree-Based Clustering},
}

\begin{document}
\maketitle

\begin{abstract}
	Accurate identification of synergistic treatment combinations and their underlying biological mechanisms is critical across many disease domains, especially cancer. In translational oncology research, preclinical systems such as patient-derived xenografts (PDX) have emerged as a unique study design evaluating multiple treatments administered to samples from the same human tumor implanted into genetically identical mice. In this paper, we propose a novel Bayesian probabilistic tree-based framework for PDX data to investigate the hierarchical relationships between treatments by inferring treatment cluster trees, referred to as treatment trees (\Rx-tree). The framework motivates a new metric of mechanistic similarity between two or more treatments accounting for inherent uncertainty in tree estimation; treatments with a high estimated similarity have potentially high mechanistic synergy. Building upon Dirichlet Diffusion Trees, we derive a closed-form marginal likelihood encoding the tree structure, which facilitates computationally efficient posterior inference via a new two-stage algorithm. Simulation studies demonstrate superior performance of the proposed method in recovering the tree structure and treatment similarities. Our analyses of a recently collated PDX dataset produce treatment similarity estimates that show a high degree of concordance with known biological mechanisms across treatments in five different cancers. More importantly, we uncover new and potentially effective combination therapies that confer synergistic regulation of specific downstream biological pathways for future clinical investigations. Our accompanying code, data, and shiny application for visualization of results are available at: \url{https://github.com/bayesrx/RxTree}.
\end{abstract}

\keywords{Approximate Bayesian Computation \and Dirichlet Diffusion Trees \and Patient Derived Xenograft \and Precision Medicine, Tree-Based Clustering}

\section{Introduction}\label{sec:intro}
According to the World Health Organization, cancer is one of the leading causes of death globally, with $\sim$10 million deaths in 2020 \citep{who_cancer}. Despite multiple advances over the years, systematic efforts to predict efficacy of cancer treatments have been stymied due to multiple factors, including patient-specific heterogeneity and treatment resistance \citep{pmid29115304,pmid33707180}. Given that the evolution of tumors relies on a limited number of biological mechanisms, there has been a recent push towards combining multiple therapeutic agents, referred to as ``combination therapy'' \citep{pmid23803949,pmid33707180}. This is driven by the core hypothesis that combinations of drugs act in synergistic manner, with each drug compensating for the drawbacks of other drugs. However, despite higher response rates and efficacy in certain instances \citep{pmid28410237}, combination therapy can lead to undesired drug-drug interactions, lower efficacy, or severe side effects \citep{DrugDrugInter}. Consequently, it is highly desirable to advance the understanding of underlying mechanisms that confer synergistic drug effects and identify potential favorable drug-drug interaction mechanisms for further investigations.

Given that not all possible drug combinations can be tested on patients in actual clinical trials, cancer researchers rely on preclinical ``model'' systems to guide the discovery of the most effective combination therapies (note, models have a different contextual meaning here). In translational oncology, preclinical models assess promising treatments and compounds, before they are phased into human clinical trials. The traditional mainstay of such preclinical models has been cell-lines, wherein cell cultures derived from human tumors are grown in an \textit{in vitro} controlled environment. However, it has been argued that they do not accurately reflect the true behavior of the host tumor and, in the process of adapting to \textit{in vitro} growth, lose the original properties of the host tumor, thus leading to limited clinical relevance and successes \citep{pmid22508028,pmid31919403}. To overcome these challenges, there has been a push towards more clinically relevant model systems that maintain a high degree of fidelity to human tumors. One such preclinical model system is \ul{P}atient-\ul{D}erived \ul{X}enograft (PDX) wherein tumor fragments obtained from cancer patients are directly transplanted into genetically identical mice \citep{pmid25185190,pmid28499452}. Compared to traditional oncology models such as cell-lines \citep{Yoshida2020}, PDX models maintain key cellular and molecular characteristics, and are thus more likely to mimic human tumors and facilitate precision medicine. More importantly, accumulating evidence suggests responses (e.g. drug sensitivity) to standard therapeutic regimens in PDXs closely correlate with patient clinical data, making PDX an effective and predictive experimental model across multiple cancers \citep{TOPP2014656,Nunes1560}.

\begin{figure}[!htb]
    \includegraphics[width=0.95\linewidth]{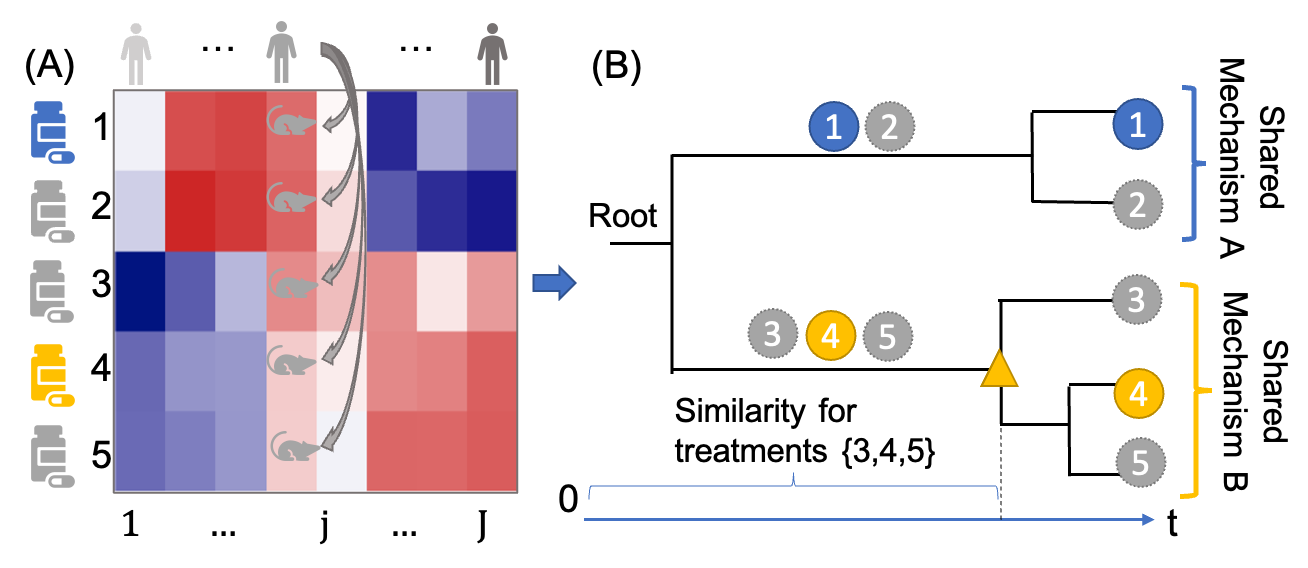}
    \caption{PDX experimental design and tree-based representation. Panel \textbf{A}: an illustrative PDX dataset with five treatments (row) and eight patients (column). Mice in a given column are implanted with tumors from the same patient and receive different treatments (across rows). The level of tumor responses are shown along a color gradient. Panel \textbf{B}: a tree structure that clusters the treatments and quantifies the similarity among mechanisms. Two treatments (1 and 4) are assumed to have different but known biological mechanisms (in different colors); the rest three treatments (2,3, and 5) have unknown mechanisms (in gray). The tree suggests two treatment groups are present ($\{1,2\}$ and $\{3,4,5\}$) that may correspond to two different known mechanisms. The horizontal position of ``$\triangle$'' represents the divergence time (defined in Section \ref{sec:DDT_JtDensity}) and the mechanism similarity for treatments $\{3,4,5\}$. In a real data analysis, the tree (topology and divergence times) is unknown and is to be inferred from PDX data.}
	\label{fig:toyEg}
\end{figure}

\paragraph{PDX experimental design and key scientific questions.} Briefly, in a standard PDX preclinical experiment, a set of common treatments are tested, and each treatment is given to multiple mice with tumors implanted from the same (matched) patient (see conceptual schema in Figure \ref{fig:toyEg}(A)). Treatment responses (e.g. tumor size) are then evaluated, resulting in a data matrix (treatments $\times$ patients) as depicted in the heatmap in Figure \ref{fig:toyEg}(A). The PDX experimental protocol, often referred to as ``co-clinical trials,'' mirrors a real human clinical trial using mouse ``avatars'' \citep{pmid25895610}. Thus the protocol serves as a scalable platform to: (a) identify underlying plausible biological mechanisms responsible for tumor growth and resistance, and (b) evaluate the effectiveness of drug combinations based on mechanistic understanding \citep{pmid25466492}. In this context, the (biological) mechanism refers to the specific mechanism of action of a treatment, which usually represents a specific target, such as an enzyme or a receptor \citep{MoA}. 
From the perspective of treatment responses as data, responses are the consequences of the downstream biological pathways from the corresponding interaction  between a treatment and the target/mechanism. 

Ideally, treatments with the same target/mechanism should induce similar responses and engender mechanism-related clustering among treatments. Evidently then, a sensible clustering of treatments would not only partition treatments into clusters but also explicate how the clusters relate to one another; in other words, a hierarchy among treatment clusters is more likely to uncover plausible mechanisms for combinations of treatments with ``similar'' responses when compared to ``flat'' clusters (e.g., $k$-means clustering). Such response-based identification of potential synergistic effects from combinations of treatments will augment understanding from known mechanistic synergy.  
In our application, using tree-based clustering, we assume known entities at the leaves, i.e., the different treatments. The treatments are assumed to act upon potentially distinct biological pathways, resulting in different levels of responses across the treated mice. In this paper, we use PDX response data on the leaves to infer a hierarchy over treatments that may empirically characterize the similarity in the targeted mechanistic pathways. The primary statistical goals are to (i) define and estimate a general metric measuring the similarity within any subset comprising two or more treatments, and (ii) facilitate (i) by conceptualizing and inferring an unknown hierarchy among treatments.

\paragraph{Tree-based representations for PDX data.} To this end, we consider a tree-based construct to explore the hierarchical relationships between treatments, referred to as  {\it treatment tree (\Rx-tree}, in short). We view such a tree structure as a representation of clustering of treatments based on mechanisms that confer synergistic effects, wherein similarities between mechanisms are captured through branch lengths. Hierarchy among treatments can be interpreted through branch lengths (from the root) that are potentially reflective of different cancer processes; this would then help identify common mechanisms and point towards treatment combinations disrupting oncological processes if administered simultaneously. 

We will focus on rooted trees. The principal ingredients of a rooted tree comprise a root node, terminal nodes (or, leaves), internal nodes and branch lengths. In the context of the \Rx-tree for PDX data, the leaves are observed treatment responses, whereas internal nodes and branch lengths are unobserved. Internal nodes are clusters of treatments, and lengths of branches between nodes are indicative of strengths of mechanism similarities. The root is a single cluster consisting of all treatments. This leads to the following interpretation: at the root all treatments share a common target or mechanism; length of path from the root to the internal node (sum of branch lengths) at which two treatments split into different clusters measures mechanism similarity between the two treatments. Thus treatments that stay clustered ``longer'' have higher mechanism similarities. 

Throughout, we will use `tree' when describing methodology for an abstract tree (acyclic graphs with distinguished root node) and `treatment tree' or `\Rx-tree' when referring to the latent tree within the application context.

\paragraph{An illustrative example.} A conceptual \Rx-tree and its interpretation is illustrated in Figure \ref{fig:toyEg} where five treatments (1 to 5) are applied on eight patients' PDXs (Figure \ref{fig:toyEg}(A)) with the corresponding (unknown true) \Rx-tree (Figure \ref{fig:toyEg}(B)) based on the PDX data. Assume two treatment groups based on different mechanisms -- treatments $\{1,2\}$ and treatments $\{3,4,5\}$; further, suppose treatment $4$ is  approved by the Food and Drug Administration (FDA). The heatmap in Panel (A) visualizes the distinct levels of response profiles to the five treatments so that treatments closer in the tree are more likely to have similar levels of responses. The \Rx-tree captures the mechanism similarity by arranging treatments $\{1,2\}$ and $\{3,4,5\}$ to stay in their respective subtrees longer and to separate the two sets of treatment early in the tree. Based on the \Rx-tree, treatments $\{3,5\}$ share high mechanism similarity values with treatment $4$; treatment $5$ is the closest to the treatment $4$, suggesting the most similar synergistic mechanism among all the evaluated treatments $1$ to $5$.

\paragraph{Existing methods and modeling background.} 
The Pearson correlation is a popular choice to assess mechanism similarity between treatments \citep{pmid21273060}, but is inappropriate to examine multi-way similarity.
A tree-structured approach based on a (binary) dendrogram obtained from hierarchical clustering of cell-line data using the cophenetic distance \citep{cophenetic_1962} was adopted in \citet{pmid32523045}; their approach, however, failed to account for uncertainty in the dendrogram, which is highly sensitive to measurement error in the response variables as well distance metrics (we show this via simulations and in real data analyses). In this paper, we consider a model for PDX data parameterized by a tree-structured object representing the \Rx-tree. The model is derived from the Dirichlet diffusion tree (DDT) \citep{Neal2003} generative model for (hierarchically) clustered data. The DDT engenders a data likelihood and a prior distribution on the tree parameter with support in the space of rooted binary trees. We can then use the posterior distribution to quantify uncertainty about the latent \Rx-tree.

\paragraph{Summary of  novel contributions and organization of the article.} Our approach based on the DDT model for PDX data results in three main novel contributions:
\begin{enumerate}
\itemsep 0em
    \item [(a)]{\it Derivation of a closed-form likelihood that encodes the tree structure.} The DDT specification results in a joint distribution on PDX data, treatment tree parameters and other model parameters. By marginalizing over unobserved data that correspond to internal nodes of the tree, we obtain a new multivariate Gaussian likelihood with a special tree-structured covariance matrix, which completely characterizes the treatment tree (Proposition \ref{prop1} and Lemma \ref{lemma1}). 
    \item [(b)] {\it Efficient two-stage algorithm for posterior sampling}. Motivated by the form of marginal data likelihood in (a), we decouple the Euclidean and tree parameters and propose a two-stage algorithm that combines an approximate Bayesian computation (ABC) procedure (for Euclidean parameters) with a Metropolis-Hasting (MH) step (for tree parameters). We demonstrate via multiple simulation studies the superiority of our hybrid approach over approaches based on classical single-stage MH algorithms (Sections \ref{sec::simulation_computational_aspect} and \ref{sec::simulation_similarity_aspect}). 
    \item [(c)] {\it Corroborating existing, and uncovering new, synergistic combination therapies.} We define and infer a new similarity measure that accounts for inherent uncertainty in estimating a latent hierarchy among treatments. As a result, the {\it maximum a posteriori} \Rx-tree and the related mechanism similarity show high concordance with known existing biological mechanisms for monotherapies and uncover new and potentially useful combination therapies (Sections \ref{sec:MonoTherapy} and \ref{sec:CmbTherapy}). 
\end{enumerate}
Of particular note is contribution (c), where we leverage a recently collated PDX dataset from the \ul{N}ovartis \ul{I}nstitutes for \ul{B}ioMedical \ul{R}esearch - \ul{PDX} \ul{E}ncyclopedia [NIBR-PDXE, \citep{pmid26479923}] that interrogated multiple targeted therapies across five different cancers. Our pan-cancer analyses of the NIBR-PDXE dataset show a high degree of concordance with known existing biological mechanisms across different cancers; for example, a high mechanistic similarity is suggested between two agents currently in clinical trials: CGM097 and HDM201 in breast cancer and colorectal cancer, known to target the same gene MDM2 \citep{pmid32651541}. In addition, our model uncovers new and potentially effective combination therapies. For example, exploiting knowledge of the combination therapy of a class of agents targeting the PI3K-MAPK-CDK pathway axes -- PI3K-CDK for breast cancer, PI3K-ERBB3 for colorectal cancer and BRAF-PI3K for melanoma  -- confers possible synergistic regulation for prioritization in future clinical studies.

The rest of the paper is organized as follows: we first review our probabilistic formulation for PDX data based on the DDT model and present the marginal data likelihood and computational implications in Section \ref{sec:DDT}. In Section \ref{sec::posterior_algorithm}, we derive the posterior inference algorithm based on a two-stage algorithm. In Section \ref{sec:Simulation}, we conduct two sets of simulations to evaluate the operating characteristics of the model and algorithm. A detailed analysis of the NIBR-PDXE dataset, results, biological interpretations and implications are summarized in Section \ref{sec:DataAnalysis}. The paper concludes by discussing implications of the findings, limitations, and future directions. 

\section{Modeling \texorpdfstring{\Rx-tree}{TEXT} via Dirichlet Diffusion Trees} \label{sec:DDT}
Given a PDX experiment with $I$ correlated treatments and $J$ independent patients, we focus on the setting with $1\times 1 \times 1$ design (one animal per PDX model per treatment) with no replicate response for each treatment and patient. A PDX experiment produces an observed data matrix $\Xb_{I\times J}=[\bX_1, \ldots, \bX_I]^\transp$ where $\bX_{i}=[X_{i1},\ldots, X_{iJ}]^\transp$ is data under treatment $i$ across $J$ patients; let the observed response column for each patient be  $\bX_{\cdot,j}=[x_{1j},\ldots,x_{Ij}]^\transp \in \mathbb{R}^I, j=1,\ldots,J$. 

In this paper, the observed treatment responses are continuous and we model the responses through a generative model that results in a Gaussian likelihood with a structured covariance:
\begin{align}\label{eq:iidNorm_noScl}
    \bX_{\cdot,j}|\bm{\Sigma}^{\mathcal{T}},\sigma^2 \stackrel{iid}{\sim}{\sf N}_I(0,{\bm{\Sigma}^{\mathcal{T}}}), \quad j=1,\ldots,J,
\end{align}
where the $\bm{\Sigma}^{\mathcal{T}}$ is a tree-structured covariance matrix that encodes the tree $\cT$. 
In particular, $\bSigma^{\cT}=\left\{\bm \Sigma^{\mathcal T}_{i,i'},i,i'=1,\ldots, I\right\}$ encodes the tree $\cT$ through two constraints \citep{Lapointe1991, McCullagh2006StructuredCM}: 
\begin{align}
\label{eq:symMat} \bSigma^{\cT}_{i',i}=\bSigma^{\cT}_{i,i'}\geq 0; \thickspace & \bSigma^{\cT}_{i,i}\geq \bSigma^{\cT}_{i,i'},\\ 
\label{eq:ultraIneq}\bSigma^{\cT}_{i,i'}\geq \min\{\bm \Sigma^{\mathcal T}_{i,i{''}}\ , \ \bm \Sigma^{\mathcal T}_{i',i{''}}\} & \text{ for all } i\neq i' \neq i{''}.
\end{align}
Each element $\bSigma^{\cT}_{i,i}$ is the covariance between treatments $i$ and $i'$ and measures their similarity. The inequality \eqref{eq:symMat} imposes the symmetry of covariance matrix and ensures the divergence of all leaves. The tree structure is characterized by the ultrametric inequality \eqref{eq:ultraIneq} that ensures $\bSigma^{\cT}$ bijectively maps to a tree $\cT$; for more details on the relationship between the covariance $\bSigma^{\cT}$ and the tree $\cT$ see \citet{McCullagh2006StructuredCM} and \citet{pmid22081761}. Of note, mean parameterized models (e.g. mixed effects models) are inappropriate for uncovering the tree parameter under the given data structure since the latent tree is completely encoded in covariance matrix $\bSigma^\cT$. 


A Bayesian formulation requires an explicit prior distribution on $\bSigma^{\cT}$ which satisfies constrains \eqref{eq:symMat} and \eqref{eq:ultraIneq}; this requirement is far from straightforward since the set of tree-structured matrices is complicated (e.g., it is not a manifold \citep{McCullagh2006StructuredCM}). We instead consider the Dirichlet Diffusion tree (DDT) model \citep{Neal2003} for hierarchically clustered data which provides two useful ingredients:
\begin{enumerate}
    \item a prior is implicitly specified on the latent treatment tree, comprising the root, internal nodes, leaves, and branch lengths;
    \item upon integrating out the internal nodes, a tractable Gaussian likelihood on PDX data with tree-structured covariance is specified. 
\end{enumerate}

We first provide a brief description of the DDT model proposed by \citet{Neal2003} and its joint density on data and tree (Section \ref{sec:DDT_JtDensity}). Subsequently, we derive an expression for the likelihood and demonstrate how it can be profitably employed to develop a generative model for PDX data and carry out \Rx-tree estimation (Section \ref{sec::marginal} and \ref{sec:Decouple}).


\subsection{The Generative Process of DDT} \label{sec:DDT_JtDensity}
\begin{figure}[!htb]
        \centering
        \includegraphics[width=\linewidth]{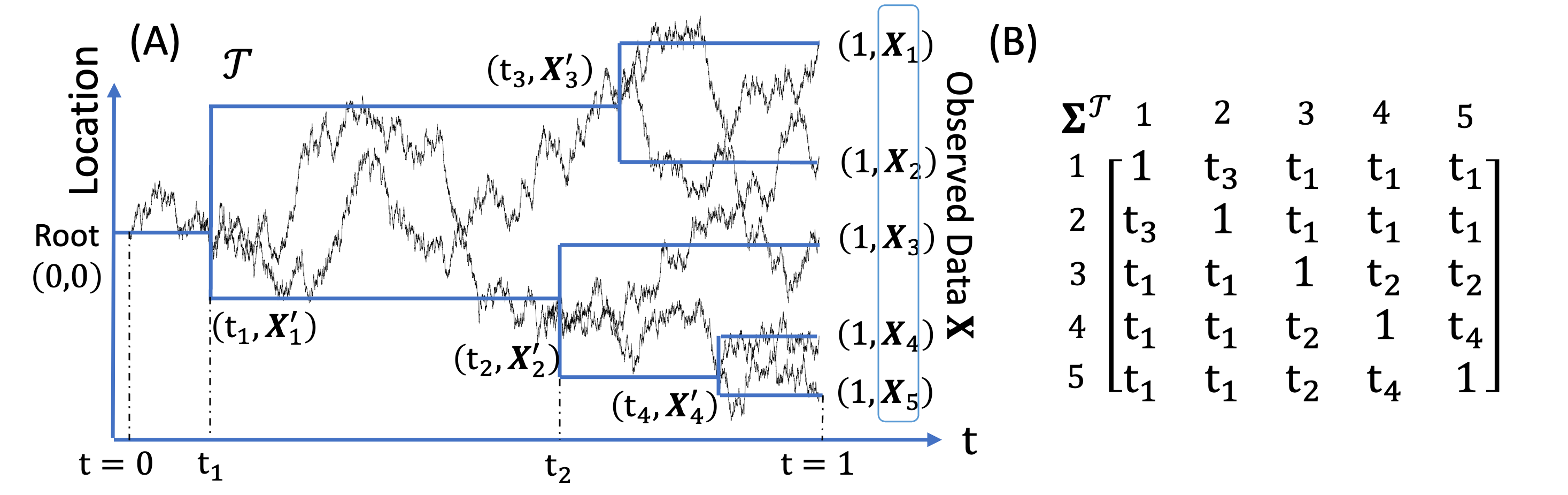}
    \caption{\small (A) A binary tree with $I=5$ leaves underlying the diffusion dynamics. The observed response vector $\bX_i,i=1,\ldots,I$ is generated by the Brownian motion up to $t=1$. The unobserved response vector $\bX_d',d=1,\ldots,(I-1)$ at the divergence is generated by the Brownian motion at time $t_d$. (B) A tree-structured matrix $\bm{\Sigma}^\cT$ that encapsulates the tree $\cT$. See the Proposition \ref{prop1} for the definition of $\bm{\Sigma}^\cT$.} 
    \label{fig:DDT_Bone}
    \end{figure}

The DDT prescribes a fragmentary, top-down mechanism to generate a binary tree (acyclic graph with a preferred node or vertex referred to as the root), starting from a root, containing $J$-dimensional observed responses $\bX_i$ at $I$ leaves/terminal nodes; each node in the tree has either 0 or 2 children excepting the root which has a solitary child. This prescription manipulates dynamics of a system of $I$ independent Brownian motions $B_1,\ldots,B_I$ on $\mathbb{R}^J$ in a common time interval $t \in [0,1]$. As shown in Figure \ref{fig:DDT_Bone}(A), all Brownian motions $B_i(t)$ start at the same point at time $t=0$, location of which is the root $\bm 0 \in \mathbb R^J$, and diverge at time points in $[0,1]$ and locations in $\mathbb R^J$ before stopping at the time $t=1$ at locations $\bX_i$. The Brownian trajectories and their divergences engender the tree structure as shown in Figure \ref{fig:DDT_Bone}(A).

Specifics on when and how the Brownian motions diverge are as follows: the first Brownian motion $B_1(t)$ starts at $t=0$ and generates $\bX_{1}$ at $t=1$; a second independent Brownian motion $B_2(t)$ starts at the same point at $t=0$, branches out from the first Brownian motion at some time $t$, after which it generates $\bX_{2}$ at time $1$. The probability of divergence in a small interval $[t,t+dt]$ is given by a \emph{divergence function} $t\mapsto a(t)$, assumed as in \citet{Neal2003} to be of the form $a(t)=c(1-t)^{-1}$ for some divergence parameter $c>0$. Inductively then, the vector of observed responses to treatment $i$, $\bX_{i}$, is generated by $B_i(t)$, which follows the path of previous ones. If at time $t$, $B_i(t)$ has not diverged and meets the previous divergent point, it will follow one of the existing path with the probability proportional to the number of data points that have previously traversed along each path. Eventually, given $B_i(t)$ has not diverged at time $t$, it will do so in $[t,t+dt]$ with probability $a(t)dt/m$, where $m$ is the number of data points that have previously traversed the current path. 

From the illustration in panel (A) of Figure \ref{fig:DDT_Bone}, we note that $B_3$ diverges from the $B_1$ and $B_2$ at time $t_1$ at location $\bX_1'$ and at $t=1$ is at location $\bX_3$, which is the $J$-dimensional response vector for treatment 3; this creates a solitary branch of length $t_1$ from the root and an unobserved internal node at location $\bX_1'$. Continuing, given three Brownian motions $B_1$, $B_2$ and $B_3$, $B_4$ does not diverge before $t_1$ and meet the previous divergent point $t_1$. $B_4$ chooses to follow the path of $B_3$ with probability $1/3$ at $t_1$ and finally diverges from $B_3$ at time $t_2 >t_1$ at location $\bX_2'$; this results in observation $\bX_4$ for treatment 4 and an unobserved internal node at $\bX_2'$, and so on. As a consequence, the binary tree that arises from the DDT comprises of:
\begin{enumerate}
\itemsep 0em
    \item [(i)]an unobserved root at the origin in $\mathbb R^J$ at time $t=0$;
    \item [(ii)] observed data $\Xb=[\bX_1,\ldots,\bX_I]^\transp \in \mathbb R^{I \times J}$ situated at the leaves of the tree;
    \item [(iii)] unobserved internal nodes $\Xb^I=[\bX_1',\ldots, \bX'_{I-1}]^\transp \in \mathbb R^{(I-1) \times J}$;
    \item [(iv)] unobserved times $\bm t=(t_1,\ldots,t_{I-1})^\transp \in [0,1]^{I-1}$ that characterize lengths of branches;
    \item [(v)] unobserved topology  $\mathcal T$ that links (i)-(iv) into a tree structure, determined by the number of data points $\bX_i$ that have traversed through each segment or branch. 
\end{enumerate}
Conceptually, observed data at the leaves $\bX_1,\ldots,\bX_I$ collectively form the observed PDX responses generated through a process involving a few parameters: tree-related parameters $(\mathcal T, \bm t)$ and the locations of internal nodes $\bX'_i$. The tree $\mathcal T$ clusters $I$ treatments as a hierarchy of $(I-1)$ levels (excluding the last level containing leaves). At level $0<d\leq I-1$ of the hierarchy, characterized by the pair $(\bX'_d,t_d)$,  the $I$ treatments are clustered into $d+1$ groups; a measure of similarity (or dissimilarity) between treatment clusters at levels $d$ and $d+1$ is given by the branch length $t_{d+1}-t_d$. 

We now give a brief description of how the joint density of $(\Xb,\Xb^I,\bm t, \mathcal T)$ can be derived; for more details we direct the reader to \citet{Neal2003} and \citet{PYDT}. For a fixed $c>0$ that governs the divergence function $a(t)=c(1-t)^{-1}$, probabilities associated with the independent Brownian motions $B_1,\ldots,B_I$ induce a joint (Lebesgue) density on the generated tree. Note that the binary tree arising from the DDT is encoded by the triples $\{(t_d,\bX'_{d},\bX_{i}), d=1,\ldots,I-1; i=1,\ldots,I\}$. An internal node at $ \bX'_d$ contains $l_d$ and $r_d$ leaves below to its left and right with $m_d=l_d+r_d$. If each of the Brownian motions is scaled by $\sigma^2>0$, then given $\mathcal{T}$ and a branch with endpoints $(t_u,\bX'_{u})$ and $(t_v,\bX'_{v})$ with $0<t_u<t_v<1$, from properties of a Brownian motion we see that $\bX'_{v} \sim N_J(\bX'_{u},\sigma^2 (t_v-t_u)\bm{I}_J)$, and the (Lebesgue) density of $\cT$ can be expressed as the product of contributions from its branches. Then the joint density of all nodes, times and the tree topology is given by
\begin{align}\label{eq:joint_llh}
    P(\Xb,\Xb^I,\bm{t},\mathcal{T}|c,\sigma^2)=\Pi_{[u,v]\in \mathcal{S}(\mathcal{T})} 
    \frac{(l_v-1)!(r_v-1)!}{(l_v+r_v-1)!}
    c(1-t_v)^{cJ_{l_v,r_v}-1}
    N_J(\bX'_{u},\sigma^2 (t_v-t_u)\bm{I}_J)
\end{align}
where $\mathcal{S}(\mathcal{T})$ is the collection of branches and  $\Xb^I_{(I-1)\times J}=[\bX'_{1},\ldots,\bX'_{(I-1)}]^\transp$ are unobserved locations of the internal nodes. On each branch $[u,v]$, the first term $\frac{(l_v-1)!(r_v-1)!}{(l_v+r_v-1)!}$ represents the chance the branch containing $l_v$ and $r_v$ leaves to its left and right respectively; $c(1-t_v)^{cJ_{l_v,r_v}-1}$ represents the probability of diverging at $t_v$ with $l_v$ and $r_v$ leaves, where $J_{l+r}=H_{l_v+r_v-1} - H_{l_v-1} - H_{r_v-1}$ with $H_n=\sum_{i=1}^n 1/i$ is the $n$th harmonic number. 

The joint density is hence parameterized by $(c,\sigma^2)$, where $c$ plays a crucial role in determining the topology $\mathcal{T}$: through the divergence function $a(t)$, it determines the propensity of the Brownian motion to diverge from its predecessors; consequently, a small $c$ engenders later divergence and a higher degree of similarity among treatments in PDX. The latent tree has two components: (i) topology $\mathcal T$ and (ii) vector of divergence times $\bm t$ determining branch lengths. 
We refer to $(c,\sigma^2)$ as the \emph{Euclidean parameters} and $(\mathcal T,\bm t)$ as \emph{tree parameters}. 
    
\subsection{Prior on tree and closed-form likelihood}\label{sec::marginal}
The joint density in \eqref{eq:joint_llh} factors into a prior $P(\bm{t},\mathcal{T}|c,\sigma^2)$ on the tree parameter through $(\mathcal T,\bm t)$ and a density $P(\Xb,\Xb^I|\bm{t},\mathcal{T},c,\sigma^2)$ that is a product of $J$-dimensional Gaussians on the internal nodes and leaves. The prior distribution on the latent tree  is thus implicitly defined through the Brownian dynamics and is parameterized by $(\mathcal T,\bm t)$ with hyperparameters $(c,\sigma^2)$. In \eqref{eq:joint_llh} the product is over the set of branches $\mathcal S(\mathcal T)$, and the contribution to the prior $P(\mathcal T,\bm t|c,\sigma^2)$ from each branch $[u,v]$ is $\frac{(l_v-1)!(r_v-1)!}{(l_v+r_v-1)!}c(1-t_v)^{cJ_{l_v,r_v}-1}$, which is free of $\sigma^2$; on the other hand, the contribution to $P(\Xb,\Xb^I|\bm{t},\mathcal{T},c,\sigma^2)$ from $[u,v]$ is the $J$-dimensional $N_J(\bX'_{u},\sigma^2 (t_v-t_u)\bm{I}_J)$, which is independent of $c$. The likelihood function based on the observed $\Xb$ is thus obtained by integrating out the unobserved internal nodes $\Xb^I$ from $P(\Xb,\Xb^I|\bm{t},\mathcal{T},\sigma^2)$. Accordingly, our first contribution is to derive a closed-form likelihood function for efficient posterior computations; to our knowledge, this task is currently achieved only through sampling-based or variational methods \citep{Neal2003,PYDT}. 

Denote as ${\sf MN}_{I \times J}(M,U,V)$ the matrix normal distribution of an $I \times J$ random matrix with mean matrix $M$, row covariance $U$, and column covariance $V$, and let $\bm I_k$ denote the $k \times k$ identity matrix. Evidently, $\Xb$ follows a matrix normal distribution since Gaussian laws of the Brownian motions imply that $[\Xb,\Xb^I]=[\bX_{1},\ldots,\bX_{I},\bX'_{1},\ldots,\bX'_{(I-1)}]^\transp$ follow a matrix normal distribution. 
	
\begin{prop}\label{prop1}
    Under the assumption that the root is located at the origin in $\mathbb R^J$, the data likelihood $\Xb|\sigma^2,\mathcal{T},\bm{t} \sim {\sf MN}_{I \times J} (\bm{0},\sigma^2 {\bm \Sigma}^{\cT},\bm{I}_J)$, where ${\bm \Sigma^{\mathcal{T}}}=\Big(\bm \Sigma^{\mathcal T}_{i,i'}\Big)$ is an $I \times I$ tree-structured covariance matrix satisfying \eqref{eq:symMat} and \eqref{eq:ultraIneq} with $\bm \Sigma^{\mathcal T}_{i,i}=1$ and $\bm \Sigma^{\mathcal T}_{i,i'}=t_d$, for $i \neq i'$ where $i,i'=1,\ldots,I$ and $d=1,\ldots,I-1$. 
\end{prop}

Proposition \ref{prop1} asserts that use of the DDT model leads to a centered Gaussian likelihood on PDX data $\bX$ with a tree-structured covariance matrix. Proposition \ref{prop1} also implies that each patient independently follows the normal distribution of \eqref{eq:iidNorm_noScl} with an additional scale parameter $(\sigma^2)$ from the Brownian motion: 
\begin{align}\label{eq:iidNorm}
    \bX_{\cdot,j}|\bm{\Sigma}^{\mathcal{T}},\sigma^2 \stackrel{iid}{\sim}{\sf N}_I(0,{\sigma^2 \bm{\Sigma}^{\mathcal{T}}}), \quad j=1,\ldots,J.
\end{align}
By setting $\bm \Sigma^{\mathcal T}_{i,i'}=t_{i,i'}$ as the divergence time of $i$ and $i'$, $\bm \Sigma^{\mathcal T}$ satisfies \eqref{eq:symMat} and \eqref{eq:ultraIneq} and encodes the tree $\cT$. For example, consider a three-leaf tree with $\bm \Sigma^{\mathcal T}_{i,i'}=t_{i,i'}$, inequality \eqref{eq:ultraIneq} implies that for the three leaves, say, $i, i'$ and $i''$, one of the following conditions must hold: (i) $t_{i',i''} \geq t_{i,i'} = t_{i,i''}$; (ii) $t_{i,i''} \geq t_{i,i'} = t_{i',i''}$; (iii) $t_{i,i'} \geq t_{i,i''} = t_{i',i''}$. We then obtain a tree containing 1) a subtree of two leaves with a higher similarity and 2) a singleton clade with a lower similarity between the singleton leaf and the two leaves in the first subtree. In particular, if $t_{i',i''} \geq t_{i,i'} = t_{i,i''}$ holds, the three-leaf tree has leaf $i$ diverging earlier before the subtree of $(i',i'')$.

\subsection{Decoupling Tree and Euclidean Parameters for Efficient Sampling.}\label{sec:Decouple} In the full joint density in \eqref{eq:joint_llh} the Euclidean and tree parameters are confounded across row and column dimensions of $\Xb$, and this may result in slow mixing of chains using traditional MCMC algorithms \citep{pmid23646991}. State-of-the-art posterior inference on $(c,\sigma^2,\mathcal{T},\bm{t})$ can be broadly classified into sampling-based approaches \citep[e.g.,][]{PYDT} and deterministic approaches based on variational message passing \citep[e.g.,][VMP]{MSP_DDT}. Variational algorithms can introduce approximation errors to the joint posterior via factorization assumptions (e.g.,mean-field) and choice of algorithm is typically determined by the speed-accuracy trade-off tailored for particular applications. On the other hand, in classical MCMC-based algorithms for DDT we observed slow convergence in the sampling chains for $c$ and $\sigma^2$ with high autocorrelations for the corresponding chains, owing to possibly the high mutual dependence between $c$ in the divergence function and the tree topology $\cT$ , resulting in slow local movements in the joint parameter space of model and tree parameters (Simulation II in Section \ref{sec::simulation_computational_aspect}).

Notwithstanding absence of the parameter $c$ in the Gaussian likelihood, the dependence, and information about, $c$ is implicit: the distribution of divergence times $\bm t$ that populate $\bm\Sigma^{\mathcal{T}}$ are completely determined by the divergence function $t \mapsto c(1-t)^{-1}$. In other words, $c$ can indeed be estimated from treatment responses $\{\bX_{\cdot,j}\}$ using the likelihood. From a sampling perspective, however, form of the likelihood obtained by integrating out the internal nodes $\Xb^I$, suggests an efficient two-stage sampling strategy that resembles the classical collapsed sampling \citep{10.2307/2290921} strategy in MCMC literature: first draw posterior samples of $(c,\sigma^2)$ and then proceed to draw posterior samples of $(\mathcal{T},\bm t)$ conditioned on each sample of $(c,\sigma^2)$.

\section{\texorpdfstring{\Rx-tree}{TEXT} Estimation and Posterior Inference}\label{sec::posterior_algorithm}
In line with the preceding discussion, we consider a two-stage sampler for Euclidean and tree parameters. While in principle MCMC techniques could be used in both stages, we propose to use a hybrid ABC-MH algorithm. Specifically, we use an approximate Bayesian computation (ABC) scheme to draw weighted samples of $(c,\sigma^2)$ in the first stage followed by a Metropolis-Hastings (MH) step  that samples $(\cT,\bm t)$ given ABC samples of $(c,\sigma^2)$ in the second stage. Motivation for using ABC in the first stage stems from: (i) availability of informative statistics; (ii) generation of better quality samples of the tree (compared to a single-stage MH); and (iii) better computational efficiency. We refer to Section \ref{sec::simulation_computational_aspect} for more details.

\subsection{Hybrid ABC-MH Algorithm}\label{sec:AlgOverivew}

ABC is a family of inference techniques that are designed to estimate the posterior density $\pr(\theta|\cD)$ of parameters $\theta$ given data $\cD$ when the corresponding likelihood $\pr(\cD|\theta)$ is intractable but fairly simple to sample from. Summarily, ABC approximates $\pr(\theta|\cD)$ by $\pr(\theta|\bm S_{obs})$ where $\bS_{obs}$ is a $d$-dimensional summary statistic that ideally captures most information about $\theta$. In the special case where $\bS_{obs}$ is a vector of sufficient statistics, it is well known that $\pr(\theta \mid \cD) = \pr(\theta \mid \bS_{obs})$. To generate a sample from the partial posterior distribution $\pr(\theta \mid \bS_{obs})$, ABC with rejection sampling proceeds by: (i) simulating $N^\syn$ values $\theta_l, l=1, \ldots, N^\syn$ from the prior distribution $\pr(\theta)$; (ii) simulating datasets $D_l$ from $\pr(\cD|\theta_l)$; (iii) computing summary statistics $\bS_{l},l=1, \ldots, N^\syn$ from $\cD_l$; (iv) retaining a subset of $\{\theta_{l_s},s=1\ldots,k\}$ of size $k<N^\syn$ that corresponds to `small' $\|\bS_{l_s}-\bS_{obs}\|$ values based on some threshold. Given pairs $\{(\theta_{l_s},\bS_{l_s})\}$, the task of estimating the partial posterior translates to a problem of conditional density estimation, e.g., based on Nadaraya-Waston type estimators and local regression adjustment variants to correct for the fact that $\bS_{l_s}$ may not be exactly $\bS_{obs}$; see \citet{sisson2018handbook} for a comprehensive review. To implement ABC, the choice of summary statistics is central. 

We detail the specialization of ABC to the marginal posterior distributions of $c$ and $\sigma^2$ in Section \ref{sec:stage1}. Given any pair of $(c,\sigma^2)$, we can sample trees from a density function up to an unknown normalizing constant based on an existing MH algorithm \citep{PYDT}. Our proposal is to condition on the posterior median of $(c,\sigma^2)$ of ABC-weighted samples from the first stage, when sampling the trees in the second stage; clearly, other choices are also available. This strategy produced comparable MAP trees and inference of other tree-derived results relative to tree samples based on full ABC samples of $c$ and $\sigma^2$.

Pseudo code for the two-stage algorithm is presented in the Supplementary Material Algorithm S1.  We briefly describe below its key components.
	
	\subsubsection{Stage 1: Sampling Euclidean Parameters \texorpdfstring{$(c,\sigma^2)$}{TEXT} using ABC}\label{sec:stage1}
	
	Accuracy and efficiency of the ABC procedure is linked to two competing desiderata on the summary statistics: (i) informative, or ideally sufficient; (ii) low-dimensional. 
	
	\textbf{Summary statistic for $\sigma^2$}. From the closed-form likelihood in Equation \eqref{eq:iidNorm}, a sufficient statistic of $\sigma^2\bm \Sigma^\cT$ is easily available, using which we construct a summary statistics for $\sigma^2$.
	\begin{lemma}\label{lemma1}
	With $\Xb$ as the observed data, the statistic $\bm{T}:=\sum_{j}\bX_{\cdot,j}\bX_{\cdot,j}^\transp$ is sufficient for $\sigma^2 {\bm \Sigma^{\mathcal{T}}}$ and follows a Wishart distribution $W_I(J,\sigma^2 \bm{\Sigma}^{\cT})$, where $\bX_{\cdot,j}=[x_{1j},\ldots,x_{Ij}] \in \mathbb{R}^I$. Then with $S^{(\sigma^2)}:=\frac{tr(\bm{T})}{IJ}$ we have $E[S^{(\sigma^2)}]=\sigma^2$ and $ Var[S^{(\sigma^2)}]=\frac{2\sigma^4tr(({\bm{\Sigma}^{\mathcal{T}}})^2)}{I^2J}$.
	\end{lemma}
	Due to the normality of $\Xb$ in \eqref{eq:iidNorm}, and the Factorization theorem \citep{CaseBerg}, we see that $\bm{T}$ is complete and sufficient for $\sigma^2 \bm \Sigma^\mathcal{T}$ and $\bm{T}\sim W_I(J,\sigma^2{\bm\Sigma^{\mathcal{T}}})$. Well-known results about the trace and determinant of $\Xb$ (see for e.g. \citet{doi:10.1080/03610928008827921}) provide the stated results on the mean and variance of $tr(\bm{T})$. Owing to its unbiasedness, we choose $S^{(\sigma^2)} ={tr(\bm{T})}/{IJ}$ as the summary statistic for $\sigma^2$ and examine its performance through simulations in Section \ref{sec:Simulation}; other choices are assessed in the Supplementary Material Section S4.1.

    \textbf{Summary statistic for $c$}. Based on the matrix normal distribution of Proposition \ref{prop1}, the divergence parameter $c$ does not appear in the observed data likelihood. Any statistic based on the entire observed data set $\Xb$ is sufficient, but not necessarily informative about $c$. In DDT, the prior distribution of the vector of branching times $\bm t$ is governed by divergence parameter $c$ via the divergence function $a(t;c)$. Thus an informative summary statistic for $c$ can be chosen by assessing its information about $\bm t$. For example, tighter observed clusters indicate small $c$ (e.g., $c<1$), where the level of tightness is indicated by the branch lengths from leaves to their respective parents. We construct summary statistics for $c$ based on a dendrogram estimated via hierarchical clustering of $\Xb$ based on pairwise distances $\delta_{i,i'}:=\|\bm X_{i}-\bm X_{i'}\|, i \neq i'$. The summary statistics $\bm{S}^{(c)}$ we choose is a ten-dimensional concatenated vector comprising the 10th, 25th, 50th, 75th and 90th percentiles of empirical distribution of: (i) $\delta_{i,i'}$; (ii) branch lengths associated with leaves of the dendrogram. Other candidate summary statistics for $c$ are examined in Supplementary Material Section S4.1.
    
    \subsubsection{Stage 2: Sampling Tree Parameters \texorpdfstring{$(\cT,\bm t)$}{TEXT} using Metropolis-Hastings}\label{sec:MH}
    For the second stage, we proceed by choosing a representative value $(c_0,\sigma^2_0)$ chosen from the posterior sample of $(c,\sigma^2)$, which in our case is the posterior median. Then a Metropolis-Hastings (MH) algorithm to sample from $\pr((\mathcal{T},\bm t)|c_0,\sigma^2_0, \Xb)$; recall that the \Rx tree is characterized by both the topology $\cT$ and divergence times $\bm t$. In particular, after initialization (e.g., the dendrogram obtained from hierarchical clustering), we first generate a candidate tree $(\mathcal{T}',\bm t')$ from the current tree $(\mathcal{T},\bm t)$ in two steps: (i) detaching a subtree from the original tree; (ii) reattaching the subtree back to the remaining tree. Acceptance probabilities for a candidate tree can be computed exactly and directly using the explicit likelihood in \eqref{eq:iidNorm}, without which they would have to be calculated iteratively \citep{Neal2003,PYDT}. See Supplementary Material Section S2.2 for details of the proposal function and the acceptance probabilities. 

\begin{remark}
In order to use the explicit likelihood in \eqref{eq:iidNorm} from Proposition \ref{prop1} to generate observed data $\Xb$, a tree-structured covariance $\bm \Sigma^{\mathcal T}$ needs to be specified, whose entries in-turn depend on the parameter $c$ through the divergence function. It is not straightforward to fix or sample a $\bm \Sigma^{\mathcal T}$ since its entries need to satisfy the inequalities \eqref{eq:ultraIneq}. It is easier to generate data $\Xb$ directly using the DDT generative mechanism in the ABC stage, and this is the approach we follow and is described in Supplementary Section S2. 

Summarily, there are three main advantages to using the explicit likelihood from Proposition \ref{prop1}: (i) decoupling of Euclidean and tree parameters to enable an efficient two-stage sampling algorithm; (ii) direct and exact computation of tree acceptance probabilities in MH stage; (iii) determination of informative sufficient statistic for $\sigma^2$ (Lemma \ref{lemma1}). 
\end{remark}

\subsection{Posterior Summary of \texorpdfstring{\Rx-Tree}{TEXT}, \texorpdfstring{$(\cT,\bm t)$}{TEXT}}\label{sec:PostTre}
While quantifying uncertainty concerning the tree parameters $(\cT,\bm t)$ is of main interest, we note that, from definition of the DDT, this is influenced by uncertainty in the model parameters. In particular, the first stage of ABC-MH produces weighted samples and we calculate the posterior median by fitting an intercept-only quantile regression with weights (see details in the Supplementary Material Section S2.1). For the \Rx-tree, we consider global and local tree posterior summaries that capture uncertainty in the latent hierarchy among all and subsets of treatments. 

Flexible posterior inference is readily available based on $L$ posterior samples of $(\cT,\bm t)$ from the MH step. It is possible to construct correspond tree-structured covariance matrices $\bm \Sigma^\cT$ from sample $(\cT,\bm t)$. Instead, we compute:
\begin{enumerate}[leftmargin=*]
    \item[(a)] a global \textit{maximum a posteriori} (MAP) estimate of the \Rx-tree that represents the overall hierarchy underlying the treatment responses;
    \item [(b)] local uncertainty estimates of co-clustering probabilities among a subset $\cA \subset \{1,\ldots,I\}$ of treatments based on posterior samples of the corresponding subset of divergence times. 
\end{enumerate}

\paragraph{Posterior co-clustering probability functions.} We elaborate on the local summary (b). Suppose $\cA=\{i,i',i''\}$ consists of three treatments. Given a tree topology $\cT$, note that at every $t \in [0,1]$ a clustering of all $I$ treatments is available and the clustering changes only at times $0<t_1<\cdots<t_{I-1}$. Consequently, for a given tree topology $\cT$ drawn from its posterior, we can compute for every level $t \in [0,1]$ a posterior probability that $i,i'$ and $i''$ belong to the same cluster. Such a posterior probability can be approximated using Monte Carlo on the $L$ posterior samples. Accordingly, we define the estimated posterior co-clustering probability (PCP) function associated with $\cA$ as,
\begin{align*}
\text{PCP}_\cA(t) & =\frac{\sum_{l=1}^L\mathbb{I}_{[0,t^{(l)}_{i,i',i''})}(t)}{L}, \label{eq:PCP}
\end{align*}
where $\mathbb{I}_B$ is the indicator function on the set $B$ and $t_{i,i',i''}^{(l)}$ is the divergence time of $\cA=\{i,i',i''\}$ in the $l$-th tree sample. Essentially, the $\text{PCP}_\cA(t)$ can be viewed as the proportion of tree samples with $\{i,i',i''\}$ having the most recent common ancestor later than $t$.

For every subset $\cA$, the function $[0,1] \ni t \mapsto \text{PCP}_\cA(t) \in [0,1]$ is non-increasing starting at 1 and ending at 0, and reveals propensity among treatments in $\cA$ to cluster as one traverses down an (estimate of) \Rx-tree starting at the root: a curve that remains flat and drops quickly near 1 indicates higher relative similarity among the treatments in $\cA$ relative to the rest of the treatments. A scalar summary of $\text{PCP}_\cA(t)$ is the area under its curve known as integrated PCP $\text{iPCP}_\cA$, which owing to the definition of $\text{PCP}_\cA(t)$, can be interpreted as the expected (or average) chance of co-clustering for treatments in $\cA$. 

\begin{figure}[!htb]
        \includegraphics[width=0.95\linewidth]{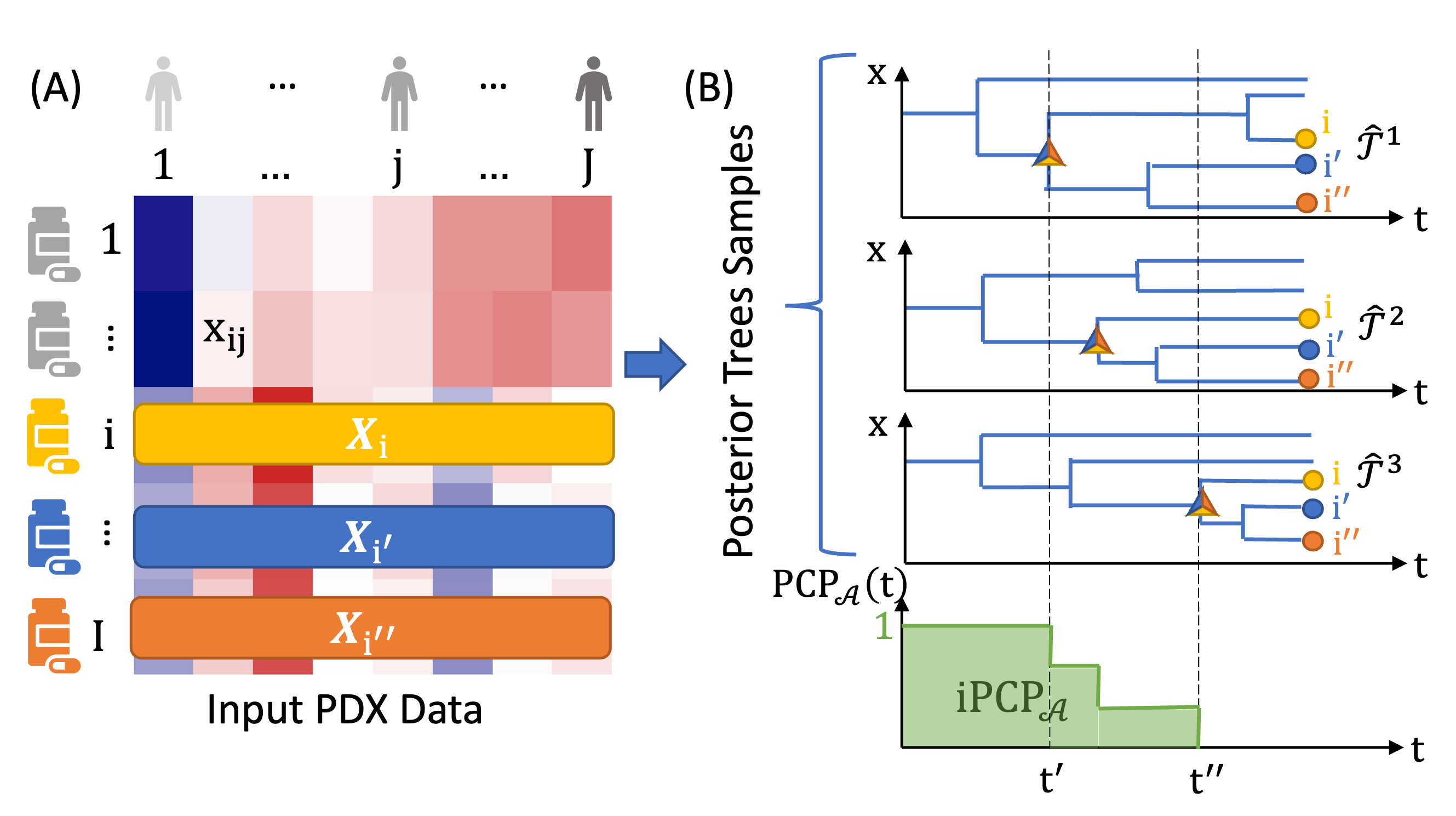}
        \caption{Posterior tree summaries. (A) The input PDX data with $I$ treatments and $J$ patients, and treatments $\cA=\{i,i',i{''}\}$ are of interest. (B) $\text{PCP}_\cA(t)$ and $\text{iPCP}_\cA$ for treatments $\cA$ based on $L=3$ posterior trees. The relevant divergence times are represented by a ``$\triangle$'' in each posterior tree sample. For example, at time $t'$, the treatments in $\cA$ diverge in one out of the three trees. Because $\text{PCP}_\cA(t')$ is defined by the proportion of posterior tree samples in which $\cA$ has \textit{not} diverged up to and including $t'$,  it drops from $1$ to $2/3$.}
	    \label{fig:iPCP}
    \end{figure}

Figure \ref{fig:iPCP} illustrates an example of a three-way $\text{iPCP}_\cA$ with $\cA=\{i,i',i{''}\}$ for a PDX data with $I$ treatments and $J$ patients (Figure \ref{fig:iPCP}(A)). Given $L=3$ posterior trees samples (Figure \ref{fig:iPCP}(B)) drawn from the PDX data, we first calculate the whole $\text{PCP}_\cA(t)$ function by moving the time $t$ from $0$ to $1$. Starting from time $t=0$, no treatment diverges at time $t=0$ and the $\text{PCP}_\cA(t)$ is $1$. At time $t'$, treatments diverge in one out of the three posterior trees and $\text{PCP}_\cA(t)$ therefore drops from $1$ to $2/3$. Moving the time toward $t=1$, treatments diverge in all trees and the $\text{PCP}_\cA(t)$ drops to $0$. The $\text{iPCP}_\cA$ then can be obtained by the area under the $\text{PCP}_\cA(t)$. 

\begin{remark}
In the special case of $\cA=\{i,i'\}$ for two treatments, the definition of iPCP$_\cA$ can be related to the cophenetic distance \citep{cophenetic_1962, pmid23323711} and, moreover, extends definition of the cophenetic distance to multiple trees. Given two treatments $i$ and $i'$ in a single tree, let $t_{d}$ be the time at which their corresponding Brownian paths diverge. Then $\text{PCP}_\cA(t)=\mathbb{I}_{[0,t_d)}(t)$  and $\text{iPCP}_\cA=t_d$; this implies that the cophenetic distance is $2(1-{t_d})$ and thus $\text{iPCP}_\cA$ and the cophenetic distances uniquely determines the same tree structure. For $L>1$ trees, a Carlo average of divergence times of $L$ trees leads to the corresponding $\text{iPCP}_\cA$. 
\end{remark}
\begin{remark}
Given $I$ treatments, since pairwise cophenetic distances from one tree determines a tree \citep{Lapointe1991, McCullagh2006StructuredCM}, one might consider summarizing and represent posterior trees in terms of an $I \times I$ matrix $\bm \Sigma$ consisting of entries $\text{iPCP}_{\{i,i'\}}$ for every pair of treatments of $(i,i')$, estimated from the posterior sample of trees. However, $\bm \Sigma$ need not to be a tree-structured matrix that uniquely encodes a tree. It is possible to project $\bm \Sigma$ on to the space of tree-structured matrices (see for e.g., \citet{pmid22081761}) but the projection might result in a non-binary tree structure. We discuss this issue and its resolution in  Supplementary Material Section S3. 
\end{remark}

\section{Simulations}\label{sec:Simulation}

Accurate characterization of similarities among any subset of treatments is central to our scientific interest in identifying the promising treatment subsets for further investigation. In addition, we have introduced a two-stage algorithm to improve our ability to efficiently draw tree samples from the posterior distribution (similarly for the Euclidean parameters). To demonstrate the modeling and computational advantages, we conduct two sets of simulations. The first simulation shows that the proposed model estimates the similarity (via iPCP) better than alternatives, even when the true data generating mechanisms deviate from DDT assumptions in terms of the form of divergence function, prior distribution for the unknown tree, and normality of the responses. The second simulation illustrates the computational efficiency of the proposed two-stage algorithm in producing higher quality posterior samples of Euclidean parameters, resulting in more accurate subsequent estimation of an unknown tree and iPCPs, two key quantities to our interpretation of real data results.

\subsection{Simulation I: Estimating Treatment Similarities}
\label{sec::simulation_similarity_aspect}
We first show that iPCPs estimated by DDT are closer to the true similarities (operationalized by functions of elements in the true divergence times in $\bm{\Sigma}^\cT$) under different true data generating mechanisms that may follow or deviate from the DDT model assumptions in three distinct aspects (the form of divergence function, the prior distribution over the unknown tree, and normality). 

\paragraph{Simulation setup.} 
We simulate data by mimicking the PDX breast cancer data (see Section \ref{sec:DataAnalysis}) with $I=20$ treatments and $J=38$ patients. We set the true scale parameter as the posterior median $\sigma_0^2$ and the true tree $\cT_0$ as the MAP tree that are estimated from the breast cancer data; We consider four scenarios to represent different levels of deviation from the DDT model assumptions:
    \begin{itemize}
    \item[(i)] No deviation of the true data generating mechanism from the fitted DDT models: given $\sigma^2_0$ and $\cT_0$, simulate data based on the DDT marginal data distribution (Equation \eqref{eq:iidNorm});
    
    \noindent The true data generating mechanism deviates from the fitted DDT in terms of:
    \item[(ii)] divergence function: same as in (i), but the true tree is a random tree from DDT with misspecified divergence function, $a(t;r)=\frac{r}{(1-t)^2}, r=0.5$;
    \item[(iii)] prior for tree topology: same as in (i), but the true tree is a random tree from the coalescence model (generated by function \verb"rcoal" in R package \verb"ape"), and,
    \item[(iv)] marginal data distribution: same as in (i), but the marginal likelihood is a centered multivariate $t$ distribution with degree-of-freedom four and scaled by $\sigma^2_0\bSigma^{\mathcal{T}_0}$.
    \end{itemize}
For each of four true data generating mechanisms above, we simulate $B=50$ replicate data sets. In the following, we use the DDT model and the two-stage algorithm for all estimation regardless of the true data generating mechanisms. For DDT, we ran the two-stage algorithm where the second stage is implemented with five parallel chains. For each chain, we ran $10,000$ iterations, discarded first $9,000$ trees and combined five chains with a total of $5,000$ posterior tree samples. 
	
First, we compute the iPCPs for all pairs of treatment combinations following the definition of $\textrm{iPCP}_\cA$ where $\cA = \{i,i'\}, 1\leq i<i'\leq I$. Two alternative approaches to defining and estimating similarities between treatments are considered: (i) similarity derived from agglomerative hierarchical clustering, and (ii) empirical Pearson correlation of the two vectors of responses $\bX_{i}$ and $\bX_{i'}$, for $i\neq i'$. In particular, for (i), we considered five different linkage methods (Ward, Ward's D2, single, complete and Mcquitty) with Euclidean distances. Given an estimated dendrogram from hierarchical clustering, the similarity for a pair of treatments is defined by first normalizing the sum of branch lengths from the root to leaf as $1$, and then calculating the area under of the co-clustering curve (AUC) obtained by cutting the dendrogram at various levels from 0 to 1. For three- or higher-way comparisons, (i) can still produce an AUC based on a dendrogram obtained from hierarchical clustering, while the empirical Pearson correlation in (ii) is undefined hence not viable as a comparator beyond assessing pairwise treatment similarities.

\paragraph{Performance metrics.} 
For treatment pairs $\cA = \{i,i'\}$, to assess the quality of estimated treatment similarities for each of the methods above (DDT-based, hierarchical-clustering-based, and empirical Pearson correlation), we compare the estimated values against the true branching time $\Sigma_{i,i'}^{\cT_0}$; similarly when assessing recovery of three-way treatment similarities, e.g., $\cA=\{i,i',i''\}$, $\Sigma^{\cT_0}_{i,i',i''}$ is defined as the time when $\{i,i',i''\}$ first branches in the true tree $\cT_0$. In particular, for replication data set $b=1, \ldots, B$, let $\hat{\Sigma}^{(b)}_{i,i'}$ generically represent the pairwise similarities for treatment subsets $(i,i')$ that can be based on DDT, hierarchical clustering or empirical pairwise Pearson correlation. For three-way comparisons, let $\hat{\Sigma}^{(b)}_{i,i',i''}$ generically represent the three-way similarities for treatment subset $(i,i',i'')$ that can be based on DDT, or hierarchical clustering.

We assess the goodness of recovery by computing $\sqrt{\sum_{i,i'}(\hat{\Sigma}^{(b)}_{i,i'} - \Sigma^{\cT_0}_{i,i'})^2}$, the Frobenious norm of the matrix in recovering the entire $\Sigma^{\cT_0}$.  We compute $\max_{i,i',i''} |\hat{\Sigma}^{(b)}_{i,i',i''} - \Sigma^{\cT_0}_{i,i',i''}|$, the max-norm of the matrix in recovering the true three-way similarities.  For a given method and treatment subset $\cA$, the above procedure results in $B$ values, the distribution of which can be compared across methods; smaller values indicate better recovery of the true similarities.

Alternatively, for each method and each treatment subset, we also compute the Pearson correlation between the estimated similarities and the true branching times across replicates for pairwise or three-way treatment subsets:  \(\widehat{\text{Cor}}\left((\hat{\Sigma}^{(b)}_{i,i'},\Sigma^{\cT_0}_{i,i'}), b=1, \ldots, B=50\right)\), for treatments $i<i'$ and \(\widehat{\text{Cor}}\left((\hat{\Sigma}^{(b)}_{i,i',i''},\Sigma^{\cT_0}_{i,i',i''}), b=1, \ldots, B=50\right)\), for treatments $i<i'<i''$. We refer to this metric as ``Correlation of correlations'' (the latter uses the fact that the entries in the true $\Sigma^{\cT_0}$ being correlations; see Equation (\ref{eq:iidNorm})); higher values indicate better recovery of the true similarities.

\begin{figure}[!htb]
    \centering
    \includegraphics[width=\linewidth]{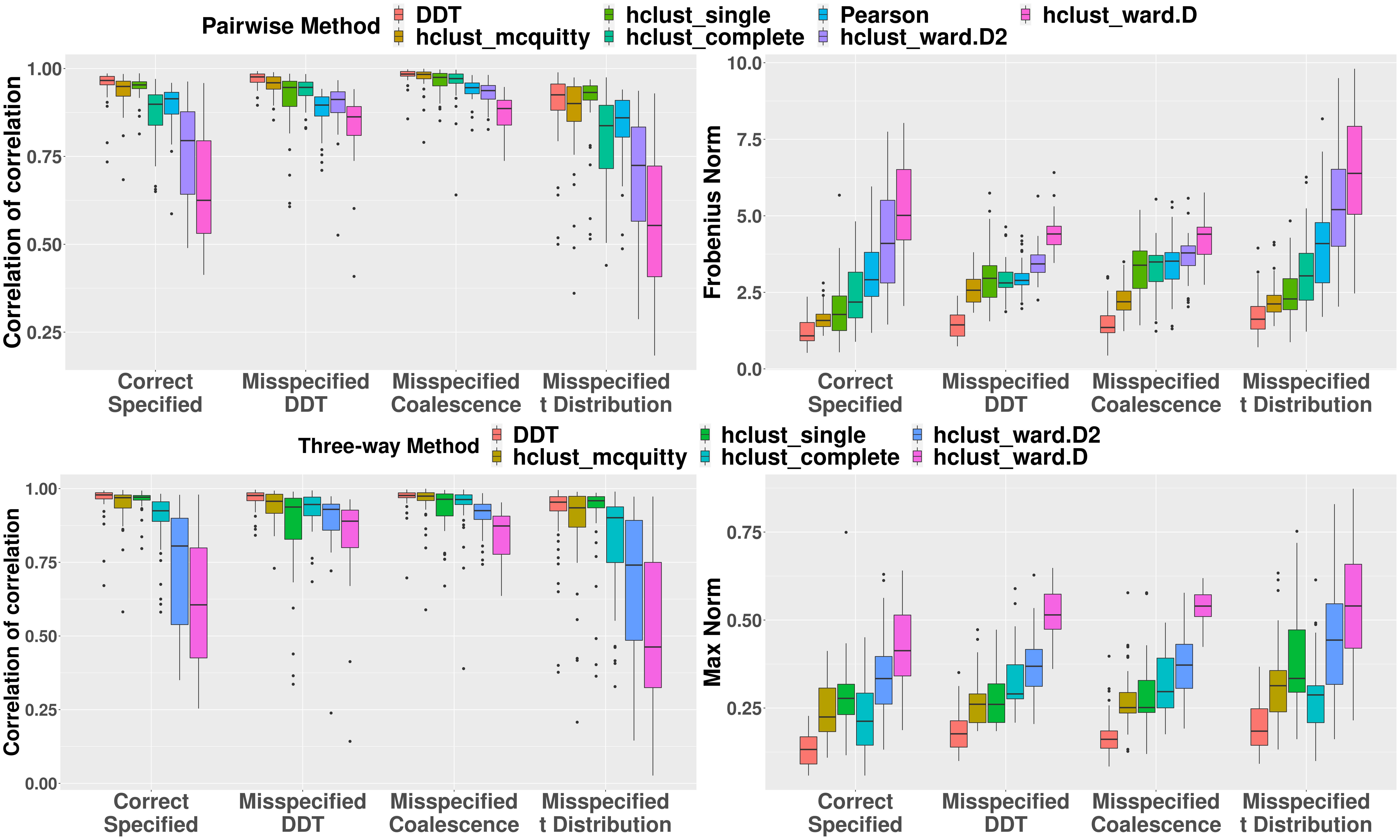}
    \caption{Simulation studies for comparing the quality of estimated treatment similarities based on DDT, hierarchical clustering, and empirical Pearson correlation. Two performance metrics are used: (Left) Correlation of correlation (higher values are better); (Right) Matrix distances with Frobenius norm for pairwise similarity and max norm for three-way similarity (lower values are better). DDT captures both true pairwise (upper panels) and three-way (lower panels) similarity best under four levels of misspecification scenarios. 
    }
    \label{fig:PCP}
    \end{figure}

\paragraph{Simulation results.} We observe that DDT better estimates the treatment similarities even under misspecified models. In particular, under scenarios where the true data generating mechanisms deviate from the fitted DDT model assumptions (ii-iv), the DDT captures the true pairwise and three-way treatment similarities the best by higher values in correlation of correlations (left panels, Figure \ref{fig:PCP}) and lower matrix/array distances (right panels, Figure \ref{fig:PCP}). In particular, the fitted DDT with divergence function $a(t)=c/(1-t)$ under Scenario i, ii and iii performed similarly well indicating the relative insensitivity to the DDT modeling assumptions with respect to divergence function and the tree generative model. Under Scenario iv where the marginal likelihood assumption deviates from Gaussian with heavier tails, the similarity estimates from all methods deteriorate relative to Scenarios i-iii. Comparing between methods, the similarities derived from hierarchical clustering with single linkage is comparable to DDT model when evaluated by correlation of correlation, but worse than DDT when evaluated by the matrix norm.

\subsection{Simulation II: Comparison with Single-Stage MCMC Algorithms}\label{sec::simulation_computational_aspect}

We have also conducted extensive simulation studies that focus on the computational aspect of the proposed algorithms and demonstrate the advantage of the proposed two-stage algorithm in producing higher quality posterior samples of the unknown tree than classical single-stage MCMC algorithms. In particular, we demonstrate that the proposed algorithm produces (i) MAP trees that are closer to the true tree than alternatives (hierarchical clustering, single-stage MH with default hierarchical clustering or the true tree at initialization) and (ii) more accurate estimation of pairwise treatment similarities compared to single-stage MCMC algorithms. See Supplementary Material Section S5 for further details. 

\paragraph{Additional simulations and sensitivity analyses.} Aside from the simulations above focusing on the tree structure and the divergence time, Supplementary Material S4 offers additional details for Euclidean parameters including the parameter inference and algorithm diagnostics.  In particular, we empirically show that current ${\bm S}^{(c)}$ and $S^{(\sigma^2)}$ outperform other candidate summary statistics in terms of bias in Section S4.1. In Section S4.2, we present additional simulation results that demonstrate that the two-stage algorithm (i) enjoys stable effective sample size (ESS) for $(c,\sigma^2)$; (ii) leads to similar or better inference on $(c,\sigma^2)$, as ascertained using credible intervals. Section S4.3 checks the convergence of MH and the goodness of fit for ABC .

\section{Treatment Trees in Cancer using PDX Data} \label{sec:DataAnalysis}

\subsection{Dataset Overview and Key Scientific Questions} 
We leverage a recently collated PDX dataset from the \ul{N}ovartis \ul{I}nstitutes for \ul{B}ioMedical \ul{R}esearch - \ul{PDX} \ul{E}ncyclopedia [NIBR-PDXE, \citep{pmid26479923}] that interrogated multiple targeted therapies across different cancers and established that PDX systems provide a more accurate measure of the response of a population of patients than traditional preclinical models. Briefly, the NIBR-PDXE consists of $>1,000$ PDX lines across a range of human cancers and uses a $1\times 1 \times 1$ design (one animal per PDX model per treatment); i.e., each PDX line from a given patient was treated simultaneously with multiple treatments allowing for direct assessments of treatment hierarchies and responses. In this paper, we focus on our analyses on a subset of PDX lines with complete responses across five common human cancers: Breast cancer (BRCA), Cutaneous Melanoma (CM, skin cancer), Colorectal cancer (CRC), Non-small Cell Lung Carcinoma (NSCLC), and Pancreatic Ductal Adenocarcinoma (PDAC). After re-scaling data and missing data imputation, different numbers of treatments, $I$, and PDX models, $J$, presented in the five cancers were, $(I,J)$: BRCA, $(20,38)$; CRC, $(20,40)$; CM, $(14,32)$; NSCLC, $(21,25)$; and PDAC, $(20,36)$. (See Supplementary Material Table S7 for treatment names and Section S6.1 for details of pre-processing procedures.) 

In our analysis, we used the best average response (BAR) as the main response, by taking the untreated group as the reference group and using the tumor size difference before and after administration of the treatment(s) following \citet{doi:10.1080/01621459.2020.1828091}. Positive values of BAR indicate the treatment(s) shrunk the tumor more than the untreated group with higher values indicative of (higher) treatment efficacy. The treatments included both drugs administered individually with established mechanisms (referred to as ``monotherapy'') and multiple drugs combined with potentially unknown synergistic effects (referred to as ``combination therapy''). Our key scientific questions were as follows: (a) identify plausible biological mechanisms that characterize treatment responses for monotherapies within and between cancers; (b) evaluate the effectiveness of combination therapies based on biological mechanisms. Due to a potentially better outcome and lower resistance, combination therapy with synergistic mechanism is highly desirable \citep{pmid28410237}. 

\noindent \underline{DDT model setup.}  
For all analyses we followed the setup in the Section \ref{sec::simulation_similarity_aspect} and obtained $N^{\syn}=600,000$ synthetic datasets from the ABC algorithm (Section \ref{sec:stage1}) with prior $c\sim {\sf Gamma}(2,2)$ and $1/\sigma^2\sim {\sf Gamma}(1,1)$ and took the first 0.5\% ($d=0.5\%$) closest data in terms of $\bm{S}^{(c)}$ and $S^{(\sigma^2)}$. We calculated the posterior median of $(c,\sigma^2)$ as described in Section \ref{sec:PostTre}. For the second-stage MH, we ran five chains of the two-stage algorithm with $(c,\sigma^2)$ fixed at the posterior median by 10,000 iterations and discarded the first 9,000 trees, which resulted in 5,000 posterior trees in total. Finally, we calculated the \Rx-tree (MAP) and iPCP based on 5,000 posterior trees for all subsequent analyses and interpretations. All computations were divided on multiple different CPUs (see the Supplementary Table S5 for the full list of CPUs). For the BRCA data with $I=20$ and $J=38$, we divided the ABC stage into $34$ compute cores with a total of $141$ CPU hours and maximum $4.7$ hours in real time. For the MH stage and the single-stage MCMC, we split the computation on $5$ compute cores with a total of $8.6$ and $12$ CPU hours, and a maximum $1.7$ and $2.5$ hours in real time, respectively. 

Our results are organized as follows: we provide a summary of the \Rx-tree estimation and treatment clusters in Section \ref{sec:rxtrees} followed by specific biological and translational interpretations in Sections \ref{sec:MonoTherapy} and \ref{sec:CmbTherapy} for monotherapy and combination therapy, respectively. Additional results can be accessed and visualized using our companion \texttt{R}-shiny application (see Supplementary Material Section S6.4 for details). 

\subsection{\texorpdfstring{\Rx-tree}{TEXT} Estimation and Treatment Clusters } \label{sec:rxtrees}
\begin{figure}[!htb]
        \centering
        \includegraphics[width=\linewidth]{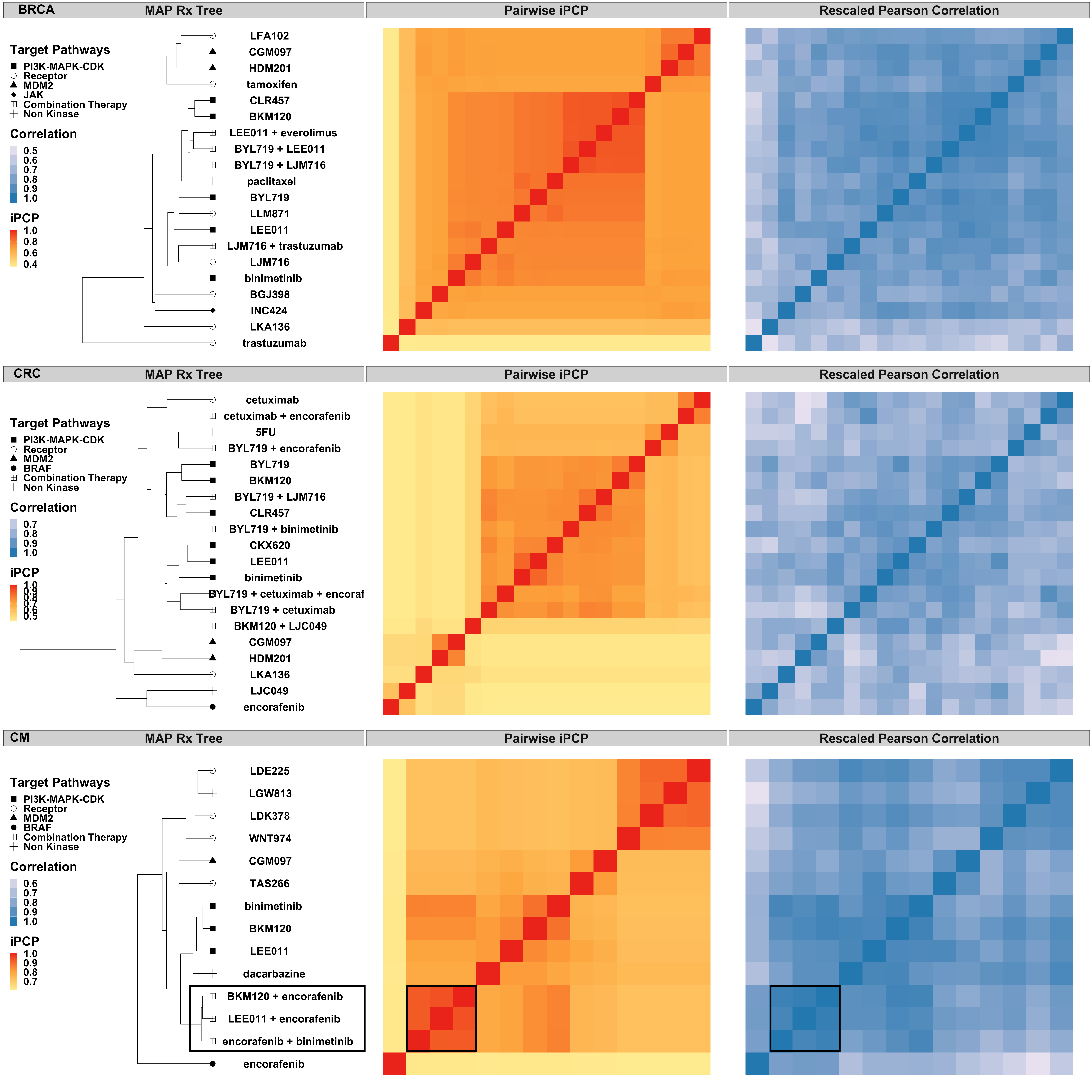}
    \caption{The \Rx-tree and iPCP for breast cancer (BRCA, top row), colorectal cancer (CRC, middle row) and melanoma (CM, lower row). Three panels in each row represent: (left) estimated \Rx-tree (MAP); distinct external target pathway information is shown in distinct shapes for groups of treatments on the leaves; (middle) estimated pairwise iPCP, i.e., the posterior mean divergence time for pairs of entities on the leaves (see the result paragraph for definition for any subset of entities); (right) scaled Pearson correlation for each pair of treatments. Note that the MAP visualizes the hierarchy among treatments; the iPCP is not calculated based on the MAP, but based on posterior tree samples (see definition in Section \ref{sec:PostTre}) 
    }
    \label{fig:ggtree}
    \end{figure}

We focus our discussion on three cancers: BRCA, CRC and CM here -- see Supplementary Materials Section S6.3 for NSCLC and PDAC. In Figure \ref{fig:ggtree}, \Rx-tree, pairwise iPCP and (scaled) Pearson correlation are shown in the left, middle and right panels, respectively. Focusing on the left two panels, we observe that the \Rx-tree and the pairwise iPCP matrix show the similar clustering patterns. For example, three combination therapies in CM form a tight subtree and are labeled by a box in the \Rx-tree of Figure \ref{fig:ggtree} and a block with higher values of iPCP among three combination therapies also shows up in the corresponding iPCP matrix with a box labeled. In our analysis, the treatments predominantly target six oncogenic pathways that are closely related to the cell proliferation and cell cycle: (i) phosphoinositide 3-kinases, PI3K; (ii) mitogen-activated protein kinases, MAPK; (iii) cyclin-dependent kinases, CDK; (iv) murine double minute 2, MDM2; (v) janus kinase, JAK; (vi) serine/threonine-protein kinase B-Raf, BRAF. We label targeting pathways above for monotherapies with solid dots and further group PI3K, MAPK and CDK due to the common downstream mechanisms \citep[e.g.,][]{pmid29547722, pmid30974877}. Roughly, the \Rx-tree from our model clusters monotherapies targeting oncogenic processes above and largely agrees with common and established  biology mechanisms. For example, all PI3K-MAPK-CDK inhibitors (solid square) belong to a tighter subtree in three cancers; two MDM2 monotherapies (solid triangle) are closest in both BRCA and CRC. While visual inspection of the MAP \Rx-tree agrees with known biology, iPCP further quantifies the similarity by assimilating the information across multiple trees from our MCMC samples. For the ensuing interpretations in Sections  \ref{sec:MonoTherapy} and \ref{sec:CmbTherapy}, we focus on iPCP and verify our model through monotherapies with known biology, since our a priori hypothesis is that monotherapies that share the same downstream pathways should exhibit higher iPCP values. Furthermore, we extend our work to identify combination therapies with synergy and discover several combination therapies for each cancer.

\subsection{Biological Mechanisms in Monotherapy}\label{sec:MonoTherapy}
Our estimation procedure exhibits a high level of concordance between known biological mechanisms and established monotherapies for multiple key signalling pathways. From the \Rx-tree in Figure \ref{fig:ggtree}, aside from the oncogenic process (solid dots) introduced above, monotherapies also target receptors (hollow circles) or other non-kinase targets (e.g. tubulin; crosses). We summarize our key findings and interpretations for each of signaling pathways and their regulatory axes, namely, PI3K-MAPK-CDK and MDM2 from cell cycle regulatory pathways, human epidermal growth factor receptor 3 (ERBB3) from receptor pathways, and tubulin from non-kinase pathways along with their implications in monotherapy across different cancers.

\begin{figure}[!htb]
        \centering
        \includegraphics[width=\linewidth]{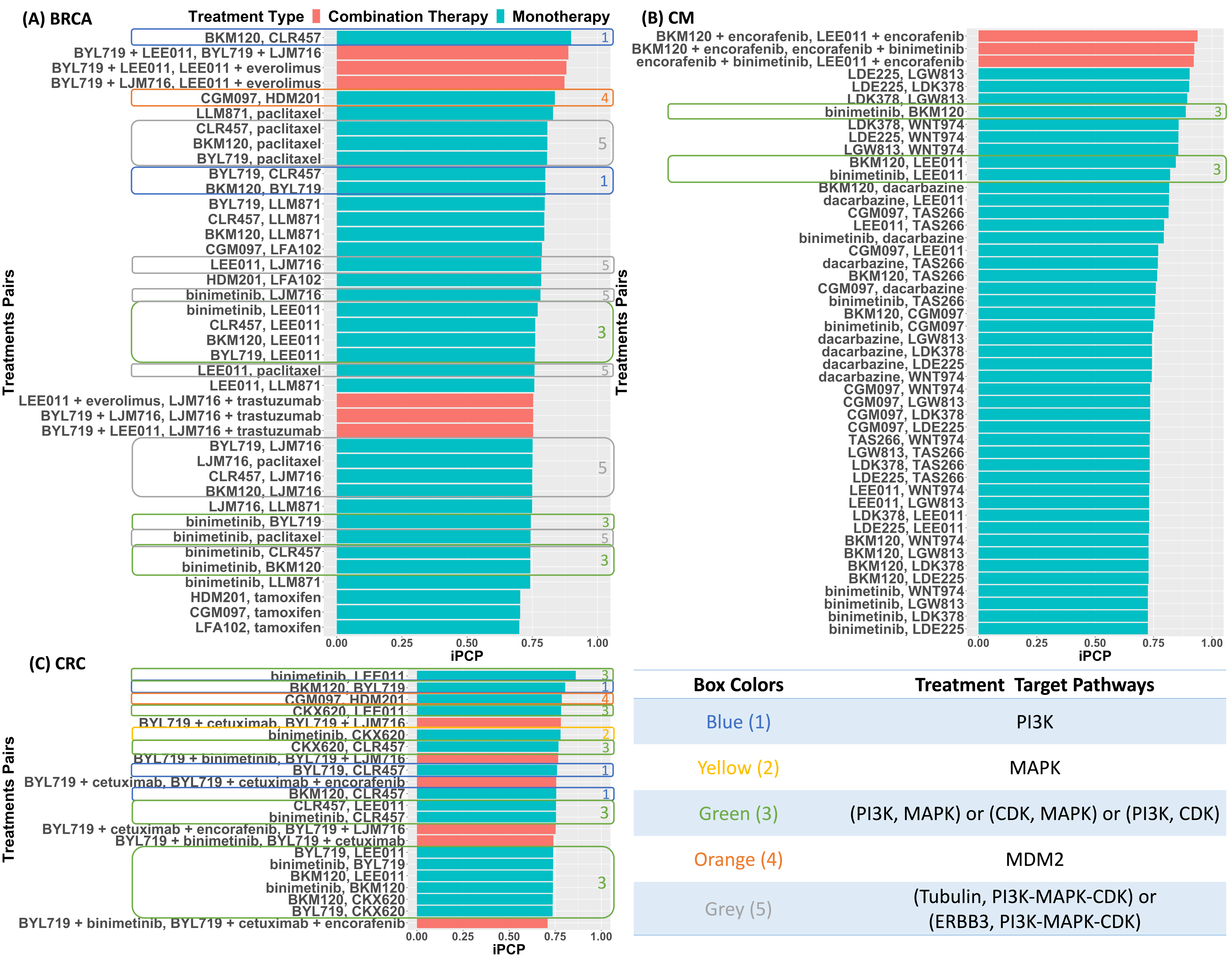}
    \caption{Bar plot of iPCPs for pairs of combination therapies (red bars) and pairs of monotherapies (green bars): (A) breast cancer, (B) melanoma and (C) colorectal cancer. The bar plots are sorted by the iPCP values (high to low); pairs of treatments are shown only if the estimated iPCP is greater than $0.7$. Monotherapies have different known targets which are listed in the bottom-right table (see Section \ref{sec:MonoTherapy} for more details and discussion  on monotherapies).
    }
    \label{fig:pairIPCP} 
    \end{figure}

\noindent \underline{PI3K-MAPK-CDK inhibitors.}
For treatments targeting PI3K, MAPK and CDK, treatments have the same target share high iPCP. In the NIBR-PDXE dataset, three PI3K inhibitors (BKM120, BYL719 and CLR457), two MAPK inhibitors (binimetinib and CKX620) and one CDK inhibitor (LEE011) were tested, but different cancers contain different numbers of treatments. Specifically, all three PI3K inhibitors present in BRCA and CRC, but only BKM120 is tested in CM; CRC contains two MAPK inhibitors while BRCA and CM only have binimetinib; LEE011 is tested in all three cancers. In Figure \ref{fig:pairIPCP}, BKM120, BYL719 and CLR457 share high pairwise iPCPs (box (1)) and all target PI3K for BRCA and CRC (BRCA, (BKM120, CLR457): 0.8986, (BKM120, BYL719): 0.8002, (BYL719, CLR457): 0.8002; CRC, (BKM120, CLR457): 0.7555, (BKM120, BYL719): 0.8041, (BYL719, CLR457): 0.7597); MAPK (box (2)) inhibitors, binimetinib and CKX620, show a high pairwise iPCP in CRC (0.7792). Asides from the pairwise iPCPs, our model also suggests high multi-way iPCPs among PI3K inhibitors in BRCA (0.8002) and CRC (0.7513). Among these inhibitors, PI3K inhibitor of BYL719 was approved by FDA for breast cancer; MAPK inhibitor of binimetinib was approved by FDA for BRAF mutant melanoma in combination with encorafenib; and CDK inhibitor of LEE011 was approved for breast cancer. 
	
Our model suggests treatments targeting different pathways also share high iPCP values across different cancers. Monotherapies targeting different cell cycle regulatory pathways (PI3K, MAPK and CDK) exhibit high iPCPs. CDK inhibitor, LEE011, and MAPK inhibitors share high pairwise iPCP values in BRCA ((LEE011, binimetinib): 0.7709), CRC ((LEE011, binimetinib): 0.8617, (LEE011, CKX620): 0.7820) and CM ((LEE011, binimetinib): 0.8210) in the Figure \ref{fig:pairIPCP} with box (3). High iPCP among MAPK and CDK inhibitors agree with biology, since it is known that CDK and MAPK collaboratively regulate downstream pathways such as Ste5 \citep{pmid29547722}. High pairwise iPCP values between PI3K and MAPK inhibitors were observed in box (3) in the Figure \ref{fig:pairIPCP}. Specifically, our model suggests high pairwise iPCPs as follows: (i) BRCA, (binimetinib, BKM120): 0.7427, (binimetinib, BYL719): 0.7441, (binimetinib, CLR457): 0.7427)); (ii) CRC, (binimetinib, BKM120): 0.7374, (binimetinib, BYL719): 0.7388, (binimetinib, CLR457): 0.7541, (CKX620, BKM120): 0.7366, (CKX620, BYL719): 0.7357, (CKX620, CLR457): 0.7676)); (iii) CM, (binimetinib, BKM120): 0.8882. Aside from the pairwise iPCPs above, high multi-way iPCPs in BRCA (0.7422), CRC (0.7300) and CM (0.8882) also show the similar information. From the existing literature, both PI3K and MAPK can be induced by ERBB3 phosphorylation \citep{pmid22178756} and it is not surprising to see high iPCPs between PI3K and MAPK inhibitors.

\noindent \underline{ERBB3 and tubulin inhibitors.}
Our model also found high iPCP values among ERBB3, tubulin and PI3K-MAPK-CDK inhibitors in BRCA. ERRB3 inhibitor, LJM716, exhibits high pairwise iPCP values with PI3K (BKM120: 0.7501, BYL719: 0.7513, CLR457: 0.7500), MAPK (binimetinib: 0.7811), CDK (LEE011: 0.7847) and tubulin (paclitaxel: 0.7505) inhibitors in the Figure \ref{fig:pairIPCP} with box (5). Since PI3K and MAPK are downstream pathways of ERBB3 \citep{pmid22178756} and CDK works closely with PI3K and MAPK \citep{pmid30974877, pmid29547722}, high iPCPs between ERBB3 inhibitor and PI3K-MAPK-CDK inhibitors are not surprising. For ERBB3 and tubulin, ERBB3 is a critical regulator of microtubule assembly \citep{pmid33583260} and tubulin plays an important role in building microtubules. Since microtubules form the skeletons of cells and are essential for cell division \citep{pmid25788699,pmid31353978}, tubulin inhibitor, paclitaxel, kills cancer cell by interfering cell division and is an FDA-approved treatment. In congruence with the above results, tubulin inhibitor paclitaxel also shares high iPCPs with PI3K (BKM120: 0.8076, BYL719: 0.8063, CLR457: 0.8076), MAPK (binimetinib: 0.7433), CDK (LEE011: 0.7587) and ERBB3 (LJM716: 0.7505) in the Figure \ref{fig:pairIPCP} with box (5). In addition, another CDK4 inhibitor BPT also inhibits tubulin \citep{pmid25950473} and PI3K inhibitor BKM120 inhibits the formation of microtubule \citep{pmid28276440}. Both offer additional reasons for high iPCP between tubulin and PI3K-MAPK-CDK inhibitors.

\noindent \underline{MDM2 inhibitors.} 
We found two drugs: CGM097 and HDM201 share high iPCP values in BRCA (0.8365) and CRC (0.7860) in the Figure \ref{fig:pairIPCP} with box (4). Since CGM097 and HDM201 target the same pathway, MDM2, high iPCPs suggest a high similarity between CGM097 and HDM201 and show consistent results between our model and underlying biological mechanism. MDM2 negatively regulates the tumor suppressor, p53 \citep{pmid24389645} and if MDM2 is suppressed by inhibitors, p53 is able to prevent tumor formation. Both CGM097 and HDM201 entered phase I clinical trial \citep{pmid32651541} for wild-type p53 solid tumors and leukemia, respectively.

\subsection{Implications in Combination Therapy}\label{sec:CmbTherapy}
Based on the concordance between the monotherapy and biology mechanism, we further investigate combination therapies to identify mechanisms with synergistic effect. In NIBR-PDXE, $21$ combination therapies were tested and only one of them includes three monotherapies (BYL719 + cetuximab + encorafenib in CRC) and the rest contain two monotherapies. Out of $21$ combination therapies, only three do not target any cell cycle (PI3K, MAPK, CDK, MDM2, JAK and BRAF) pathways (see Supplementary Material Table S8 for the full list of combination therapies). From the \Rx-tree in Figure \ref{fig:ggtree}, combination therapies tend to form a tighter subtree and are closer to monotherapies targeting PI3K-MAPK-CDK, which implies that the mechanisms under combination therapies are similar to each other and are closer to the PI3K-MAPK-CDK pathways. We identified several combination therapies with known synergistic effects and  provide a brief description for  each of the cancers in the following paragraphs.
	
\noindent \underline{Breast cancer.}
Four combination therapies were tested in BRCA and three therapies targeting PI3K-MAPK-CDK (BYL719 + LJM716, BYL719 + LEE011 and LEE011 + everolimus) form a subtree in \Rx-tree with a high three-way iPCP (0.8719). Among these combination therapies, PI3K-CDK inhibitor, BYL719 + LEE011, suggests a possible synergistic regulation \citep{pmid25002028, pmid28704762, pmid31101835}. Based on the high iPCP between BYL719 + LEE011 and the rest two therapies, we suggest synergistic effect for combination therapies targeting PI3K-ERBB3 (BYL719 + LJM716), and CDK-MTOR (LEE011 + everolimus) for future investigation. 

\noindent \underline{Colorectal cancer.} 
Our model suggests a high three-way iPCP (0.7437) among PI3K-EGFR (BYL719 + cetuximab), PI3K-EGFR-BRAF (BYL719 + cetuximab + encorafenib) and PI3K-ERBB3 (BYL719 + LJM716) inhibitors. Since the triple therapy (BYL719 + cetuximab + encorafenib) enters the phase I clinical trial with synergy \citep{CRC_tripleCmb}, our model proposes the potential synergistic effect for PI3K-ERBB3 based on iPCP for future investigation. Of note, we found a modest iPCP (0.6280) between the FDA-approved combination therapy EGFR-BRAF (cetuximab + encorafenib) and PI3K-EGFR-BRAF (BYL719 + cetuximab + encorafenib) and the modest iPCP can be explained by an additional drug-drug interaction between BYL719 and encorafenib in triple-combined therapy \citep{pmid28363909}. 
	
\noindent \underline{Melanoma.} 
In NIBR-PDXE, three combination therapies were tested in CM, and all of them consist one monotherapy targeting PI3K-MAPK-CDK and the other one targeting BRAF. A tight subtree is observed in the \Rx-tree and our model also suggests a high iPCP (0.9222) among three combination therapies. Since PI3K, MAPK and CDK work closely and share a high iPCP (0.8204) among monotherapies in CM, a high iPCP (0.9222) among three combination therapies is not surprising. Since two combination therapies of BRAF-MAPK (dabrafenib + trametinib and encorafenib + binimetinib) are approved by FDA for BRAF-mutant metastatic melanoma \citep{pmid29573941, pmid30219628, pmid31166680}, we recommend the synergy for BRAF-PI3K (encorafenib + BKM120) and BRAF-CDK (encorafenib + LEE011) inhibitors.

\paragraph{Comparison to alternative approaches.} Unlike the probabilistic generative modeling approach proposed in this paper, standard distance-based agglomerative hierarchical clustering and Pearson correlation can also be applied to the PDX data to estimate the similarity. However, simple pairwise similarities can be potentially noisy and the uncertainty in the estimation is not fully incorporated due to the absence of a generative model. As we showed in the Section \ref{sec::simulation_similarity_aspect} (Simulation I) that agglomerative hierarchical clustering and the Pearson correlation leads to inferior recovery of the true branching times and the true tree structure under different data generating mechanisms mimicking the real data. As further evidence, we compute pairwise similarities based on Pearson correlation (other distance metrics show similar patterns) in the right panel of Figure \ref{fig:ggtree}. By mapping the original Pearson correlation $\rho \in [-1,1]$ through a linear function $\frac{\rho+1}{2}$, we make the range of iPCP and Pearson correlation comparable. We observe that pairwise iPCP estimated through the DDT model is less noisy than Pearson correlation. For example, both iPCP and Pearson correlation in CM show higher similarities among combination therapy framed by a box, but iPCP exhibits a clearer pattern than Pearson correlation.

\section{Summary and Discussion}\label{sec:discussion}
In translational oncology research, PDX studies have emerged as a unique study design that evaluates multiple treatments when applied to samples from the same human tumor implanted into genetically identical mice. PDX systems are promising tools for large-scale screening to evaluate a large number of FDA-approved and novel cancer therapies. However, there remain scientific questions concerning how distinct treatments may be synergistic in inducing similar efficacious responses, and how to identify promising subsets of treatments for further clinical evaluation. To this end, in this paper, we propose a probabilistic framework to learn treatment trees (\Rx -trees) from PDX data to identify promising treatment combinations and plausible biological mechanisms that confer synergistic effect(s). In particular, in a Bayesian framework based on the Dirichlet Diffusion Tree, we estimate a {\it maximum a posteriori} rooted binary tree with the treatments on the leaves and propose a posterior uncertainty-aware similarity measure (iPCP) for any subset of treatments.  The divergence times of the DDT encode the tree topology and are profitably interpreted within the context of an underlying plausible biological mechanism of treatment actions. 

From the class of probabilistic models with an unknown tree structure component, we have chosen the DDT mainly owing to the availability of a closed-form marginal likelihood that directly links the tree topological structure to the covariance structure of the observed PDX data, which additionally decouples the Euclidean and tree parameters; to the best of our knowledge this method has not been proposed or explored hitherto for the DDT. The decoupling leads to efficient posterior inference via a two-stage algorithm that confers several advantages. The algorithm generates posterior samples of Euclidean parameters through approximate Bayesian computation and passes the posterior medians to a second stage classical Metropolis-Hastings algorithm for sampling from the conditional posterior distribution of the tree given all other quantities. Through simulation studies, we show that the proposed two-stage algorithm generates better posterior tree samples and captures the true similarity among treatments better than alternatives such as single-stage MCMC and naive Pearson correlations. The posterior samples of trees are summarized by iPCP, which we propose to measure the empirical mechanistic similarity for multiple treatments incorporating uncertainty. 

Using the proposed methodology on NIBR-PDXE data, we estimate \Rx-trees and iPCPs for five cancers. Among the monotherapies, iPCP is highly concordant with known biology across different cancers. For example, CGM097 and HDM201 show a high iPCP value among treatments in breast cancer and melanoma, which corroborates known mechanisms, since both monotherapies target the same biological pathway, MDM2, and have been further validated in recent clinical trials \citep{pmid32651541}. The proposed iPCP can also suggest improvements upon an existing combination therapy. We first identify a combination therapy with known synergy (not based on the our data) and then determine which additional therapies (monotherapies or combination therapies) have high iPCPs when considered together with the existing combination therapy. Based on the NIBR-PDXE data, for each cancer, we suggest potential synergies between PI3K-ERBB3 and CDK-MTOR for breast cancer, PI3K-ERBB3 for colorectal cancer, and BRAF-PI3K and BRAF-CDK for melanoma that could be potentially explored in future translational studies. 

The present analysis shows the promise in guiding the potential treatment strategies by building trees upon NIBR-PDXE dataset using treatment responses, but assume independent patients without using the underlying patient-specific genomic information that is available in the NIBR-PDXE.
By including patient-specific genomic information, we may further improve our ability to identify synergistic treatments that may be specific to a subset of patients. One approach to utilize genomic information could be to extend the DDT model to incorporate patient-specific genomic information in the mean structure or the column covariance of the marginal likelihood of Equation \eqref{eq:joint_llh}. Models with non-Gaussian marginal likelihood and non-binary treatment tree in principle can be defined by considering generative tree models based on general diffusion processes \citep{10.5555/3044805.3045094,PYDT}. Both extensions raise significant, non-trivial methodological and computation issues (e.g., deriving tractable likelihoods; finding low-dimensional summary statistics for new parameters) and constitute the foundation for future work. 

\paragraph{Code and data availability} We also provide a general purpose code in \texttt{R} that accompanies this manuscript along with all the necessary documentation and datasets required to replicate our results (see \url{https://github.com/bayesrx/RxTree}). 
Furthermore, to aid access and visualization of the results, we have also developed an \texttt{R}-shiny application (see Supplementary Material Section S6.4).

\bibliographystyle{unsrtnat}
\bibliography{DDT}

\begin{thebibliography}{52}
\providecommand{\natexlab}[1]{#1}
\providecommand{\url}[1]{\texttt{#1}}
\expandafter\ifx\csname urlstyle\endcsname\relax
  \providecommand{\doi}[1]{doi: #1}\else
  \providecommand{\doi}{doi: \begingroup \urlstyle{rm}\Url}\fi

\bibitem[Ferlay et~al.(2020)Ferlay, Ervik, Lam, Colombet, Mery, Piñeros,
  Znaor, Soerjomataram, and Bray]{who_cancer}
J~Ferlay, M~Ervik, F~Lam, M~Colombet, L~Mery, M~Piñeros, A~Znaor,
  I~Soerjomataram, and F~Bray.
\newblock \emph{Global Cancer Observatory: Cancer Today}.
\newblock Lyon, France: International Agency for Research on Cancer, 2020.
\newblock Available from: \url{https://gco.iarc.fr/today}, accessed 05.28.2021.

\bibitem[Dagogo-Jack and Shaw(2018)]{pmid29115304}
I.~Dagogo-Jack and A.~T. Shaw.
\newblock {{T}umour heterogeneity and resistance to cancer therapies}.
\newblock \emph{Nat Rev Clin Oncol}, 15\penalty0 (2):\penalty0 81--94, 02 2018.

\bibitem[Groisberg and Subbiah(2021)]{pmid33707180}
R.~Groisberg and V.~Subbiah.
\newblock {{C}ombination therapies for precision oncology: the ultimate
  whack-a-mole game}.
\newblock \emph{Clin Cancer Res}, 27\penalty0 (10):\penalty0 2672--2674, May
  2021.

\bibitem[Sawyers(2013)]{pmid23803949}
C.~L. Sawyers.
\newblock {{P}erspective: combined forces}.
\newblock \emph{Nature}, 498\penalty0 (7455):\penalty0 S7, Jun 2013.

\bibitem[Bayat~Mokhtari et~al.(2017)Bayat~Mokhtari, Homayouni, Baluch,
  Morgatskaya, Kumar, Das, and Yeger]{pmid28410237}
R.~Bayat~Mokhtari, T.~S. Homayouni, N.~Baluch, E.~Morgatskaya, S.~Kumar,
  B.~Das, and H.~Yeger.
\newblock {{C}ombination therapy in combating cancer}.
\newblock \emph{Oncotarget}, 8\penalty0 (23):\penalty0 38022--38043, Jun 2017.

\bibitem[Sun et~al.(2016)Sun, Sanderson, and Zheng]{DrugDrugInter}
W.~Sun, P.~E. Sanderson, and W.~Zheng.
\newblock {{D}rug combination therapy increases successful drug repositioning}.
\newblock \emph{Drug Discov Today}, 21\penalty0 (7):\penalty0 1189--1195, 07
  2016.

\bibitem[Tentler et~al.(2012)Tentler, Tan, Weekes, Jimeno, Leong, Pitts,
  Arcaroli, Messersmith, and Eckhardt]{pmid22508028}
J.~J. Tentler, A.~C. Tan, C.~D. Weekes, A.~Jimeno, S.~Leong, T.~M. Pitts, J.~J.
  Arcaroli, W.~A. Messersmith, and S.~G. Eckhardt.
\newblock {{P}atient-derived tumour xenografts as models for oncology drug
  development}.
\newblock \emph{Nat Rev Clin Oncol}, 9\penalty0 (6):\penalty0 338--350, Apr
  2012.

\bibitem[Bhimani et~al.(2020)Bhimani, Ball, and Stebbing]{pmid31919403}
J.~Bhimani, K.~Ball, and J.~Stebbing.
\newblock {{P}atient-derived xenograft models-the future of personalised cancer
  treatment}.
\newblock \emph{Br J Cancer}, 122\penalty0 (5):\penalty0 601--602, 03 2020.

\bibitem[Hidalgo et~al.(2014)Hidalgo, Amant, Biankin, Budinská, Byrne, Caldas,
  Clarke, de~Jong, Jonkers, Mælandsmo, Roman-Roman, Seoane, Trusolino, and
  Villanueva]{pmid25185190}
M.~Hidalgo, F.~Amant, A.~V. Biankin, E.~Budinská, A.~T. Byrne, C.~Caldas,
  R.~B. Clarke, S.~de~Jong, J.~Jonkers, G.~M. Mælandsmo, S.~Roman-Roman,
  J.~Seoane, L.~Trusolino, and A.~Villanueva.
\newblock {{P}atient-derived xenograft models: an emerging platform for
  translational cancer research}.
\newblock \emph{Cancer Discov}, 4\penalty0 (9):\penalty0 998--1013, Sep 2014.

\bibitem[Lai et~al.(2017)Lai, Wei, Lin, Qin, Cheng, and Li]{pmid28499452}
Y.~Lai, X.~Wei, S.~Lin, L.~Qin, L.~Cheng, and P.~Li.
\newblock {{C}urrent status and perspectives of patient-derived xenograft
  models in cancer research}.
\newblock \emph{J Hematol Oncol}, 10\penalty0 (1):\penalty0 106, 05 2017.

\bibitem[Yoshida(2020)]{Yoshida2020}
Go~J. Yoshida.
\newblock Applications of patient-derived tumor xenograft models and tumor
  organoids.
\newblock \emph{Journal of Hematology {\&} Oncology}, 13\penalty0 (1):\penalty0
  4, Jan 2020.
\newblock ISSN 1756-8722.
\newblock \doi{10.1186/s13045-019-0829-z}.
\newblock URL \url{https://doi.org/10.1186/s13045-019-0829-z}.

\bibitem[Topp et~al.(2014)Topp, Hartley, Cook, Heong, Boehm, McShane, Pyman,
  McNally, Ananda, Harrell, Etemadmoghadam, Galletta, Alsop, Mitchell, Fox,
  Kerr, Hutt, Kaufmann, {Australian Ovarian Cancer Study}, Swisher, Bowtell,
  Wakefield, and Scott]{TOPP2014656}
Monique~D. Topp, Lynne Hartley, Michele Cook, Valerie Heong, Emma Boehm, Lauren
  McShane, Jan Pyman, Orla McNally, Sumitra Ananda, Marisol Harrell, Dariush
  Etemadmoghadam, Laura Galletta, Kathryn Alsop, Gillian Mitchell, Stephen~B.
  Fox, Jeffrey~B. Kerr, Karla~J. Hutt, Scott~H. Kaufmann, {Australian Ovarian
  Cancer Study}, Elizabeth~M. Swisher, David~D. Bowtell, Matthew~J. Wakefield,
  and Clare~L. Scott.
\newblock Molecular correlates of platinum response in human high-grade serous
  ovarian cancer patient-derived xenografts.
\newblock \emph{Molecular Oncology}, 8\penalty0 (3):\penalty0 656--668, 2014.
\newblock ISSN 1574-7891.
\newblock \doi{https://doi.org/10.1016/j.molonc.2014.01.008}.
\newblock URL
  \url{https://www.sciencedirect.com/science/article/pii/S1574789114000209}.

\bibitem[Nunes et~al.(2015)Nunes, Vrignaud, Vacher, Richon, Li{\`e}vre,
  Cacheux, Weiswald, Massonnet, Chateau-Joubert, Nicolas, Dib, Zhang, Watters,
  Bergstrom, Roman-Roman, Bi{\`e}che, and Dangles-Marie]{Nunes1560}
Manoel Nunes, Patricia Vrignaud, Sophie Vacher, Sophie Richon, Astrid
  Li{\`e}vre, Wulfran Cacheux, Louis-Bastien Weiswald, Gerald Massonnet, Sophie
  Chateau-Joubert, Andr{\'e} Nicolas, Colette Dib, Weidong Zhang, James
  Watters, Donald Bergstrom, Sergio Roman-Roman, Ivan Bi{\`e}che, and Virginie
  Dangles-Marie.
\newblock Evaluating patient-derived colorectal cancer xenografts as
  preclinical models by comparison with patient clinical data.
\newblock \emph{Cancer Research}, 75\penalty0 (8):\penalty0 1560--1566, 2015.
\newblock ISSN 0008-5472.
\newblock \doi{10.1158/0008-5472.CAN-14-1590}.
\newblock URL \url{https://cancerres.aacrjournals.org/content/75/8/1560}.

\bibitem[Clohessy and Pandolfi(2015)]{pmid25895610}
J.~G. Clohessy and P.~P. Pandolfi.
\newblock {{M}ouse hospital and co-clinical trial project--from bench to
  bedside}.
\newblock \emph{Nat Rev Clin Oncol}, 12\penalty0 (8):\penalty0 491--498, Aug
  2015.

\bibitem[Lunardi and Pandolfi(2015)]{pmid25466492}
A.~Lunardi and P.~P. Pandolfi.
\newblock {{A} co-clinical platform to accelerate cancer treatment
  optimization}.
\newblock \emph{Trends Mol Med}, 21\penalty0 (1):\penalty0 1--5, Jan 2015.

\bibitem[Grant et~al.(2010)Grant, Combs, and Acosta]{MoA}
R.L. Grant, A.B. Combs, and D.~Acosta.
\newblock {E}xperimental models for the investigation of toxicological
  mechanisms.
\newblock In Charlene~A. McQueen, editor, \emph{Comprehensive Toxicology
  (Second Edition)}, pages 203--224. Elsevier, Oxford, second edition edition,
  2010.
\newblock ISBN 978-0-08-046884-6.
\newblock \doi{https://doi.org/10.1016/B978-0-08-046884-6.00110-X}.
\newblock URL
  \url{https://www.sciencedirect.com/science/article/pii/B978008046884600110X}.

\bibitem[Krumbach et~al.(2011)Krumbach, Schüler, Hofmann, Giesemann, Fiebig,
  and Beckers]{pmid21273060}
R.~Krumbach, J.~Schüler, M.~Hofmann, T.~Giesemann, H.~H. Fiebig, and
  T.~Beckers.
\newblock {{P}rimary resistance to cetuximab in a panel of patient-derived
  tumour xenograft models: activation of {M}{E}{T} as one mechanism for drug
  resistance}.
\newblock \emph{Eur J Cancer}, 47\penalty0 (8):\penalty0 1231--1243, May 2011.

\bibitem[Sokal and Rohlf(1962)]{cophenetic_1962}
Robert~R. Sokal and F.~James Rohlf.
\newblock The comparison of dendrograms by objective methods.
\newblock \emph{Taxon}, 11\penalty0 (2):\penalty0 33--40, 1962.
\newblock ISSN 00400262.
\newblock URL \url{http://www.jstor.org/stable/1217208}.

\bibitem[Narayan et~al.(2020)Narayan, Molenaar, Teng, Cornelissen, Roelofs,
  Menezes, Dik, Lagerweij, Broersma, Petersen, Marin~Soto, Brands, van Kuiken,
  Lecca, Lenos, In~'t Veld, van Wieringen, Lang, Sulman, Verhaak, Baumert,
  Stalpers, Vermeulen, Watts, Bailey, Slotman, Versteeg, Noske, Sminia,
  Tannous, Wurdinger, Koster, and Westerman]{pmid32523045}
R.~S. Narayan, P.~Molenaar, J.~Teng, F.~M.~G. Cornelissen, I.~Roelofs,
  R.~Menezes, R.~Dik, T.~Lagerweij, Y.~Broersma, N.~Petersen, J.~A. Marin~Soto,
  E.~Brands, P.~van Kuiken, M.~C. Lecca, K.~J. Lenos, S.~G. J.~G. In~'t Veld,
  W.~van Wieringen, F.~F. Lang, E.~Sulman, R.~Verhaak, B.~G. Baumert, L.~J.~A.
  Stalpers, L.~Vermeulen, C.~Watts, D.~Bailey, B.~J. Slotman, R.~Versteeg,
  D.~Noske, P.~Sminia, B.~A. Tannous, T.~Wurdinger, J.~Koster, and B.~A.
  Westerman.
\newblock {{A} cancer drug atlas enables synergistic targeting of independent
  drug vulnerabilities}.
\newblock \emph{Nat Commun}, 11\penalty0 (1):\penalty0 2935, 06 2020.

\bibitem[Neal({2003})]{Neal2003}
Radford Neal.
\newblock {{Density modeling and clustering using {D}irichlet diffusion
  trees}}.
\newblock \emph{{Bayesian Statistics}}, {7}:\penalty0 {619--629}, {2003}.

\bibitem[Gao et~al.(2015)Gao, Korn, Ferretti, Monahan, Wang, Singh, Zhang,
  Schnell, Yang, Zhang, Balbin, Barbe, Cai, Casey, Chatterjee, Chiang, Chuai,
  Cogan, Collins, Dammassa, Ebel, Embry, Green, Kauffmann, Kowal, Leary, Lehar,
  Liang, Loo, Lorenzana, Robert~McDonald, McLaughlin, Merkin, Meyer, Naylor,
  Patawaran, Reddy, Röelli, Ruddy, Salangsang, Santacroce, Singh, Tang,
  Tinetto, Tobler, Velazquez, Venkatesan, Von~Arx, Wang, Wang, Wiesmann, Wyss,
  Xu, Bitter, Atadja, Lees, Hofmann, Li, Keen, Cozens, Jensen, Pryer, Williams,
  and Sellers]{pmid26479923}
H.~Gao, J.~M. Korn, S.~Ferretti, J.~E. Monahan, Y.~Wang, M.~Singh, C.~Zhang,
  C.~Schnell, G.~Yang, Y.~Zhang, O.~A. Balbin, S.~Barbe, H.~Cai, F.~Casey,
  S.~Chatterjee, D.~Y. Chiang, S.~Chuai, S.~M. Cogan, S.~D. Collins,
  E.~Dammassa, N.~Ebel, M.~Embry, J.~Green, A.~Kauffmann, C.~Kowal, R.~J.
  Leary, J.~Lehar, Y.~Liang, A.~Loo, E.~Lorenzana, E.~Robert~McDonald, M.~E.
  McLaughlin, J.~Merkin, R.~Meyer, T.~L. Naylor, M.~Patawaran, A.~Reddy,
  C.~Röelli, D.~A. Ruddy, F.~Salangsang, F.~Santacroce, A.~P. Singh, Y.~Tang,
  W.~Tinetto, S.~Tobler, R.~Velazquez, K.~Venkatesan, F.~Von~Arx, H.~Q. Wang,
  Z.~Wang, M.~Wiesmann, D.~Wyss, F.~Xu, H.~Bitter, P.~Atadja, E.~Lees,
  F.~Hofmann, E.~Li, N.~Keen, R.~Cozens, M.~R. Jensen, N.~K. Pryer, J.~A.
  Williams, and W.~R. Sellers.
\newblock {H}igh-throughput screening using patient-derived tumor xenografts to
  predict clinical trial drug response.
\newblock \emph{Nat. Med.}, 21\penalty0 (11):\penalty0 1318--1325, Nov 2015.

\bibitem[Konopleva et~al.(2020)Konopleva, Martinelli, Daver, Papayannidis, Wei,
  Higgins, Ott, Mascarenhas, and Andreeff]{pmid32651541}
M.~Konopleva, G.~Martinelli, N.~Daver, C.~Papayannidis, A.~Wei, B.~Higgins,
  M.~Ott, J.~Mascarenhas, and M.~Andreeff.
\newblock {{M}{D}{M}2 inhibition: an important step forward in cancer therapy}.
\newblock \emph{Leukemia}, 34\penalty0 (11):\penalty0 2858--2874, 11 2020.

\bibitem[Lapointe and Legendre(1991)]{Lapointe1991}
Fran{\c{c}}ois-Joseph Lapointe and Pierre Legendre.
\newblock The generation of random ultrametric matrices representing
  dendrograms.
\newblock \emph{Journal of Classification}, 8\penalty0 (2):\penalty0 177--200,
  Dec 1991.
\newblock ISSN 1432-1343.
\newblock \doi{10.1007/BF02616238}.
\newblock URL \url{https://doi.org/10.1007/BF02616238}.

\bibitem[McCullagh(2006)]{McCullagh2006StructuredCM}
P.~McCullagh.
\newblock Structured covariance matrices in multivariate regression models.
\newblock Technical report, Department of Statistics, University of Chicago,
  2006.

\bibitem[Bravo et~al.(2009)Bravo, Wright, Eng, Keles, and Wahba]{pmid22081761}
H.~C. Bravo, S.~Wright, K.~H. Eng, S.~Keles, and G.~Wahba.
\newblock {{E}stimating tree-structured covariance matrices via mixed-integer
  programming}.
\newblock \emph{J Mach Learn Res}, 5:\penalty0 41--48, 2009.

\bibitem[Knowles and Ghahramani(2015)]{PYDT}
D.~A. Knowles and Z.~Ghahramani.
\newblock {{P}itman-{Y}or diffusion trees for {B}ayesian hierarchical
  clustering}.
\newblock \emph{IEEE Trans Pattern Anal Mach Intell}, 37\penalty0 (2):\penalty0
  271--289, Feb 2015.

\bibitem[Turner et~al.(2013)Turner, Sederberg, Brown, and
  Steyvers]{pmid23646991}
B.~M. Turner, P.~B. Sederberg, S.~D. Brown, and M.~Steyvers.
\newblock {{A} method for efficiently sampling from distributions with
  correlated dimensions}.
\newblock \emph{Psychol Methods}, 18\penalty0 (3):\penalty0 368--384, Sep 2013.

\bibitem[Knowles et~al.(2011)Knowles, Gael, and Ghahramani]{MSP_DDT}
David~A. Knowles, Jurgen~Van Gael, and Zoubin Ghahramani.
\newblock Message passing algorithms for dirichlet diffusion trees.
\newblock \emph{International Conference on Machine Learning (ICML)}, 2011.

\bibitem[Liu(1994)]{10.2307/2290921}
Jun~S. Liu.
\newblock The collapsed {G}ibbs sampler in {B}ayesian computations with
  applications to a gene regulation problem.
\newblock \emph{Journal of the American Statistical Association}, 89\penalty0
  (427):\penalty0 958--966, 1994.
\newblock ISSN 01621459.
\newblock URL \url{http://www.jstor.org/stable/2290921}.

\bibitem[Sisson et~al.(2018)Sisson, Fan, and Beaumont]{sisson2018handbook}
Scott~A Sisson, Yanan Fan, and Mark Beaumont.
\newblock \emph{Handbook of approximate {B}ayesian computation}.
\newblock CRC Press, 2018.

\bibitem[Casella and Berger(2001)]{CaseBerg}
George Casella and Roger Berger.
\newblock \emph{Statistical Inference}.
\newblock {Duxbury Resource Center}, June 2001.
\newblock ISBN 0534243126.

\bibitem[Mathai(1980)]{doi:10.1080/03610928008827921}
A.M. Mathai.
\newblock Moments of the trace of a noncentral {W}ishart matrix.
\newblock \emph{Communications in Statistics - Theory and Methods}, 9\penalty0
  (8):\penalty0 795--801, 1980.
\newblock \doi{10.1080/03610928008827921}.
\newblock URL \url{https://doi.org/10.1080/03610928008827921}.

\bibitem[Cardona et~al.(2013)Cardona, Mir, Rosselló, Rotger, and
  Sánchez]{pmid23323711}
G.~Cardona, A.~Mir, F.~Rosselló, L.~Rotger, and D.~Sánchez.
\newblock {{C}ophenetic metrics for phylogenetic trees, after {S}okal and
  {R}ohlf}.
\newblock \emph{BMC Bioinformatics}, 14:\penalty0 3, Jan 2013.

\bibitem[Rashid et~al.(2020)Rashid, Luckett, Chen, Lawson, Wang, Zhang, Laber,
  Liu, Yeh, Zeng, and Kosorok]{doi:10.1080/01621459.2020.1828091}
Naim~U. Rashid, Daniel~J. Luckett, Jingxiang Chen, Michael~T. Lawson,
  Longshaokan Wang, Yunshu Zhang, Eric~B. Laber, Yufeng Liu, Jen~Jen Yeh,
  Donglin Zeng, and Michael~R. Kosorok.
\newblock High-dimensional precision medicine from patient-derived xenografts.
\newblock \emph{Journal of the American Statistical Association}, 0\penalty0
  (0):\penalty0 1--15, 2020.
\newblock \doi{10.1080/01621459.2020.1828091}.
\newblock URL \url{https://doi.org/10.1080/01621459.2020.1828091}.

\bibitem[Repetto et~al.(2018)Repetto, Winters, Bush, Reiter, Hollenstein,
  Ammerer, Pryciak, and Colman-Lerner]{pmid29547722}
M.~V. Repetto, M.~J. Winters, A.~Bush, W.~Reiter, D.~M. Hollenstein,
  G.~Ammerer, P.~M. Pryciak, and A.~Colman-Lerner.
\newblock {{C}{D}{K} and {M}{A}{P}{K} synergistically regulate signaling
  dynamics via a shared multi-site phosphorylation region on the scaffold
  protein {S}te5}.
\newblock \emph{Mol Cell}, 69\penalty0 (6):\penalty0 938--952, 03 2018.

\bibitem[Kurtzeborn et~al.(2019)Kurtzeborn, Kwon, and Kuure]{pmid30974877}
K.~Kurtzeborn, H.~N. Kwon, and S.~Kuure.
\newblock {{M}{A}{P}{K}/{E}{R}{K} {S}ignaling in regulation of renal
  differentiation}.
\newblock \emph{Int J Mol Sci}, 20\penalty0 (7), Apr 2019.

\bibitem[Balko et~al.(2012)Balko, Miller, Morrison, Hutchinson, Young,
  Rinehart, Sánchez, Jee, Polyak, Prat, Perou, Arteaga, and
  Cook]{pmid22178756}
J.~M. Balko, T.~W. Miller, M.~M. Morrison, K.~Hutchinson, C.~Young,
  C.~Rinehart, V.~Sánchez, D.~Jee, K.~Polyak, A.~Prat, C.~M. Perou, C.~L.
  Arteaga, and R.~S. Cook.
\newblock {{T}he receptor tyrosine kinase {E}rb{B}3 maintains the balance
  between luminal and basal breast epithelium}.
\newblock \emph{Proc Natl Acad Sci U S A}, 109\penalty0 (1):\penalty0 221--226,
  Jan 2012.

\bibitem[Wu et~al.(2021)Wu, Islam, Lee, Takase, Guo, Andrews, Buzhdygan,
  Mathew, Li, Arai, Lo, Ramirez, and Lok]{pmid33583260}
L.~Wu, M.~R. Islam, J.~Lee, H.~Takase, S.~Guo, A.~M. Andrews, T.~P. Buzhdygan,
  J.~Mathew, W.~Li, K.~Arai, E.~H. Lo, S.~H. Ramirez, and J.~Lok.
\newblock {{E}rb{B}3 is a critical regulator of cytoskeletal dynamics in brain
  microvascular endothelial cells: implications for vascular remodeling and
  blood-brain-barrier modulation}.
\newblock \emph{J Cereb Blood Flow Metab}, page 271678X20984976, Feb 2021.

\bibitem[Gunning et~al.(2015)Gunning, Ghoshdastider, Whitaker, Popp, and
  Robinson]{pmid25788699}
P.~W. Gunning, U.~Ghoshdastider, S.~Whitaker, D.~Popp, and R.~C. Robinson.
\newblock {{T}he evolution of compositionally and functionally distinct actin
  filaments}.
\newblock \emph{J Cell Sci}, 128\penalty0 (11):\penalty0 2009--2019, Jun 2015.

\bibitem[Haider et~al.(2019)Haider, Rahaman, Yar, and Kamal]{pmid31353978}
K.~Haider, S.~Rahaman, M.~S. Yar, and A.~Kamal.
\newblock {{T}ubulin inhibitors as novel anticancer agents: an overview on
  patents (2013-2018)}.
\newblock \emph{Expert Opin Ther Pat}, 29\penalty0 (8):\penalty0 623--641, Aug
  2019.

\bibitem[Mahale et~al.(2015)Mahale, Bharate, Manda, Joshi, Jenkins,
  Vishwakarma, and Chaudhuri]{pmid25950473}
S.~Mahale, S.~B. Bharate, S.~Manda, P.~Joshi, P.~R. Jenkins, R.~A. Vishwakarma,
  and B.~Chaudhuri.
\newblock {{A}ntitumour potential of {B}{P}{T}: a dual inhibitor of {C}{D}{K}4
  and tubulin polymerization}.
\newblock \emph{Cell Death Dis}, 6:\penalty0 e1743, May 2015.

\bibitem[Bohnacker et~al.(2017)Bohnacker, Prota, Beaufils, Burke, Melone,
  Inglis, Rageot, Sele, Cmiljanovic, Cmiljanovic, Bargsten, Aher, Akhmanova,
  Díaz, Fabbro, Zvelebil, Williams, Steinmetz, and Wymann]{pmid28276440}
T.~Bohnacker, A.~E. Prota, F.~Beaufils, J.~E. Burke, A.~Melone, A.~J. Inglis,
  D.~Rageot, A.~M. Sele, V.~Cmiljanovic, N.~Cmiljanovic, K.~Bargsten, A.~Aher,
  A.~Akhmanova, J.~F. Díaz, D.~Fabbro, M.~Zvelebil, R.~L. Williams, M.~O.
  Steinmetz, and M.~P. Wymann.
\newblock {{D}econvolution of {B}uparlisib's mechanism of action defines
  specific {P}{I}3{K} and tubulin inhibitors for therapeutic intervention}.
\newblock \emph{Nat Commun}, 8:\penalty0 14683, 03 2017.

\bibitem[Zhao et~al.(2014)Zhao, Yu, and Hu]{pmid24389645}
Y.~Zhao, H.~Yu, and W.~Hu.
\newblock {{T}he regulation of {M}{D}{M}2 oncogene and its impact on human
  cancers}.
\newblock \emph{Acta Biochim Biophys Sin (Shanghai)}, 46\penalty0 (3):\penalty0
  180--189, Mar 2014.

\bibitem[Vora et~al.(2014)Vora, Juric, Kim, Mino-Kenudson, Huynh, Costa,
  Lockerman, Pollack, Liu, Li, Lehar, Wiesmann, Wartmann, Chen, Cao,
  Pinzon-Ortiz, Kim, Schlegel, Huang, and Engelman]{pmid25002028}
S.~R. Vora, D.~Juric, N.~Kim, M.~Mino-Kenudson, T.~Huynh, C.~Costa, E.~L.
  Lockerman, S.~F. Pollack, M.~Liu, X.~Li, J.~Lehar, M.~Wiesmann, M.~Wartmann,
  Y.~Chen, Z.~A. Cao, M.~Pinzon-Ortiz, S.~Kim, R.~Schlegel, A.~Huang, and J.~A.
  Engelman.
\newblock {{C}{D}{K} 4/6 inhibitors sensitize {P}{I}{K}3{C}{A} mutant breast
  cancer to {P}{I}3{K} inhibitors}.
\newblock \emph{Cancer Cell}, 26\penalty0 (1):\penalty0 136--149, Jul 2014.

\bibitem[Bonelli et~al.(2017)Bonelli, Digiacomo, Fumarola, Alfieri, Quaini,
  Falco, Madeddu, La~Monica, Cretella, Ravelli, Ulivi, Tebaldi, Calistri,
  Delmonte, Ampollini, Carbognani, Tiseo, Cavazzoni, and
  Petronini]{pmid28704762}
M.~A. Bonelli, G.~Digiacomo, C.~Fumarola, R.~Alfieri, F.~Quaini, A.~Falco,
  D.~Madeddu, S.~La~Monica, D.~Cretella, A.~Ravelli, P.~Ulivi, M.~Tebaldi,
  D.~Calistri, A.~Delmonte, L.~Ampollini, P.~Carbognani, M.~Tiseo,
  A.~Cavazzoni, and P.~G. Petronini.
\newblock {{C}ombined inhibition of {C}{D}{K}4/6 and
  {P}{I}3{K}/{A}{K}{T}/m{T}{O}{R} pathways induces a synergistic anti-tumor
  effect in malignant pleural mesothelioma cells}.
\newblock \emph{Neoplasia}, 19\penalty0 (8):\penalty0 637--648, Aug 2017.

\bibitem[Yuan et~al.(2019)Yuan, Wen, Yost, Xing, Yan, Han, Mortimer, and
  Yim]{pmid31101835}
Y.~Yuan, W.~Wen, S.~E. Yost, Q.~Xing, J.~Yan, E.~S. Han, J.~Mortimer, and J.~H.
  Yim.
\newblock {{C}ombination therapy with {B}{Y}{L}719 and {L}{E}{E}011 is
  synergistic and causes a greater suppression of p-{S}6 in triple negative
  breast cancer}.
\newblock \emph{Sci Rep}, 9\penalty0 (1):\penalty0 7509, 05 2019.

\bibitem[Geel et~al.(2014)Geel, Elez, Bendell, Faris, Lolkema, Eskens,
  Spreafico, Kavan, Delord, Schuler, Wainberg, Yamada, Yoshino, Demuth, Avsar,
  Chatterjee, Zhu, Bernards, Tabernero, and Schellens]{CRC_tripleCmb}
Robin~Van Geel, Elena Elez, Johanna~C. Bendell, Jason~Edward Faris, Martijn P.
  J.~K. Lolkema, Ferry Eskens, Anna Spreafico, Petr Kavan, Jean-Pierre Delord,
  Martin~H. Schuler, Zev~A. Wainberg, Yasuhide Yamada, Takayuki Yoshino, Tim
  Demuth, Emin Avsar, Arkendu Chatterjee, Peijuan Zhu, Rene Bernards, Josep
  Tabernero, and Jan~HM Schellens.
\newblock Phase {I} study of the selective {B}{R}{A}{F}{V}600 inhibitor
  encorafenib ({L}{G}{X}818) combined with cetuximab and with or without the
  α-specific {P}{I}3{K} inhibitor {B}{Y}{L}719 in patients with advanced
  {B}{R}{A}{F}-mutant colorectal cancer.
\newblock \emph{Journal of Clinical Oncology}, 32\penalty0
  (15\_suppl):\penalty0 3514--3514, 2014.
\newblock \doi{10.1200/jco.2014.32.15\_suppl.3514}.
\newblock URL \url{https://doi.org/10.1200/jco.2014.32.15_suppl.3514}.

\bibitem[van Geel et~al.(2017)van Geel, Tabernero, Elez, Bendell, Spreafico,
  Schuler, Yoshino, Delord, Yamada, Lolkema, Faris, Eskens, Sharma, Yaeger,
  Lenz, Wainberg, Avsar, Chatterjee, Jaeger, Tan, Maharry, Demuth, and
  Schellens]{pmid28363909}
R.~M. J.~M. van Geel, J.~Tabernero, E.~Elez, J.~C. Bendell, A.~Spreafico,
  M.~Schuler, T.~Yoshino, J.~P. Delord, Y.~Yamada, M.~P. Lolkema, J.~E. Faris,
  F.~A. L.~M. Eskens, S.~Sharma, R.~Yaeger, H.~J. Lenz, Z.~A. Wainberg,
  E.~Avsar, A.~Chatterjee, S.~Jaeger, E.~Tan, K.~Maharry, T.~Demuth, and
  J.~H.~M. Schellens.
\newblock {{A} phase {I}b dose-escalation study of encorafenib and cetuximab
  with or without alpelisib in metastatic {B}{R}{A}{F}-mutant colorectal
  cancer}.
\newblock \emph{Cancer Discov}, 7\penalty0 (6):\penalty0 610--619, 06 2017.

\bibitem[Dummer et~al.(2018{\natexlab{a}})Dummer, Ascierto, Gogas, Arance,
  Mandala, Liszkay, Garbe, Schadendorf, Krajsova, Gutzmer, Chiarion-Sileni,
  Dutriaux, de~Groot, Yamazaki, Loquai, Moutouh-de Parseval, Pickard, Sandor,
  Robert, and Flaherty]{pmid29573941}
R.~Dummer, P.~A. Ascierto, H.~J. Gogas, A.~Arance, M.~Mandala, G.~Liszkay,
  C.~Garbe, D.~Schadendorf, I.~Krajsova, R.~Gutzmer, V.~Chiarion-Sileni,
  C.~Dutriaux, J.~W.~B. de~Groot, N.~Yamazaki, C.~Loquai, L.~A. Moutouh-de
  Parseval, M.~D. Pickard, V.~Sandor, C.~Robert, and K.~T. Flaherty.
\newblock {{E}ncorafenib plus binimetinib versus vemurafenib or encorafenib in
  patients with {B}{R}{A}{F}-mutant melanoma ({C}{O}{L}{U}{M}{B}{U}{S}): a
  multicentre, open-label, randomised phase 3 trial}.
\newblock \emph{Lancet Oncol}, 19\penalty0 (5):\penalty0 603--615, 05
  2018{\natexlab{a}}.

\bibitem[Dummer et~al.(2018{\natexlab{b}})Dummer, Ascierto, Gogas, Arance,
  Mandala, Liszkay, Garbe, Schadendorf, Krajsova, Gutzmer, Chiarion~Sileni,
  Dutriaux, de~Groot, Yamazaki, Loquai, Moutouh-de Parseval, Pickard, Sandor,
  Robert, and Flaherty]{pmid30219628}
R.~Dummer, P.~A. Ascierto, H.~J. Gogas, A.~Arance, M.~Mandala, G.~Liszkay,
  C.~Garbe, D.~Schadendorf, I.~Krajsova, R.~Gutzmer, V.~Chiarion~Sileni,
  C.~Dutriaux, J.~W.~B. de~Groot, N.~Yamazaki, C.~Loquai, L.~A. Moutouh-de
  Parseval, M.~D. Pickard, V.~Sandor, C.~Robert, and K.~T. Flaherty.
\newblock {{O}verall survival in patients with {B}{R}{A}{F}-mutant melanoma
  receiving encorafenib plus binimetinib versus vemurafenib or encorafenib
  ({C}{O}{L}{U}{M}{B}{U}{S}): a multicentre, open-label, randomised, phase 3
  trial}.
\newblock \emph{Lancet Oncol}, 19\penalty0 (10):\penalty0 1315--1327, 10
  2018{\natexlab{b}}.

\bibitem[Robert et~al.(2019)Robert, Grob, Stroyakovskiy, Karaszewska,
  Hauschild, Levchenko, Chiarion~Sileni, Schachter, Garbe, Bondarenko, Gogas,
  Mandalá, Haanen, Lebbé, Mackiewicz, Rutkowski, Nathan, Ribas, Davies,
  Flaherty, Burgess, Tan, Gasal, Voi, Schadendorf, and Long]{pmid31166680}
C.~Robert, J.~J. Grob, D.~Stroyakovskiy, B.~Karaszewska, A.~Hauschild,
  E.~Levchenko, V.~Chiarion~Sileni, J.~Schachter, C.~Garbe, I.~Bondarenko,
  H.~Gogas, M.~Mandalá, J.~B. A.~G. Haanen, C.~Lebbé, A.~Mackiewicz,
  P.~Rutkowski, P.~D. Nathan, A.~Ribas, M.~A. Davies, K.~T. Flaherty,
  P.~Burgess, M.~Tan, E.~Gasal, M.~Voi, D.~Schadendorf, and G.~V. Long.
\newblock {{F}ive-year outcomes with dabrafenib plus trametinib in metastatic
  melanoma}.
\newblock \emph{N Engl J Med}, 381\penalty0 (7):\penalty0 626--636, 08 2019.

\bibitem[Heaukulani et~al.(2014)Heaukulani, Knowles, and
  Ghahramani]{10.5555/3044805.3045094}
Creighton Heaukulani, David~A. Knowles, and Zoubin Ghahramani.
\newblock Beta diffusion trees.
\newblock In \emph{Proceedings of the 31st International Conference on
  International Conference on Machine Learning - Volume 32}, ICML’14, page
  II–1809–II–1817. JMLR.org, 2014.

\end{thebibliography}


\begin{thebibliography}{17}
\providecommand{\natexlab}[1]{#1}
\providecommand{\url}[1]{\texttt{#1}}
\expandafter\ifx\csname urlstyle\endcsname\relax
  \providecommand{\doi}[1]{doi: #1}\else
  \providecommand{\doi}{doi: \begingroup \urlstyle{rm}\Url}\fi

\bibitem[Blum(2010)]{Blum2010}
Michael~G.B. Blum.
\newblock {Approximate Bayesian computation: A nonparametric perspective}.
\newblock \emph{Journal of the American Statistical Association}, 105\penalty0
  (491):\penalty0 1178--1187, 2010.
\newblock ISSN 01621459.
\newblock \doi{10.1198/jasa.2010.tm09448}.

\bibitem[Beaumont et~al.(2002)Beaumont, Zhang, and Balding]{Beaumont2025}
Mark~A. Beaumont, Wenyang Zhang, and David~J. Balding.
\newblock Approximate {B}ayesian computation in population genetics.
\newblock \emph{Genetics}, 162\penalty0 (4):\penalty0 2025--2035, 2002.
\newblock ISSN 0016-6731.
\newblock URL \url{https://www.genetics.org/content/162/4/2025}.

\bibitem[Sisson et~al.(2019)Sisson, Fan, and Beaumont]{Fan2018}
Scott~A. Sisson, Yanan Fan, and Mark Beaumont.
\newblock \emph{Handbook of Approximate Bayesian Computation}.
\newblock Chapman and Hall/CRC, 1st ed. edition, 2019.

\bibitem[Biau et~al.(2015)Biau, Cérou, and Guyader]{biau2015}
Gérard Biau, Frédéric Cérou, and Arnaud Guyader.
\newblock New insights into approximate {B}ayesian computation.
\newblock \emph{Ann. Inst. H. Poincaré Probab. Statist.}, 51\penalty0
  (1):\penalty0 376--403, 02 2015.
\newblock \doi{10.1214/13-AIHP590}.
\newblock URL \url{https://doi.org/10.1214/13-AIHP590}.

\bibitem[Knowles and Ghahramani(2015)]{PYDT}
D.~A. Knowles and Z.~Ghahramani.
\newblock {{P}itman-{Y}or diffusion trees for {B}ayesian hierarchical
  clustering}.
\newblock \emph{IEEE Trans Pattern Anal Mach Intell}, 37\penalty0 (2):\penalty0
  271--289, Feb 2015.

\bibitem[Bravo et~al.(2009)Bravo, Wright, Eng, Keles, and Wahba]{pmid22081761}
H.~C. Bravo, S.~Wright, K.~H. Eng, S.~Keles, and G.~Wahba.
\newblock {{E}stimating tree-structured covariance matrices via mixed-integer
  programming}.
\newblock \emph{J Mach Learn Res}, 5:\penalty0 41--48, 2009.

\bibitem[Neal({2003})]{Neal2003}
Radford Neal.
\newblock {{Density modeling and clustering using {D}irichlet diffusion
  trees}}.
\newblock \emph{{Bayesian Statistics}}, {7}:\penalty0 {619--629}, {2003}.

\bibitem[Murtagh and Legendre(2014)]{Murtagh2014}
Fionn Murtagh and Pierre Legendre.
\newblock Ward's hierarchical agglomerative clustering method: Which algorithms
  implement ward's criterion?
\newblock \emph{Journal of Classification}, 31\penalty0 (3):\penalty0 274--295,
  Oct 2014.
\newblock ISSN 1432-1343.
\newblock \doi{10.1007/s00357-014-9161-z}.
\newblock URL \url{https://doi.org/10.1007/s00357-014-9161-z}.

\bibitem[Gelman et~al.(2013)Gelman, Carlin, Stern, and Rubin]{gelmanbda13}
Andrew Gelman, John~B. Carlin, Hal~S. Stern, and Donald~B. Rubin.
\newblock \emph{Bayesian Data Analysis}.
\newblock Chapman and Hall/CRC, 3rd ed. edition, 2013.

\bibitem[Geyer(2011)]{geyer2011introduction}
Charles~J Geyer.
\newblock Introduction to {M}arkov chain {M}onte {C}arlo.
\newblock In Steve Brooks, Andrew Gelman, Galin~L. Jones, and Xiao-Li Meng,
  editors, \emph{Handbook of {M}arkov chain {M}onte {C}arlo}, chapter~1, pages
  3--48. Chapman and Hall/CRC, 2011.

\bibitem[Scott et~al.(2016)Scott, Blocker, Bonassi, Chipman, George, and
  McCulloch]{scott2016bayes}
Steven~L Scott, Alexander~W Blocker, Fernando~V Bonassi, Hugh~A Chipman,
  Edward~I George, and Robert~E McCulloch.
\newblock Bayes and big data: The consensus monte carlo algorithm.
\newblock \emph{International Journal of Management Science and Engineering
  Management}, 11\penalty0 (2):\penalty0 78--88, 2016.

\bibitem[Geweke(1992)]{Geweke92evaluatingthe}
John Geweke.
\newblock Evaluating the accuracy of sampling-based approaches to the
  calculation of posterior moments.
\newblock In \emph{In Bayesian Statistics}, pages 169--193. University Press,
  1992.

\bibitem[Prangle et~al.(2014)Prangle, Blum, Popovic, and Sisson]{ABCCoverage}
D.~Prangle, M.~G.~B. Blum, G.~Popovic, and S.~A. Sisson.
\newblock Diagnostic tools for approximate bayesian computation using the
  coverage property.
\newblock \emph{Australian \& New Zealand Journal of Statistics}, 56\penalty0
  (4):\penalty0 309--329, 2014.
\newblock \doi{10.1111/anzs.12087}.
\newblock URL \url{https://onlinelibrary.wiley.com/doi/abs/10.1111/anzs.12087}.

\bibitem[Cook et~al.(2006)Cook, Gelman, and Rubin]{Cook2006}
Samantha~R. Cook, Andrew Gelman, and Donald~B. Rubin.
\newblock {Validation of software for Bayesian models using posterior
  quantiles}.
\newblock \emph{Journal of Computational and Graphical Statistics}, 15\penalty0
  (3):\penalty0 675--692, 2006.
\newblock ISSN 10618600.
\newblock \doi{10.1198/106186006X136976}.

\bibitem[Billera et~al.(2001)Billera, Holmes, and Vogtmann]{BILLERA2001733}
Louis~J. Billera, Susan~P. Holmes, and Karen Vogtmann.
\newblock Geometry of the space of phylogenetic trees.
\newblock \emph{Advances in Applied Mathematics}, 27\penalty0 (4):\penalty0 733
  -- 767, 2001.
\newblock ISSN 0196-8858.
\newblock \doi{https://doi.org/10.1006/aama.2001.0759}.
\newblock URL
  \url{http://www.sciencedirect.com/science/article/pii/S0196885801907596}.

\bibitem[Rashid et~al.(2020)Rashid, Luckett, Chen, Lawson, Wang, Zhang, Laber,
  Liu, Yeh, Zeng, and Kosorok]{doi:10.1080/01621459.2020.1828091}
Naim~U. Rashid, Daniel~J. Luckett, Jingxiang Chen, Michael~T. Lawson,
  Longshaokan Wang, Yunshu Zhang, Eric~B. Laber, Yufeng Liu, Jen~Jen Yeh,
  Donglin Zeng, and Michael~R. Kosorok.
\newblock High-dimensional precision medicine from patient-derived xenografts.
\newblock \emph{Journal of the American Statistical Association}, 0\penalty0
  (0):\penalty0 1--15, 2020.
\newblock \doi{10.1080/01621459.2020.1828091}.
\newblock URL \url{https://doi.org/10.1080/01621459.2020.1828091}.

\bibitem[Giulino-Roth et~al.(2017)Giulino-Roth, van Besien, Dalton, Totonchy,
  Rodina, Taldone, Bolaender, Erdjument-Bromage, Sadek, Chadburn, Barth,
  Dela~Cruz, Rainey, Kung, Chiosis, and Cesarman]{pmid28619753}
L.~Giulino-Roth, H.~J. van Besien, T.~Dalton, J.~E. Totonchy, A.~Rodina,
  T.~Taldone, A.~Bolaender, H.~Erdjument-Bromage, J.~Sadek, A.~Chadburn, M.~J.
  Barth, F.~S. Dela~Cruz, A.~Rainey, A.~L. Kung, G.~Chiosis, and E.~Cesarman.
\newblock {{I}nhibition of {H}sp90 {S}uppresses {P}{I}3{K}/{A}{K}{T}/m{T}{O}{R}
  {S}ignaling and {H}as {A}ntitumor {A}ctivity in {B}urkitt {L}ymphoma}.
\newblock \emph{Mol Cancer Ther}, 16\penalty0 (9):\penalty0 1779--1790, 09
  2017.

\end{thebibliography}

\end{document}


\maketitle

\tableofcontents

\section{Proof of Proposition 1}\label{supp:PropProof}
\begin{figure}[!htb]
    \centering
        \includegraphics[width=\linewidth]{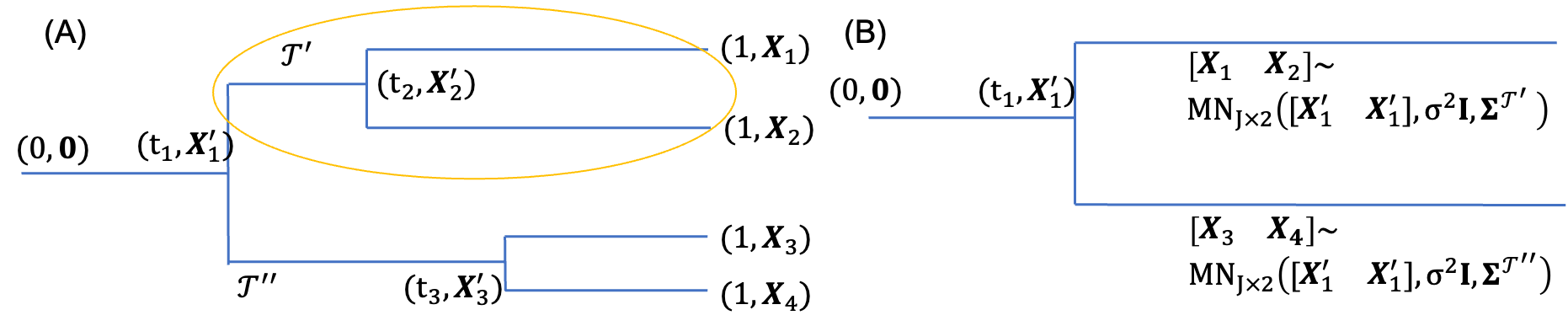}
    \caption{Merging subtrees for the integration process. (A) First step of merging upper subtree, and (B) Final step of merging all subtrees.}
    \label{fig:pfProp1}
    \end{figure}

    We provide a proof for a tree with four leaves (see Figure \ref{fig:pfProp1}) and extension to trees with a larger number of leaves follows by induction. The main idea is to merge subtrees backward and integrate out responses of internal nodes when merging subtrees. 
    
    \begin{proof}
    Consider a subtree $\mathcal{T}'$ rooted at $(t_1,\bX_{1}')$ with two leaves $(1,\bX_{1})$ and $(1,\bX_{2})$, and one internal node $(t_2,\bX_{2}')$ (see Panel (A) of Figure \ref{fig:pfProp1}). Assume that the root $(t_1,\bX_{1}')$ of the subtree is fixed, and responses $\bX_{i},\bX_{i}'\in \mathbb{R}^J, J \geq 1, i=1,2$. With $\bm t=(t_1,t_2,t_3)^\transp$, the conditional distribution for leaf responses would be $\bX_{i}|\bX_{2}',\mathcal{T},\bm t \sim N_J(\bX_{2}',(1-t_2)\sigma^2\bm{I}), i=1,2$. Since $\bX_{2}' |\bX_{1}',\mathcal{T} ,\bm t\sim N_J(\bX_{1}',(t_2-t_1)\sigma^2\bm{I})$, based on the conjugacy of the normal distribution, the marginal distribution is also normal. Conditional on $\bm t$ and $\mathcal{T}$, mean and covariance of $\bX_{i}, i=1,2$ can be derived by the law of iterated expectations and results in the distribution of the subtree $\mathcal{T}'$ with two leaves: 
    
    \begin{align*}
        E[\bX_{i}]&=E[E[\bX_{i}|\bX_{2}']]=E[\bX_{2}']=\bX_{1}', \quad i=1,2; \\
        Var[\bX_{i}]&=Var[E[\bX_{i}|\bX_{2}']] + E[Var[\bX_{i}|\bX_{2}']]=Var[\bX_{2}'] + E[(1-t_2)\sigma^2\bm{I}] = (1-t_1)\sigma^2\bm{I}_J;\\
        Cov[\bX_{1},\bX_{2}]&= Cov[E[\bX_{1}|\bX_{2}'],E[\bX_{2}|\bX_{2}']] + E[Cov[\bX_{1},\bX_{2}|\bX_{2}']] = Var[\bX_{2}']+E[0] = (t_2-t_1)\sigma^2\bm{I}_J; \thinspace
    \end{align*}
    The marginal distribution for the subtree $\mathcal{T}'$ with two leaves is 
    \begin{align*}
    \begin{bmatrix} \bX_{1} & \bX_{2}
    \end{bmatrix} \sim {\sf MN}_{J\times 2}\left(\begin{bmatrix} \bX_{1}' & \bX_{1}'
    \end{bmatrix} , \bm{I}_J, \sigma^2 {\bm \Sigma^\mathcal{T'}} \right), 
    \quad {\bm\Sigma^\mathcal{T'}} = \begin{bmatrix}1-t_1 & t_2-t_1\\ t_2-t_1 & 1-t_1\end{bmatrix}.
    \end{align*}
    
    Therefore, we can merge two leaves responses $\bX_{1}$ and $\bX_{2}$. Similarly, we can also merge the other subtree $\mathcal{T''}$ to obtain. 
    \begin{align*}
    \begin{bmatrix} \bX_{3} & \bX_{4}
    \end{bmatrix} \sim {\sf MN}_{J\times 2}\left(\begin{bmatrix} \bX_{1}' & \bX_{1}'
    \end{bmatrix} , \bm{I}_J, \sigma^2 {\bm \Sigma^\mathcal{T''}} \right), 
    \quad {\bm\Sigma^\mathcal{T''}} = \begin{bmatrix}1-t_1 & t_3-t_1\\ t_3-t_1 & 1-t_1\end{bmatrix}.
    \end{align*}
    
    Eventually, we can merge two subtrees (see Panel (B) of Figure \ref{fig:pfProp1}), $\mathcal{T}'$ and $\mathcal{T}''$. From conjugacy of the normal distribution, the resulting joint marginal distribution of $\bX_{i}, i=1,2,3,4$ is normal. The mean and the variance can be derived along identical lines as above. The only term left is the covariance, and we need to (re-)compute them for locations within and between the combined subtrees. Explicitly,
    \begin{align*}
        Cov[\bX_{1},\bX_{2}]&= Cov[E[\bX_{1}|\bX_{1}'],E[\bX_{2}|\bX_{1}']] + E[Cov[\bX_{1},\bX_{2}|\bX_{1}']] = Var[\bX_{1}']+E[(t_2-t_1)\sigma^2\bm{I}_J] = t_2\sigma^2\bm{I}_J\\
        Cov[\bX_{1},\bX_{3}]&= Cov[E[\bX_{1}|\bX_{1}'],E[\bX_{3}|\bX_{1}']] + E[Cov[\bX_{1},\bX_{3}|\bX_{1}']] = Var[\bX_{1}']+E[0] = t_1\sigma^2\bm{I}_J\thinspace.
    \end{align*}
    This ensures that
    \begin{align*}
    \Xb^\transp=\begin{bmatrix} \bX_{1} & \bX_{2} & \bX_{3} & \bX_{4}
    \end{bmatrix} \sim {\sf MN}_{J\times 4}\left(\begin{bmatrix} 0 & 0 & 0 & 0
    \end{bmatrix} , \bm{I}_J, \sigma^2 {\bm\Sigma^\mathcal{T}} \right), {\bm\Sigma^\mathcal{T}} = \begin{bmatrix}1 & t_2 & t_1 & t_1\\ t_2 & 1 & t_1 & t_1 \\ t_1 & t_1 &1 & t_3\\ t_1 & t_1 & t_3 &1\end{bmatrix},
    \end{align*}
    as required. Moreover, denote $t_{i,i'}$ as the most recent divergence time of leaves $i$ and $i'$. We observe that $t_1=t_{1,3}=t_{1,4}=t_{2,3}=t_{2,4}, t_2=t_{1,2}$, and $t_3=t_{3,4}$ and complete the Proposition 1.
    \end{proof} 
    
    
\section{Efficient Two-Stage Hybrid ABC-MH Algorithm}\label{supp:Alg}
Here we offer details of two-stage algorithm with pseudo code. In the Section \ref{supp:alg_ABC}, we describe the full algorithm of the ABC with the following posterior summary of Euclidean parameters $(c,\sigma^2)$. The Section \ref{supp:alg_MH} includes the implementation of the proposal function and the acceptance probability of MH stage. Pseudo code for the full two-stage algorithm is presented below in Algorithm \ref{algo1}

\subsection{ABC Stage and the Posterior Summary of \texorpdfstring{$c$}{TEXT} and \texorpdfstring{$\sigma^2$}{TEXT}}\label{supp:alg_ABC}
The Section 3 of the Main Paper states the main idea of ABC and we offer the full algorithm of ABC including (i) the synthetic data generation process, (ii) the regression adjustment \citep{Blum2010} of ABC, and (iii) posterior summary of the Euclidean parameters.

\begin{figure}[!htb]
    \includegraphics[width=0.95\linewidth]{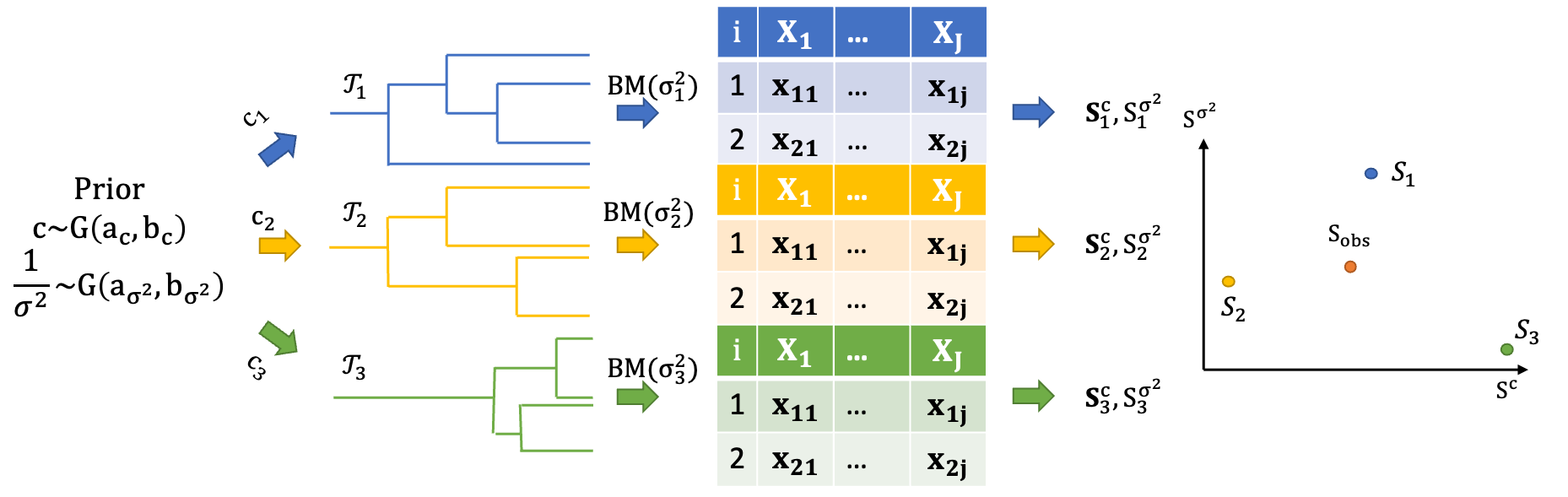}
    \caption{Schematic diagram of synthetic data generation and the calculation of summary statistics (first stage of Algorithm \ref{algo1}). $S_{\obs}$ is calculated based on the actual observed data. }
\label{fig:flowChart}
\end{figure}

\paragraph{Data generation in ABC.}
Following Section 2 in the Main Paper, a synthetic data is generated from DDT as follows: (i) given $c_l \sim {\sf Gamma}(a_c,b_c)$, generate a tree $\mathcal{T}_l$ through the divergence function $a(t)=c_l(1-t)^{-1}$, and (ii) given $\mathcal{T}_l$ and $1/\sigma^2_l \sim {\sf Gamma}(a_{\sigma^2},b_{\sigma^2})$, generate triples ${(t_j,\bX_i',\bX_i)},i'=1\ldots I-1, i=1\ldots I$ by a scaled Brownian motion upon $\mathcal{T}_l$. After discarding $(\mathcal{T}_l, t_i, \bX_i')$, the leaf locations $\bX_i$ form an $I$ by $J$ observed data matrix $\Xb_l$. In Algorithm \ref{algo1}, ABC repeats the procedure above to generate $N^{\syn}$ synthetic data (see Figure \ref{fig:flowChart}).

\paragraph{Regression adjustment in ABC.} 
Originally proposed in \cite{Beaumont2025} and later generalized by \citet{Blum2010}, regression adjustment for ABC is performed in Step 8 of Algorithm \ref{algo1}. The motivation is to use smoothing technique to weaken the effect of the discrepancy between the summary statistic calculated from synthetic data and that from the observed data. We briefly describe the the procedure of $c$. Additional details can be found in \cite{Beaumont2025} and \citet{Blum2010}. Suppose we are given the observed summary statistics $\bm{S}^{(c)}_{\obs}$ and unadjusted samples $(c^{\unadj}_l, \bm{S}^{(c)}_{l}), l=1, \ldots, k$, we can calculate the weight for each sample by 
\begin{align} \label{abc_weights}
    w_l^{(c)}=K_h(\|\bm{S}^{(c)}_{l} - \bm{S}^{(c)}_{\obs}\|)
\end{align}
, where the bandwidth $h$ is set at the largest value, such that $K_h(\max_{l=1\ldots k} \|\bm{S}^{(c)}_{l} - \bm{S}^{(c)}_{\obs}\|)=0$ to ensure non-zero importance weight for $k$ samples \citep{Fan2018} and mean integrated square error consistency \citep{biau2015}. Regression adjustment seeks to produce adjusted samples $c_l$ but maintain the sample weights and thus assumes the following model for the unadjusted samples $c^{\unadj}$ with mean-zero i.i.d errors $\epsilon_l$ where $E(\epsilon_l^2)<\infty$ for $l=1\ldots,k$:
\begin{align}\label{Reg_model}
c^{\unadj}_l&={m}(\bm{S}^{(c)}_l)+{\epsilon}_l\thinspace.
\end{align}
The estimated regression function $\hat m$ is then a kernel-based local-linear polynomial obtained as a solution of  $\text{argmin}_{\alpha,\beta}\sum_{l=1}^k[c^{\unadj}_l - (\alpha+\beta(\bm{S}^{(c)}_l-\bm{S}^{(c)}_{\obs}))]^2 w_l^{(c)}$. 
Using the empirical residuals $\hat{\epsilon}_l = c^{\unadj}_l-\hat{m}(\bm{S}^{(c)}_l)$, we then construct the adjusted values $c_l = \hat{m}(\bm{S}^{(c)}_{\obs})+\hat{\epsilon}_l$.


\paragraph{Posterior summary of Euclidean parameters $(c,\sigma^2)$.} 
The first stage of our ABC-MH algorithm produces weighted samples $\{c_\ell,w_l^{(c)}\}$, $\{\sigma^2_\ell,w_l^{(\sigma^2)}\}$, $l = 1, \ldots, k$, and we summarize the weighted samples as follows. We illustrate the calculations with $c$, and the calculations for $\sigma^2$ follow similarly. We calculate the posterior median and $95\%$ credible interval by finding the $50$, $2.5$ and $97.5\%$ quantiles, and use the posterior median for the second stage of the proposed ABC-MH algorithm when sampling the tree. In general, for calculating the $q\times 100\%$ quantile, we fit an intercept-only quantile regression of $c_\ell$ with weights $w_l^{(c)}$; this is implemented by \verb"rq" wrapped in the summary function \verb"summary.abc" in the R package \verb"abc". 

\subsection{MH Algorithm for Updating the Tree in the DDT Model.}\label{supp:alg_MH}
\begin{figure}[!htb]
    \centering
    \includegraphics[width=0.95 \linewidth]{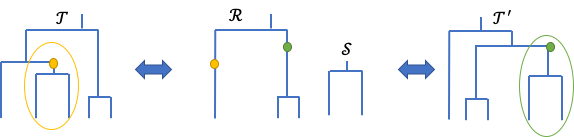}
\caption{Schematic diagram of proposing a candidate tree in MH. (Left) Current tree $\mathcal{T}$ with detach point $u$ (yellow); (Middle) Intermediate subtrees with remaining tree $\mathcal{R}$ and the detached subtree $\mathcal{S}$; (Right) The proposed tree $\mathcal{T}'$ with reattached point $v$ (green).}
\label{sfig:MH_TRupdate}
\end{figure}

In the second stage of Algorithm \ref{algo1}, we have used existing MH tree updates \citep{PYDT}. We briefly describe the proposal for generating a candidate tree $\mathcal{T}'$ from the current tree $\mathcal{T}$ and the acceptance probability. Given the current tree, a candidate tree is proposed in two steps: (i) detaching a subtree from the original tree, and  (ii) reattaching the subtree back to the remaining tree (see Figure \ref{sfig:MH_TRupdate}). In Step i, let $(\mathcal{S},\mathcal{R})$ be the output of the random detach function that divides the original tree $\mathcal{T}$ into two parts at the detaching point $u$, where $\mathcal{S}$ is the detached subtree and $\mathcal{R}$ is the remaining tree. In this paper, we generate the detaching point $u$ by uniformly selecting a node and taking the parent of the node as the detaching point. In Step ii, for the re-attaching point $v$, we follow the divergence and branching behaviors of the generative DDT model by treating subtree $\mathcal{S}$ as a single datum and adding a new datum $\mathcal{S}$ to $\mathcal{R}$. Given the point $v$, a candidate tree $\mathcal{T}'$ results by re-attaching $\mathcal{S}$ back to $\mathcal{R}$ at point $v$.  The time of re-attaching point $t_v$ is then earlier than the time of the root of $\mathcal{S}$ to avoid distortion of $\mathcal{S}$: $t_v < t(\text{root}(\mathcal{S}))$. By choosing $u$ and $v$ as above, we have described the proposal distribution from $\mathcal{T}$ to $\mathcal{T}'$, $q(v,\mathcal{R})$, which is essentially the probability of diverging at $v$ on the subtree $\mathcal{R}$. The acceptance probability is then
\begin{align}\label{acceptProb}
\min\bigg\{1,\frac{f(\mathcal{T}',\Xb)q(u,\mathcal{R})}{f(\mathcal{T},\Xb)q(v,\mathcal{R})}\bigg\}
\end{align}
, where $f(\mathcal{T},\Xb)=f(\mathcal{T},\Xb|c_0,\sigma^2_0)=P(\Xb|\mathcal{T},\sigma_0^2)P(\mathcal{T}|c_0)$, $P(\Xb|\mathcal{T},\sigma^2_0)$ is the likelihood of the tree structure (Proposition 1), $P(\mathcal{T}|c_0)$ is the prior for the tree (the first two terms in Equation (4)), and $c_0$ and $\sigma^2_0$ are representative value chosen from the posterior sample of $c$ and $\sigma^2$, respectively.

\begin{algorithm}[!htb]
    \caption{Two-stage hybrid ABC-MH algorithm}\label{algo1}
    \hspace*{\algorithmicindent} \textbf{Input}: 
    \begin{itemize}
        \item[(a)] Observed data: $\Xb=[\bX_1,\ldots,\bX_{I}]^\transp$ consisting of $I$ points in $\mathbb{R}^J$;
        \item[(b)] Summary statistics $\bm{S}^{(c)},S^{(\sigma^2)}$ defined in the Main Paper Section 3.1.1;
        \item[(c)] Synthetic data of size $N^{\syn}$ and threshold $d\in(0,1)$ with $k=\lceil N^{\syn}d \rceil$, the number of nearest synthetic data sets to retain;
        \item[(d)] Prior for model parameters: $c\sim {\sf Gamma}(a_c,b_c)$, $\frac{1}{\sigma^2}\sim {\sf Gamma}(a_{\sigma^2},b_{\sigma^2})$;
        \item[(e)] Univariate Kernel $K_h(\cdot)$ with bandwidth $h>0$ and compact support.
    \end{itemize}
    \hspace*{\algorithmicindent} \textbf{Output}: 
    \begin{itemize}
        \item[(a)] Posterior samples of $c$ and $\sigma^2$ of size $k=N^{\syn}d$;
        \item[(b)] posterior samples of $(\mathcal{T},\bm t)$.
    \end{itemize}

    \begin{algorithmic}[1]
        \Procedure{Euclidean parameters}{$c,\sigma^2$}
        \For{$l = 1...N^{\syn}$}
            \State Sample Euclidean parameters from prior $c_l \sim {\sf Gamma}(a_c,b_c),\sigma_l^2\sim {\sf Gamma}(a_{\sigma^2},b_{\sigma^2})$;
            \State Simulate data $\Xb_l$ from DDT using $(c_l,\sigma^2_l)$;
            \State Compute: $\bm{S}^{(c)}_{l}$ and $S^{(\sigma^2)}_{l}$ along with $\|\bm{S}^{(c)}_{l} - \bm{S}^{(c)}_{\obs}\|$ and $\|S^{(\sigma^2)}_{l} - S^{(\sigma^2)}_{\obs}\|$.
        \EndFor
        \State Choose $\{(c_{l_s},\sigma^2_{l_s}),s=1,\ldots,k\}$ corresponding to $k$ smallest $\|\bm{S}^{(c)}_{l} - \bm{S}^{(c)}_{\obs}\|$ and $\|S^{(\sigma^2)}_{l} - S^{(\sigma^2)}_{\obs}\|$
        \State Calculate the sample weights $w_{l_s}^{(c)}=K_h(\|\bm{S}^{(c)}_{l_s} - \bm{S}^{(c)}_{\obs}\|)$ and $w_{l_s}^{(\sigma^2)}=K_{h'}(\|S^{(\sigma^2)}_{l_s} - S^{(\sigma^2)}_{\obs}\|)$ based on Equation \eqref{abc_weights};
        \State Compute regression adjusted samples $c_{l_s}$ and $\sigma^2_{l_s}$ with weights $w_{l_s}^{(c)}$ and $w_{l_s}^{(\sigma^2)}$ with the model \eqref{Reg_model} and calculate posterior summary $c_0$ and $\sigma^2_0$ plugging the adjusted $c_{l_s}$ and $\sigma^2_{l_s}$.
        \EndProcedure
        \Procedure{Tree parameters}{$(\mathcal{T},\bm t)$}
        \State Follow the MH algorithm in Section \ref{supp:alg_MH} with fixed $c_0$ and $\sigma^2_0$ at the posterior median values and compute acceptance probabilities with Equation \ref{acceptProb}.
    \EndProcedure
    \end{algorithmic}
    \end{algorithm}

\section{Tree Projection of Pairwise iPCP Matrix}\label{supp:iPCP}
\begin{figure}[!htb]
    \centering
    \includegraphics[width=0.95 \linewidth]{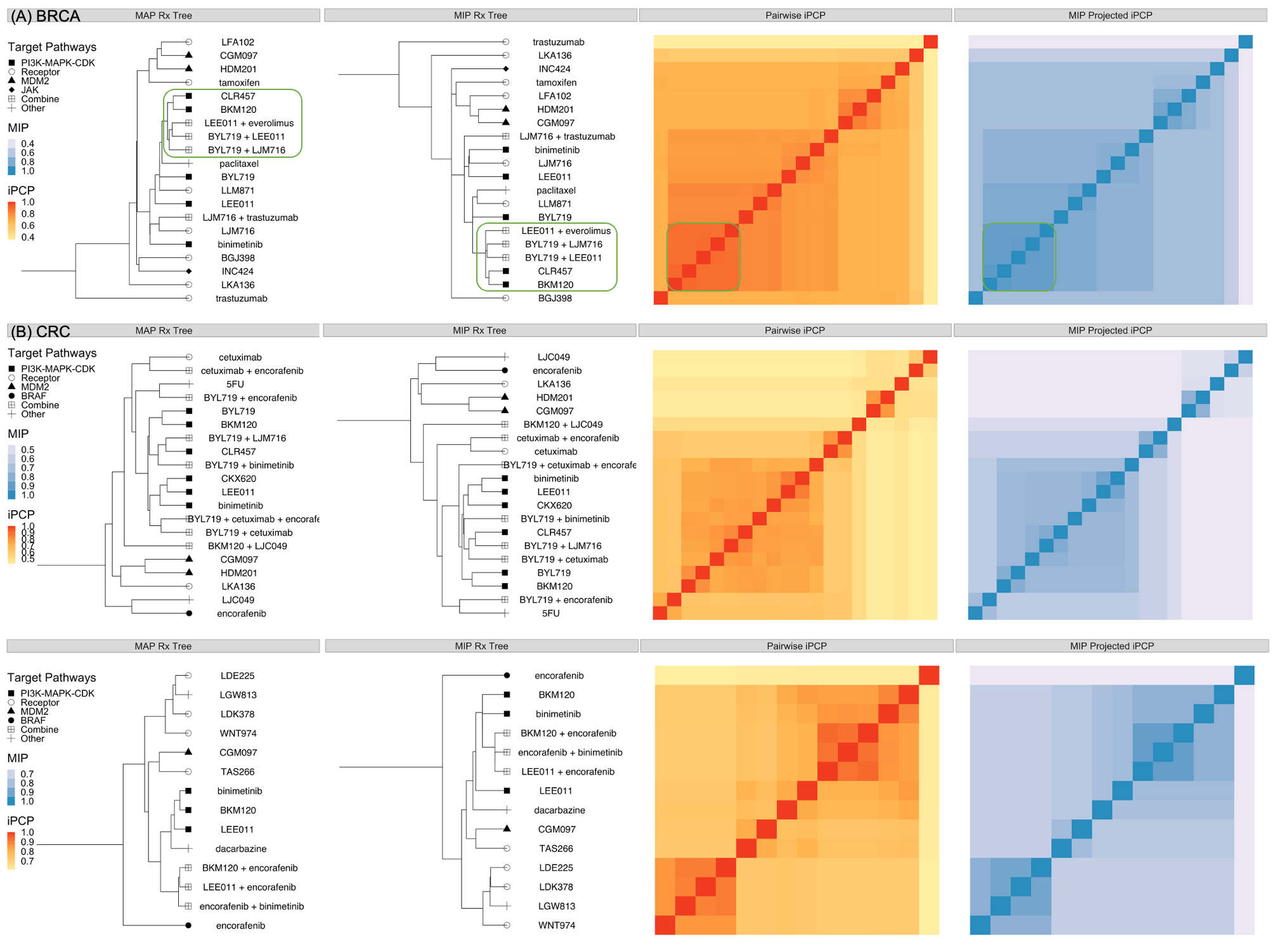}
\caption{Comparison between (Left two columns) the tree structure from the MAP and the projected iPCP matrix (MIP tree) and (Right two columns) the matrix from the original iPCP matrix and the projected iPCP matrix for (A) breast cancer, (B) colorectal cancer and (C) melanoma. The matrix from the original iPCP and the MIP projected iPCP matrix are aligned by the MIP tree.}
\label{sfig:MIP}
\end{figure}

In the Main Paper Section 3.2, we mentioned that a pairwise iPCP matrix $\bm \Sigma$ with entries $\text{iPCP}_{i,i'}, i,i'=1,\ldots,I$ need not to be a tree-structured matrix and we address the projection of $\Sigma$ on to the space of tree-structured matrices here. Given $L>1$ posterior trees with $I$ leaves and the corresponding pairwise iPCP matrix $\bm{\Sigma}=\Big(\text{iPCP}_{i,i'}\Big)$, each entry of iPCP matrix can be express as $\text{iPCP}_{i,i'}=\frac{\sum_{l=1}^L t_{i,i'}^{(l)}}{L}$, where $t_{i,i'}^{(l)}$ is the divergence time of leaves $i$ and $i'$ in the $l$-th posterior tree. Obviously, every entry of the iPCP matrix takes the element-wise Monte Carlo average over $L$ tree-structured matrix and breaks the inequalities (2) and (3) in the Main Paper. Following the work of \citet{pmid22081761}, by representing a tree as a tree-structured matrix, we can project $\Sigma$ on to the closest tree-structured matrix in terms of Frobenius norm. The projection can be formulated as a constrained mixed-integer programming (MIP) problem:
\begin{gather*}
    \argmin_{\bm{\Sigma}^\cT} \quad  \norm{\bm{\Sigma}-\bm{\Sigma}^{\cT}}_F \\
    \textrm{s.t.} \quad \Sigma^\cT_{i,i'}\geq 0; \thickspace \Sigma^\cT_{i,i}\geq \Sigma^\cT_{i,i'}; \thickspace  \Sigma^\cT_{i,i'}\geq \min(\Sigma^\cT_{i,i{''}},\Sigma^\cT_{i',i{''}}), \text{ for all } i\neq i' \neq i{''}.
\end{gather*}
We applied the projection on the pairwise iPCP matrix from the breast cancer (panel (A)), colorectal cancer (panel (B)) and melanoma (panel (C)) data of NIBR-PDXE and show the result in the Figure \ref{sfig:MIP}. In Figure \ref{sfig:MIP}, the MAP tree, the tree representation of projected iPCP matrix (MIP tree), the original iPCP matrix and the projected iPCP matrix are shown in from the left to the right columns, respectively. From the left two columns of the tree structures, we found that trees from the MAP and MIP show similar pattern and the MIP tree allows a non-binary tree structure. For example, three combination therapies and two PI3K inhibitors (CLR457 and BKM120) framed by a box form a tight subtree in both MAP and MIP tree, but the subtree in the MIP is non-binary. For the iPCP matrix, high element-wise correlation $\text{Cor}(\Sigma^\cT_{i,i'},\Sigma_{i,i'})$ between the original iPCP $\Sigma$ and the projected iPCP $\Sigma^\cT$ are presented (BRCA: 0.9987; CRC: 0.9962; CM: 0.9918).

\section{Simulation Studies of Euclidean Parameters}\label{supp:Simulation_Euclidean}
In this section, we empirically compare the Euclidean parameters of $c$ and $\sigma^2$ from ABC of the proposed two-stage algorithm and single-stage MCMC. We organize this section as follows. We first compare other candidate summary statistics of $c$ and $\sigma^2$ for ABC in Section \ref{supp:otherSS}. In Section \ref{supp:RealParamInfer}, we illustrate the superior inference performance of Euclidean parameters from ABC than single-stage MCMC through a series of simulations. Section \ref{supp:Diag_Sen} offers the diagnostic statistics and the sensitivity analysis for ABC stage of the proposed two-stage algorithm and checks the convergence of $c$ and $\sigma^2$ for the single-stage MCMC.

\paragraph{Simulation setup.} \label{sec:Simulation_setup} 
For illustrative purposes, we fixed the observed PDX data matrix with $50$ treatments ($I=50$) and $10$ PDX mice ($J=10$) in all simulation scenarios. In addition, we let $c$ and $\sigma^2$ take values from $\{0.3,0.5,0.7,1\}$ and $\{0.5,1\}$ respectively to mimic the PDX data with tight and well-separated clusters. For each pair of $(c,\sigma^2)$, $200$ replicated experiments with different tree and observed PDX data matrices were independently drawn according to the DDT generating model. We specify a prior distribution for $c\sim {\sf Gamma}(2,2)$ with shape and rate parameterization. For diffusion variance $\sigma^2$, let $1/\sigma^2\sim {\sf Gamma}(1,1)$. We compare ABC-MH of the proposed two-stage algorithm against two alternatives based on single-stage MH algorithms \citep{Neal2003} (see details in Section \ref{supp:alg_MH}). The first one initializes at the true parameter values and the true tree, referred to as $\MH_{\sf true}$. The idealistic initialization at the truth is a best case scenario in applying existing MH algorithm to inferring DDT models. The second alternative, referred to as ${\MH}_{\sf default}$, initializes $(c,\sigma^2)$ by a random draw from the prior; the unknown tree is initialized by agglomerative hierarchical clustering with Euclidean distance and squared Ward's linkage \citep{Murtagh2014} -- thus providing a fair apples-to-apples comparison. For the ABC, we generated $N^\syn$ synthetic data of $c$ and $\sigma^2$ and kept $k=\lceil N^\syn d\rceil$ nearest samples in terms of the $\|\bm{S}^{(c)}_{l} - \bm{S}^{(c)}_{\obs}\|$ and $\|S^{(\sigma^2)}_{l} - S^{(\sigma^2)}_{\obs}\|$. We varied the number of synthetic data $N^\syn$ and the threshold parameter $d\in (0,1)$ under different settings and we specified $N^\syn$ and $d$ in each of the following sections. We ran two MH algorithms with 10,000 iterations and discarded the first 7,000 iterations.


\paragraph{Performance metrics for Euclidean parameters.} We used two algorithm performance metrics to compare our algorithm to the classical single-stage MCMC algorithms. First we computed the effective sample sizes for each Euclidean parameter $c$ and $\sigma^2$ ($\ESS_c$ and $\ESS_{\sigma^2}$) given a nominal sample size (NSS) kept for posterior inference. ESS for each parameter represents the number of independent draws equivalent to NSS posterior draws of correlated ($\MH_{\sf true}$ and $\MH_{\sf default}$) or independent and unequally weighted samples (ABC stage of the proposed algorithm). We let NSS for MH algorithms be the number of consecutive posterior samples in a single chain after a burn-in period; let NSS for ABC be $k$ as in Step 6, Algorithm \ref{algo1}. For $c$ and $\sigma^2$, the ESS of MH \citep{gelmanbda13} is estimated by ${\NSS}/(1+\sum_{t=1}^\infty\hat{\rho_t})$ where $\hat{\rho_t}$ is the estimated autocorrelation function with lag $t$ \citep{geyer2011introduction}. The ESS for ABC \citep{Fan2018} is the reciprocal of the sum of squared normalized weights, $1/\sum_{l=1}^k \tilde{W}_l^2$, where $\tilde{W}_l= w_l/\sum_{l'=1}^k w_{l'}$ (see weights, $w_l$, in Equation \eqref{abc_weights}). Second, we evaluated how well did the posterior distributions recover the true $(c,\sigma^2)$. We computed the mean absolute percent bias for $c$ and $\sigma^2$: $|\mathbb{E}\{c\mid \Xb\}-c|/c$ and $|\mathbb{E}\{\sigma^2\mid \Xb\}-\sigma^2|/\sigma^2$, respectively. We also computed the empirical coverage rates of the nominal $95\%$ credible intervals (CrI) for $c$ and $\sigma^2$.

\subsection{Other Choices of Summary Statistics}\label{supp:otherSS}
Proposition 1 points towards other potential summary statistics for the first stage of Algorithm \ref{algo1} that uses ABC to produce weighted samples to approximate the posterior distributions for $c$ and $\sigma^2$. Here we consider a few such alternatives with $N^\syn=600,000$ and $d=0.5\%$ and empirically compare their performances to the summary statistics used in the Main Paper ($\bm{S}^{(c)}$ and $S^{(\sigma^2)}$) in terms of the mean absolute percent bias in recovering the true parameter values of $c$ and $\sigma^2$. 
    
\begin{figure}[!htb]
        \centering
        \includegraphics[width=\linewidth]{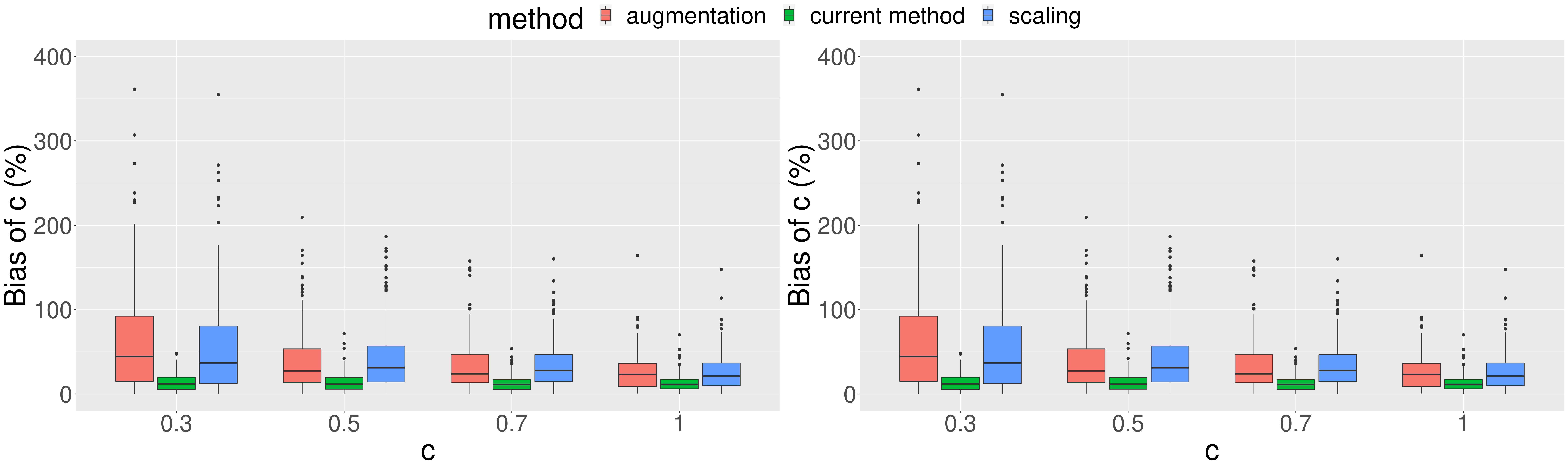}
    \caption{Comparison among different summary statistics for $c$ (red: ($\bm{Q}_{\bm{T}}, S^{(\sigma^2)})$; green: $\bm{S}^{(c)}$; blue: $\bm{Q}_{\bm{T}}/S^{(\sigma^2)}$) under different values of $\sigma^2$ in terms of the mean absolute percent bias. (Left) $\sigma^2=0.5$; (Right) $\sigma^2=1$.}
    \label{fig:otherSS_bias_c}
    \end{figure}

\paragraph{Summary statistic for $c$.} Unlike building $\bm{S}^{(c)}$ based on the inter-point distance, the off-diagonal terms of $\bm{T}=\sum_{j}\bX_{\cdot,j}\bX_{\cdot,j}^\transp$ (see the definition of $\bm{T}$ in Lemma 1 in Main Paper) is another potential summary statistic for $c$. Since the divergence parameter $c$ affects the marginal likelihood implicitly through the divergence time ${\bm t}$, the summary statistics for $\bm t$ is informative for $c$. From Proposition 1, $\bm{T}$ is sufficient for $\sigma^2\Sigma_{\mathcal{T}}$, where the off-diagonal terms of $\sigma^2\Sigma_{\mathcal{T}}$ taking the form $\sigma^2 t_d,d=1\ldots n-1$ and containing unrelated information from $\sigma^2$. Let $\bm{Q}_{\bm{T}}$ be a vector of the 10th, 25th, 50th, 75th and 90th percentiles of the off-diagonal terms of $\bm{T}$. Because $\bm{T}$ is sufficient for $\sigma^2\Sigma_{\mathcal{T}}$ and involves extra Gaussian diffusion variance parameter, we can design alternative summary statistics based on $\bm{Q}_{\bm{T}}$ through (i) augmentation, $(\bm{Q}_{\bm{T}}, S^{(\sigma^2)})$ or (ii) scaling, ${\bm{Q}_{\bm{T}}}/{S^{(\sigma^2)}}$. From Figure \ref{fig:otherSS_bias_c}, $\bm{S}^{(c)}$ proposed in the Main Paper outperformed the summary statistics from $\bm{Q}_{\bm{T}}$ by producing less biased posterior mean estimates. 

\begin{figure}[!htb]
        \centering
        \includegraphics[width=\linewidth]{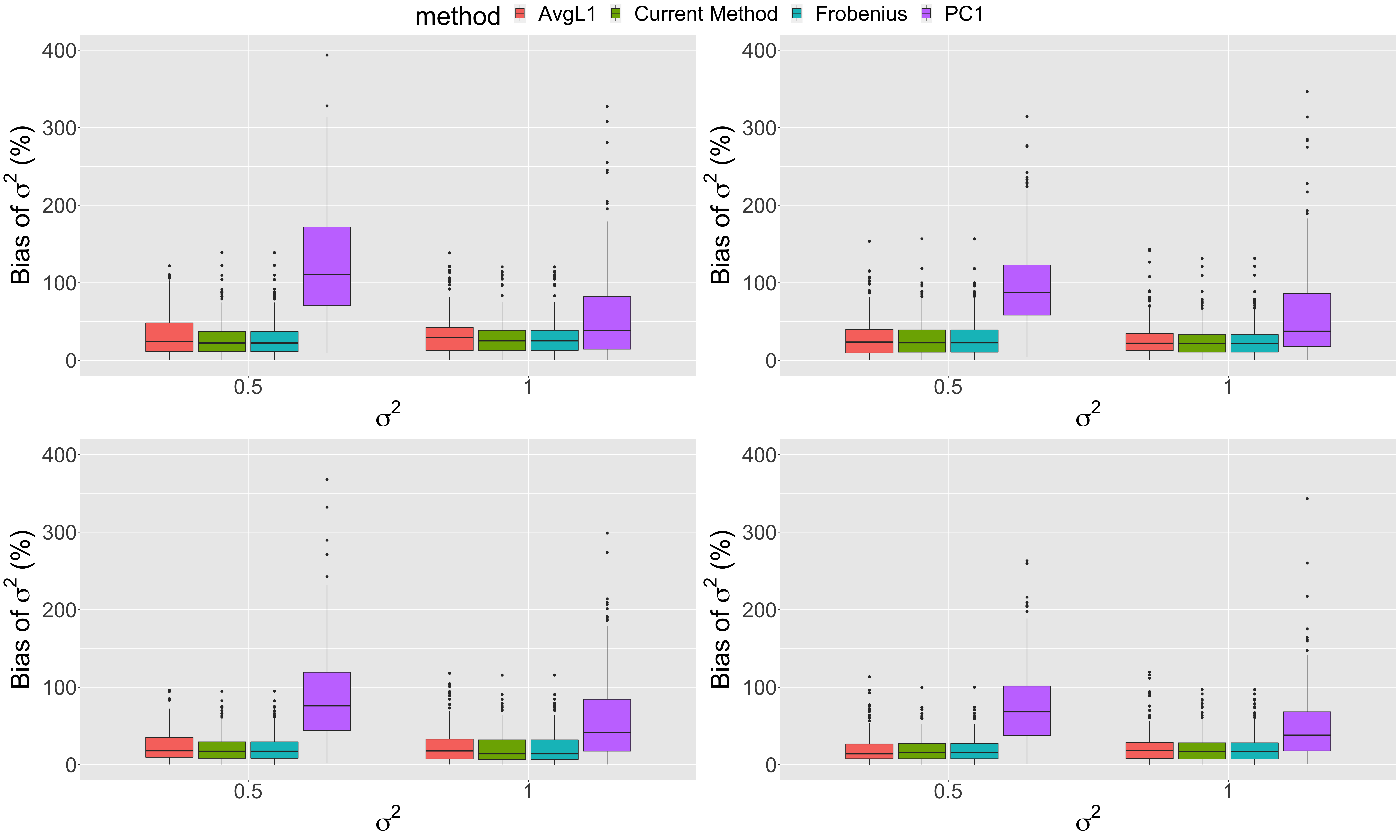}
    \caption{Comparison among different summary statistics for $\sigma^2$ under different values of $c$ in terms of the mean absolute percent bias. (Upper Left) $c=0.3$; (Upper Right) $c=0.5$; (Lower Left) $c=0.7$; (Upper Right) $c=1.0$.}
    \label{fig:otherSS_bias_s}
    \end{figure}

\paragraph{Summary statistic for $\sigma^2$.} Following Proposition 1, several matrix functionals on the data $\Xb$ or statistics $\bm{T}$ can be considered as alternatives to $S^{(\sigma^2)}$. We compare performance of three candidates:  (i) average $L_1$ norm (AvgL1) of columns: $\frac{1}{J}\sum_{j=1}^J|\bX_{\cdot,j}|_1$; (ii) Frobenius norm of $\Xb$; and, (iii) vector containing 10th, 25th, 50th, 75th and 90th percentiles of first principal component (PC1) of $\Xb$. From Figure \ref{fig:otherSS_bias_s}, the first three methods are comparable while ABC based on principal components shows larger bias due to the information loss.

\subsection{Posterior Inference of Euclidean Parameters}\label{supp:RealParamInfer}
In this section, we show that two-stage algorithm (ABC-MH) outperforms the single-stage MCMC (MH) for real parameters in terms of (i) stable effective sample size (ESS) for $(c,\sigma^2)$; (ii) similar or better inference on $(c,\sigma^2)$, as ascertained using mean absolute percent bias and nominal $95\%$ credible intervals. 

\subsubsection{Stable Effective Sample Sizes of ABC-MH}  We calculated ESS-to-NSS ratios at varying truths of $c$ and $\sigma^2$. To illustrate, we matched the NSS budget of ABC with that of MH (NSS = $3,000$) by keeping $d=0.5\%$ of $N^{\syn}=600,000$ synthetic data sets that are closest to the observed data in terms of the summary statistic for each parameter (Step 6 of Algorithm \ref{algo1}). Table \ref{stab:ESS-to-NSS} shows that the $\ESS_c/\NSS$ and $\ESS_{\sigma^2}/\NSS$ ratio from ABC is stable between $0.64$ to $0.68$ and around $0.83$ across different $c$ and $\sigma^2$ values, respectively. In contrast, the $\ESS_c/\NSS$ ratio for MH quickly deteriorates (${\MH_{\sf true}}$: $0.97$ to $0.41$; ${\MH_{\sf default}}$: $0.73$ to $0.35$) as $c$ increases from $0.3$ to $1$ and $\ESS_{\sigma^2}/\NSS$ for MH are extremely poor ($<0.06$) across different values of $c$ and $\sigma^2$. MH produced very good $\ESS_c$ under small value $c=0.3$ but poor $\ESS_c$ under $c=1$. As a result, under larger values of $c$, MH algorithms must run longer to reach a target $\ESS_c$. Although $\ESS_c$ for ABC is not as high as ${\MH_{\sf true}}$ or ${\MH_{\sf default}}$ at $c=0.3$, the stability of $\ESS_{c}$ of ABC means that a predictably constant NSS is needed for conducting posterior inference across different values of $c$. Finally, the $\ESS_{\sigma^2}$ for the diffusion variance parameter from MH algorithms are strikingly smaller than ABC, indicating ABC should be preferred.


\begin{table}[!htb]
    \caption{ESS-to-NSS ratios between ABC-MH ($d=0.5\%$), $\textrm{MH}_{\sf true}$, and $\textrm{MH}_{\sf default}$. All values here are obtained from $200$ independent replications. For each random replication at $(c,\sigma^2)$. All methods were controlled to produce identical NSS with size $3,000$.}
    \label{stab:ESS-to-NSS}
    \centering
    \begin{tabular}{llllll}
    \toprule
    & & \multicolumn{2}{c}{ESS/NSS(sd) for $c$} & \multicolumn{2}{c}{ESS/NSS(sd) for $\sigma^2$}\\ \cmidrule(r){3-4} \cmidrule(r){5-6}
    $c$ & method & $\sigma^2=0.5$     & $\sigma^2=1$  & $\sigma^2=0.5$     & $\sigma^2=1$ \\
    \midrule
    \multirow[c]{3}{*}{0.3} & ABC-MH & 0.68(0.032) & 0.67(0.027) & 0.83(0.0048) & 0.83(0.0042) \\ 
    & $\textrm{MH}_{\sf true}$ & 0.97(0.11) & 0.96(0.13) & 0.051(0.061) & 0.056(0.072) \\ 
    & $\textrm{MH}_{\sf default}$ & 0.73(0.33) & 0.67(0.34) & 0.028(0.043) & 0.038(0.08) \\ 
    \midrule
    \multirow[c]{3}{*}{0.5}& ABC-MH & 0.66(0.02) & 0.65(0.018) & 0.83(0.0047) & 0.83(0.0044) \\ 
    & $\textrm{MH}_{\sf true}$ & 0.85(0.23) & 0.83(0.24) & 0.034(0.042) & 0.045(0.067) \\ 
    & $\textrm{MH}_{\sf default}$ & 0.66(0.35) & 0.62(0.34) &  0.033(0.051) & 0.041(0.067) \\ 
    \midrule
    \multirow[c]{3}{*}{0.7}& ABC-MH & 0.65(0.017) & 0.64(0.017) & 0.83(0.0047) & 0.83(0.004) \\ 
    & $\textrm{MH}_{\sf true}$ & 0.63(0.31) & 0.67(0.32) & 0.024(0.027) & 0.029(0.038) \\ 
    & $\textrm{MH}_{\sf default}$ & 0.53(0.33) & 0.51(0.35) & 0.028(0.039) & 0.038(0.072) \\ 
    \midrule
    \multirow[c]{3}{*}{1.0}& ABC-MH & 0.65(0.017) & 0.64(0.017) & 0.83(0.0044) & 0.83(0.0041) \\ 
    & $\textrm{MH}_{\sf true}$ & 0.41(0.3) & 0.44(0.32) & 0.019(0.026) & 0.019(0.023) \\ 
    & $\textrm{MH}_{\sf default}$ & 0.35(0.29) & 0.35(0.29) & 0.022(0.026) & 0.022(0.027) \\ 
    \bottomrule
    \end{tabular}
\end{table}

\subsubsection{Superior Quality Posterior Inference of ABC-MH}
\begin{figure}[!htb]
    \centering
    \includegraphics[width=\linewidth]{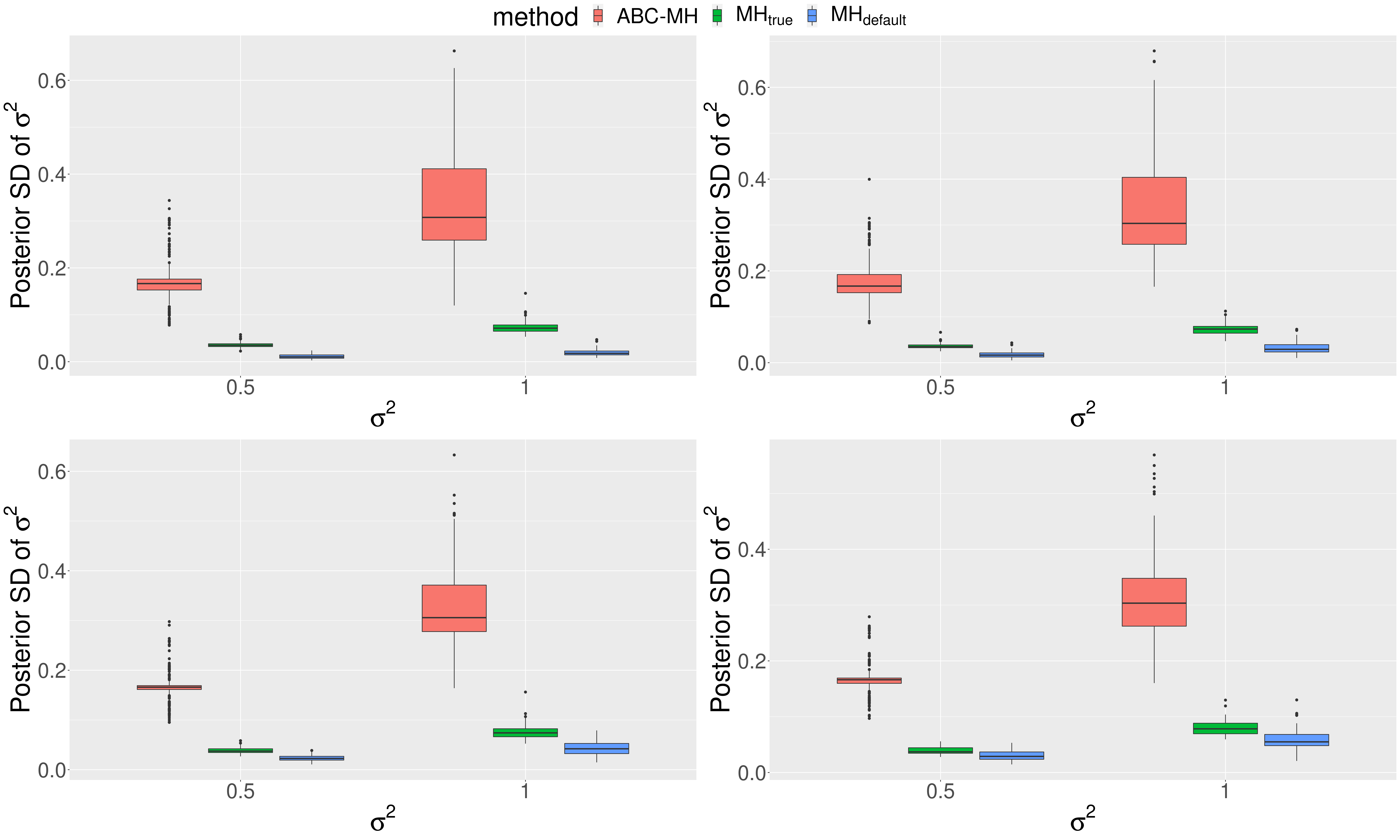}
    \caption{(Upper left) $c=0.3$; (Upper right) $c=0.5$; (Lower left) $c=0.7$; (Lower right) $c=1.0$. The posterior standard deviation of $\sigma^2$ from MH (green and blue) are close to zero across different true $c$ showing MH is stuck. Results are based on 200 replications.}
\label{sfig:Vars}
\end{figure}

Does ABC give better posterior inference with a fixed computational budget? To make fair comparisons, we fixed a total CPU time and used the same computing processor to run the ABC (1st stage of Algorithm 1) and MH algorithms. Let $t_{\sf MH}$ and $t_{\sf ABC}$ be the estimated CPU time for generating one iteration in MH and one synthetic data in ABC on the same processor. Note, $t_{\sf MH}$ includes the additional time for proposing a valid tree. By varying the number of synthetic samples, we can match the total CPU time used by ABC with that of MH algorithms which were run for $10,000$ iterations. We generated ${10,000t_{\sf MH}}/t_{\sf ABC}=17,345$ synthetic data sets and took $d=5\%$ with summary statistics $\bm{S}^{(c)}$ and $S^{(\sigma^2)}$ (see different values of $d$ in Section \ref{supp:d_sensitivity}) for ABC. Table \ref{stab:d5percent} shows that ABC produced posterior samples that confer comparable inferences about $c$ in terms of the bias and coverage of nominal $95\%$ CrIs. The posterior mean of $c$ from ABC is comparable to that from $\MH_{\sf true}$ and less biased than $\MH_{\sf default}$ for all settings. The coverage rates of the nominal $95\%$ CrIs from ABC are comparable to $\MH_{\sf true}$ but higher than $\MH_{\sf default}$. $\MH_{\sf true}$, however, is initialized at true values and is unrealistic in practice. We observed $\MH_{\sf true}$ sometimes failed to converge (Table \ref{stab:ConvPer}), stuck around the initial true values and resulted in deceptively low biases and good coverage rates. Turning to the inference of $\sigma^2$, ABC offers a much better alternative to MH algorithms in terms of smaller bias in the posterior mean and better coverage of the $95\%$ credible intervals (Table \ref{stab:d5percent}). This is primarily caused by the difficulty of MH in exploring the posterior distribution of $\sigma^2$ resulting in chains with high auto-correlations. The squeezed boxplots in Figure \ref{sfig:Vars} indicate that the chains for $\sigma^2$ in $\MH_{\sf true}$ and $\MH_{\sf default}$ were almost always slowly mixing and stuck around the initial values. In addition, unlike the serial nature of MH, ABC can be further parallelized to reduce the wall clock time to a fraction of what is required by MH using multicore processors. Although parallelizing MH with techniques such as consensus MCMC \citep[e.g.,][]{scott2016bayes} is possible, the parallelized ABC does not require data splitting and will not trade the quality of posterior inference for computational speed.

\begin{sidewaystable}
    \caption{Comparison of inferential performance for $c$ and $\sigma^2$ between ABC-MH ($d=5\%$), $\textrm{MH}_{\sf true}$, and $\textrm{MH}_{\sf default}$. All values here are obtained from $200$ independent replications. For each random replication at $(c,\sigma^2)$, all methods were run for identical total CPU time and only converged chains from MH algorithms were included.}
    \label{stab:d5percent}
    \centering
    \begin{tabular}{llllllllll}
    \toprule
    & & \multicolumn{2}{c}{Percent Bias(sd) for $c$}& \multicolumn{2}{c}{Coverage(sd) for $c$} & \multicolumn{2}{c}{Percent Bias(sd) for $\sigma^2$}& \multicolumn{2}{c}{Coverage(sd) for $\sigma^2$}\\
    \cmidrule(r){3-4} \cmidrule(r){5-6} \cmidrule(r){7-8} \cmidrule(r){9-10}
    $c$ & method & $\sigma^2=0.5$ & $\sigma^2=1$ & $\sigma^2=0.5$ & $\sigma^2=1$ & $\sigma^2=0.5$ & $\sigma^2=1$ & $\sigma^2=0.5$ & $\sigma^2=1$ \\
    \midrule
    \multirow[c]{3}{*}{0.3} & ABC-MH & 12(9.4) & 13(9.9) & 98(0.99) & 99(0.71) & 28(25) & 31(25) & 90(2.1) & 88(2.3) \\ 
    & $\textrm{MH}_{\sf true}$ & 13(9.8) & 12(9.5) & 94(2) & 95(1.9) & 9.4(7.1) & 9(6.6) & 80(3.4) & 82(3.3) \\ 
    & $\textrm{MH}_{\sf default}$ & 45(20) & 46(20) & 33(5.5) & 30(6.1) & 71(12) & 72(11) & 0(0) & 0(0)\\ 
    \midrule
    \multirow[c]{3}{*}{0.5}& ABC-MH & 15(11) & 15(11) & 92(1.9) & 93(1.8) & 28(26) & 27(22) & 90(2.2) & 94(1.7) \\ 
    & $\textrm{MH}_{\sf true}$ & 11(9) & 11(8.6) & 97(1.7) & 97(1.6) & 8.6(6.7) & 9.9(7) & 80(4) & 78(4.1) \\ 
    & $\textrm{MH}_{\sf default}$ & 33(18) & 31(19) & 60(5.5) & 57(5.7) & 54(17) & 57(16) & 1.2(1.2) & 1.3(1.3)\\ 
    \midrule
    \multirow[c]{3}{*}{0.7}& ABC-MH & 13(10) & 14(11) & 96(1.5) & 93(1.8) & 21(18) & 22(20) & 97(1.2) & 94(1.6)\\ 
    & $\textrm{MH}_{\sf true}$ & 12(9.1) & 12(9.1) & 95(2.6) & 96(2.1) & 11(8.3) & 11(8.3) & 70(5.3) & 68(4.9) \\ 
    & $\textrm{MH}_{\sf default}$ & 25(15) & 27(16) & 73(5.5) & 69(5.9) & 38(17) & 41(19) & 12(4) & 8.1(3.5) \\ 
    \midrule
    \multirow[c]{3}{*}{1.0}& ABC-MH & 14(11) & 14(13) & 95(1.5) & 94(1.6) & 19(18) & 21(18) & 98(1.1) & 96(1.5) \\ 
    & $\textrm{MH}_{\sf true}$ & 11(7.6) & 13(11) & 97(2) & 92(3.5) & 13(9.1) & 10(7.7) & 64(5.8) & 85(4.6) \\ 
    & $\textrm{MH}_{\sf default}$ & 14(11) & 16(14) & 93(3.5) & 89(3.8) & 24(15) & 27(16) & 35(6.5) & 21(5.1) \\ 
    \bottomrule
    \end{tabular}
\end{sidewaystable}

\subsection{Algorithm Diagnostics}\label{supp:Diag_Sen}
Here we examine the convergence of MH through the Geweke statistics \citep{Geweke92evaluatingthe} and the goodness of fit for ABC. Specifically, two important hyper-parameters are involved in ABC: (i) the kernel bandwidth $h$ for samples weights in Equation \ref{abc_weights} and (ii) the threshold $d$ for $k=\lceil N^\syn d \rceil$ nearest samples in the Step 6 of Algorithm \ref{algo1}. We follow the test from \citet{ABCCoverage} to justify the kernel bandwidth $h$ and conduct the sensitivity analysis for threshold $d$ to understand how threshold $d$ affects the result in terms of the inferential performance.

\subsubsection{Convergence of MH Chains in Simulations}\label{supp:ConvMH} 
In all of our simulations, we ran MH for $10,000$ iterations. Table \ref{stab:ConvPer} shows that the percentages of the converged MH chains for $200$ replications are between $12.5$ and $68.5\%$ within a total $10,000$ iterations (based on Geweke statistic). Running the chains longer will increase these percentages. In contrast, with appropriate choice of bandwidth and the fraction of synthetic samples to keep, ABC does not involve convergence issues and according to Section \ref{supp:RealParamInfer} achieves better ESS for a fixed NSS and similar or better quality posterior inference for fixed CPU time. 


\begin{table}[!htb]
  \caption{Percentage of converged chains for (i) MH initialized at true $(c,\sigma^2)$ ($\textrm{MH}_{\sf true}$), and (ii) MH initialized randomly from prior ($\textrm{MH}_{\sf default}$). All values here are obtained from 200 independent replications.}
  \label{stab:ConvPer}
  \centering
  \begin{tabular}{llllll}
    \toprule
    & & \multicolumn{2}{c}{Convergence $\%$ for $c$} & \multicolumn{2}{c}{Convergence $\%$ for $\sigma^2$}\\
    \cmidrule(r){3-4} \cmidrule(r){5-6} 
    $c$ & method & $\sigma^2=0.5$     & $\sigma^2=1$  & $\sigma^2=0.5$     & $\sigma^2=1$ \\
    \midrule
    \multirow[c]{2}{*}{0.3} & $\textrm{MH}_{\sf true}$ & 68.0 & 68.5  & 16.5 & 22.5\\ 
    & $\textrm{MH}_{\sf default}$ & 36.5 & 28.5 &12.5 & 16.0 \\ 
    \midrule
    \multirow[c]{2}{*}{0.5} & $\textrm{MH}_{\sf true}$ & 50.0 & 52.0 & 23.0 & 29.5\\ 
    & $\textrm{MH}_{\sf default}$ & 40.5 & 37.5 & 18.5 &23.5 \\ 
    \midrule
    \multirow[c]{2}{*}{0.7} & $\textrm{MH}_{\sf true}$ & 38.0 & 46.0 &26.5 &27.0 \\ 
    & $\textrm{MH}_{\sf default}$ & 33.5 & 31.0 & 14.5 & 25.0 \\ 
    \midrule
    \multirow[c]{2}{*}{1.0} & $\textrm{MH}_{\sf true}$ & 35.0 & 30.5 & 20.5 & 30.5\\ 
    & $\textrm{MH}_{\sf default}$ & 27.5 & 33.0 & 14.0 & 25.0 \\
    \bottomrule
  \end{tabular}
\end{table}

    \subsubsection{Diagnostics for ABC}\label{supp:ABC_Diag}
    We empirically justify the choice of the kernel bandwidth $h$ and the goodness of approximation in ABC algorithm by the calibration method from \citet{ABCCoverage} based on the coverage property of the credible interval. Suppose we generated pseudo-observed data $\Xb_e$ in the $e$th replication from the DDT model with parameter $(c_e,\sigma^2_e)$, where $c_e$ and $\sigma^2_e$ are random draws from the prior ($c_e\sim {\sf Gamma}(a_c,b_c), 1/{\sigma^2_e} \sim {\sf Gamma}(a_{\sigma^2},b_{\sigma^2})$) and $e=1\ldots E$. Once the tuning parameters ($N^{\syn},d,h$) are decided, Algorithm \ref{algo1} will output regression adjusted sample $(c_\ell,\sigma^2_\ell)$ with size $\ell=1,\ldots,k;k=\lceil N^{\syn}d \rceil$ based on the input data $D$. We describe diagnostics for $c$, and note that an identical description applies to $\sigma^2$ as well. According to \citet{Cook2006}, the ABC procedure produces reliable approximations of the posterior if the random variables $q^{(c)}_e:=\frac{1}{k}\sum_{l=1}^k \mathbb{I}_{\{c_{\ell} >c_e\}}$ follow a uniform distribution over the interval $(0,1)$. Accordingly, \citet{ABCCoverage} suggest a goodness-of-fit test $H_0: q_e^{(c)}\sim {\sf Unif(0,1)}$ as a diagnostic in order to calibrate ABC. If the test fails to reject the null hypothesis, the empirical quantiles can be viewed as being indistinguishable from the uniform distribution, and the credible interval from the posterior samples would show the asserted coverage. We use the Kolmogorov–Smirnov statistic to carry out the test, follow the simulation setting with $I=50$ and $J=10$, and reuse $600,000$ synthetic data sets. The synthetic data is randomly split into two non-overlapping subsets: training data with size $597,000$ and pseudo-observed data with size $E=3,000$. Again, we run the ABC part of Algorithm \ref{algo1} by treating each of the pseudo-observed data sets as the actually observed data with $N^{\syn}=597,000$ and $d=0.5\%$. We obtained statistically non-significant KS statistics for $c$ and $\sigma^2$ ($p$-values: $0.61$ for $c$, 0.71 for $\sigma^2$). The 95\% credible intervals from ABC showed 94.9\% and 95.93\% empirical coverage rates which are close to the nominal level.

\begin{figure}[!htb]
    \begin{subfigure}{0.495\textwidth}
        \centering
        \includegraphics[width=\linewidth]{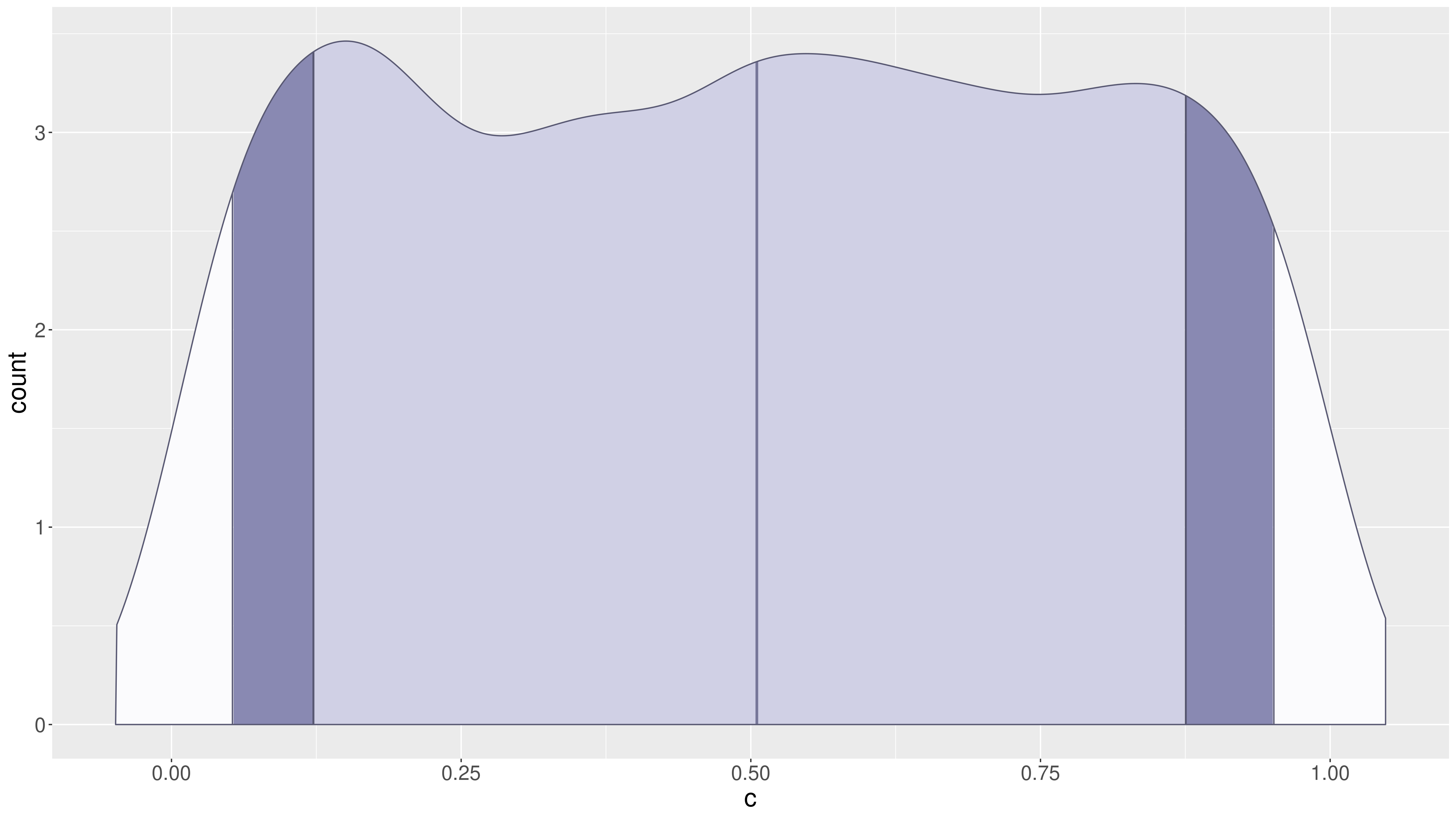}
        \caption{Empirical quantiles at true $c$ }
        \label{fig:eCDF_c}
    \end{subfigure}
    \begin{subfigure}{0.495\textwidth}
        \centering
        \includegraphics[width=\linewidth]{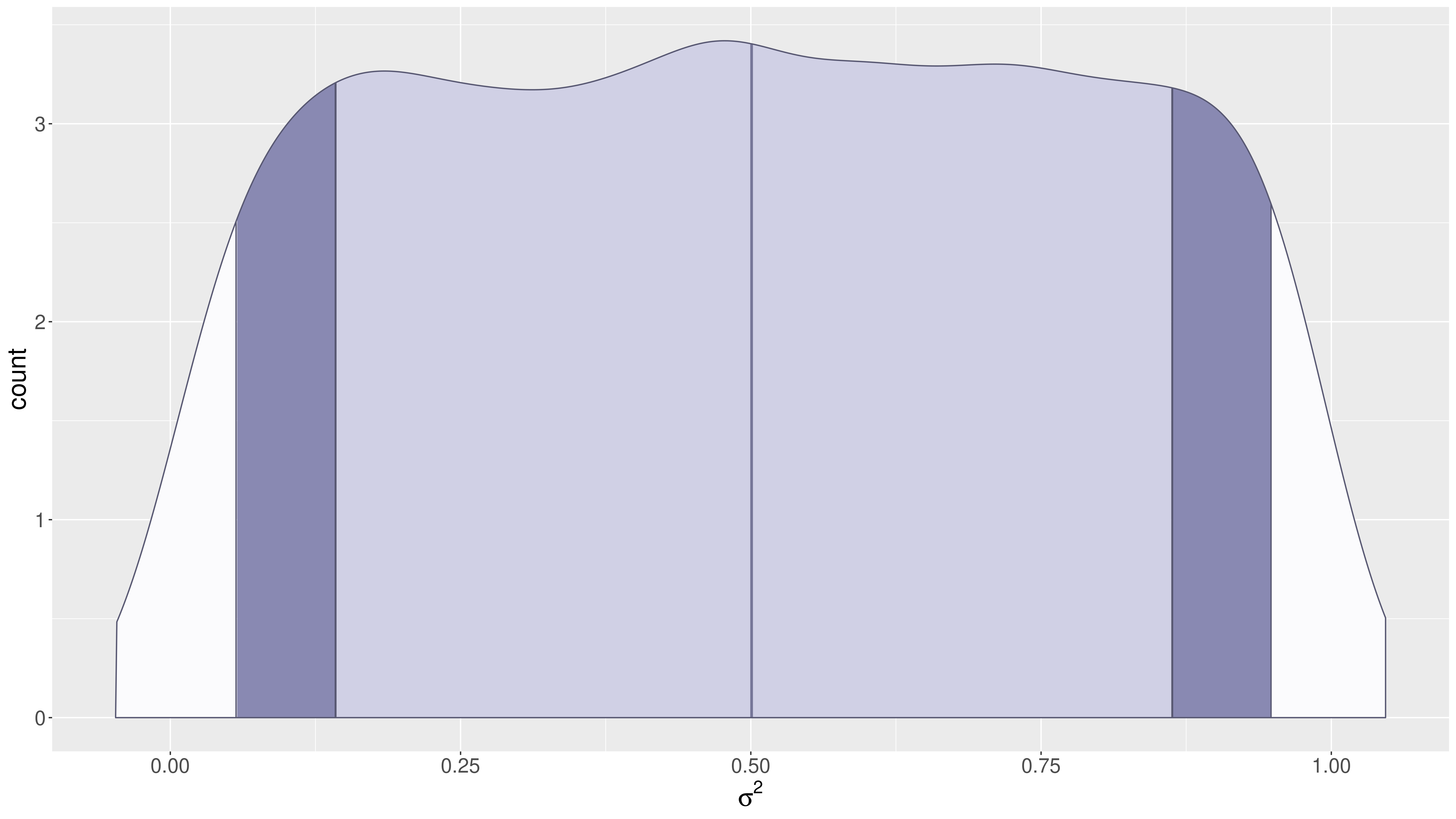}
        \caption{Empirical quantiles at true $\sigma^2$}
        \label{fig:eCDF_s}
    \end{subfigure}
\caption{The empirical quantiles at the true value follow the standard uniform distribution indicating calibrated ABC. Results are based on $3,000$ independent draws from the prior.}
\label{fig:eCDF}
\end{figure} 

\subsubsection{Sensitivity Analysis of \texorpdfstring{$k$}{TEXT} Nearest Samples}\label{supp:d_sensitivity}
In the previous section, we have used a simple diagnostic procedure to show the choice of bandwidth parameter $h$ is reasonable. Here we focus on conducting additional simulations to investigate how does varying values of $d$ in the Step 6 of Algorithm \ref{algo1} impact the inferential performance of ABC. We focus on $c$ to illustrate the main points. Similar to Table \ref{stab:d5percent} in the Section \ref{supp:RealParamInfer} where $d=5\%$, in the following we show the results for $d=0.5\%$ and $d=1\%$ in Table \ref{stab:d_sen}. First, for ABC itself, the bias in the posterior mean is similar, e.g. the mean bias is $14\%$ for all three different $d$ when $c=1.0$ and $\sigma^2=0.5$. For each pair of $(c,\sigma^2)$, the empirical coverage rate of the 95\% credible interval decreases when $d$ increases from $0.5\%$ to $5\%$. Specifically, the empirical coverage range from $92\%$ to $99\%$ for $d=5\%$, $88\%$ to $97\%$ for $d=1\%$ and $84\%$ to $94\%$ for $d=0.5\%$. This is likely caused by a smaller sample size $k=\lceil N^{\syn}d \rceil$ and a  higher posterior variance under a similar level of bias.

\begin{table}[!htb]
    \caption{Sensitivity analysis of $d$ for ABC-MH. We compare the inferential performance for $c$ among ABC-MH with $d=5\%$, ABC-MH with $d=1\%$, ABC-MH with $d=0.5\%$, $\textrm{MH}_{\sf true}$, and $\textrm{MH}_{\sf default}$. All values here are obtained from $200$ independent replications. For each random replication at $(c,\sigma^2)$, all methods were run for identical total CPU time and only converged chains from MH algorithms were included.}
    \label{stab:d_sen}
    \centering
    \begin{tabular}{llllll}
    \toprule
    & & \multicolumn{2}{c}{Percent Bias(sd)}& \multicolumn{2}{c}{Coverage(sd)}\\
    \cmidrule(r){3-4} \cmidrule(r){5-6}
    $c$ & method & $\sigma^2=0.5$ & $\sigma^2=1$ & $\sigma^2=0.5$ & $\sigma^2=1$\\
    \midrule
    \multirow[c]{5}{*}{0.3} & ABC-MH with $d=5\%$ & 12(9.4) & 13(9.9) & 98(0.99) & 99(0.71) \\
    & ABC-MH with $d=1\%$ & 13(9.8) & 14(10) & 97(1.2) & 96(1.5) \\
    & ABC-MH with $d=0.5\%$ & 14(10) & 15(11) & 94(1.6) & 92(1.9) \\
    & $\textrm{MH}_{\sf true}$ & 13(9.8) & 12(9.5) & 94(2) & 95(1.9) \\ 
    & $\textrm{MH}_{\sf default}$ & 45(20) & 46(20) & 33(5.5) & 30(6.1) \\ 
    \midrule
    \multirow[c]{5}{*}{0.5} & ABC-MH with $d=5\%$ & 15(11) & 15(11) & 92(1.9) & 93(1.8)\\
    & ABC-MH with $d=1\%$ & 15(12) & 16(12) & 88(2.3) & 90(2.1) \\
    & ABC-MH with $d=0.5\%$ & 16(12) & 16(12) & 84(2.6) & 86(2.4) \\ 
    & $\textrm{MH}_{\sf true}$ & 11(9) & 11(8.6) & 97(1.7) & 97(1.6) \\ 
    & $\textrm{MH}_{\sf default}$ & 33(18) & 31(19) & 60(5.5) & 57(5.7) \\ 
    \midrule
    \multirow[c]{5}{*}{0.7} & ABC-MH with $d=5\%$ & 13(10) & 14(11) & 96(1.5) & 93(1.8) \\
    & ABC-MH with $d=1\%$ & 13(10) & 13(11) & 94(1.7) & 90(2.1) \\
    & ABC-MH with $d=0.5\%$ & 13(11) & 14(11) & 90(2.1) & 89(2.2) \\ 
    & $\textrm{MH}_{\sf true}$ & 12(9.1) & 12(9.1) & 95(2.6) & 96(2.1) \\ 
    & $\textrm{MH}_{\sf default}$ & 25(15) & 27(16) & 73(5.5) & 69(5.9) \\ 
    \midrule
    \multirow[c]{5}{*}{1.0} & ABC-MH with $d=5\%$ & 14(11) & 14(13) & 95(1.5) & 94(1.6) \\ 
    & ABC-MH with $d=1\%$ & 14(10) & 14(13) & 88(2.3) & 92(1.9)\\ 
    & ABC-MH with $d=0.5\%$ & 14(11) & 15(13) & 86(2.4) & 86(2.4) \\ 
    & $\textrm{MH}_{\sf true}$ & 11(7.6) & 13(11) & 97(2) & 92(3.5) \\ 
    & $\textrm{MH}_{\sf default}$ & 14(11) & 16(14) & 93(3.5) & 89(3.8) \\ 
    \bottomrule
    \end{tabular}
\end{table}

\section{Additional Simulation Results of \texorpdfstring{\Rx-Trees}{TEXT}}\label{supp:TreeSim}
In this Section, we provide more simulation results for the Section 4.2 in the Main Paper. We empirically compare the the proposed two-stage ABC-MH with the single-stage MCMC in terms of the MAP tree estimation (Section \ref{supp:Two_Single_Alg_MAP}) and recovery of pairwise treatment similarities (Section \ref{supp:Two_Single_Alg_Similarity}).

\paragraph{Simulation setup.} For the following simulations, we followed the same setup as in Section \ref{supp:Simulation_Euclidean} with $I=50$ and $J=10$, and let $c$ and $\sigma^2$ take values from $\{0.3,0.5,0.7,1.0\}$ and $\{0.5,1.0\}$, respectively. For each pair of $(c,\sigma^2)$, $50$ pairs of tree and data on the leaves were independently drawn based on the DDT model. For ABC, we generated $N^\syn=600,000$ synthetic data sets from the DDT model with threshold parameter $d=0.5\%$. We assigned priors on $c\sim {\sf Gamma}(2,2)$ and $1/\sigma^2 \sim {\sf Gamma}(1,1)$ with shape and rate parameterization. We compare the proposed algorithm against two alternatives based on MH algorithms ($\MH_{\sf true}$ and ${\MH}_{\sf default}$. We ran MH algorithms (the 2nd stage of the proposed algorithm, $\MH_{\sf true}$ and ${\MH}_{\sf default}$) with 10,000 iterations and discarded the first 7,000 iterations.

\paragraph{Performance metrics.} We assess the accuracy of tree estimation using Billera–Holmes– Vogtmann (BHV) distance \citep{BILLERA2001733} between the true tree and the {\it maximum a posteriori} (MAP) tree obtained from ABC-MH, $\MH_{\sf true}$ and $\MH_{\sf default}$, or between the true tree and the dendrogram obtained from hierarchical clustering, respectively. 
For the pairwise similarities, we follow the Section 4.1 and calculate iPCPs for all pairs of treatments and evaluate the iPCPs by correlation of correlation for estimated similarities and true branching time and the Frobenius norm for the overall matrix.

\subsection{Recovery of the True Tree}\label{supp:Two_Single_Alg_MAP} 
\begin{figure}[!htb]
        \centering
        \includegraphics[width=\linewidth]{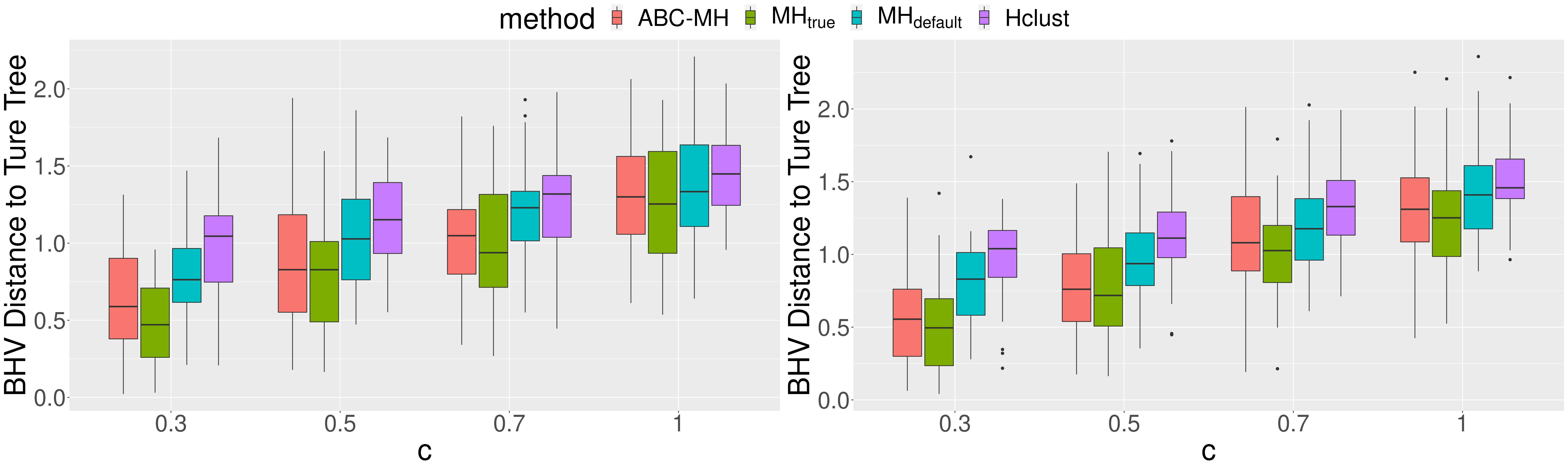}
    \caption{(Left) $\sigma^2=0.5$; (Right) $\sigma^2=1$. The BHV distance between the MAP estimate and the underlying tree for each algorithm. Results are based on 50 replications.}
    \label{fig:BHV}
    \end{figure}

The proposed two-stage algorithm decoupled the real and tree parameters, produced better inference for Euclidean parameters (See Section \ref{supp:RealParamInfer}), resulting in better inference for the unknown treatment tree. In particular, Figure \ref{fig:BHV} shows that, in terms of the BHV distance, the MAP tree estimates from ABC-MH better recovers the trees than $\MH_{\sf default}$ and hierarchical clustering with Euclidean distance and squared Ward linkage ({Hclust}). On average, MAP from $\MH_{\sf true}$ is the closest to the true underlying tree. However, $\MH_{\sf true}$ requires knowledge about the truth and is unrealistic in practice. In addition, we observed that the chains from $\MH_{\sf true}$ in fact did not mix well and were stuck at the initial values hence falsely appearing accurate. The second stage MH for sampling the tree built on the high-quality posterior samples of $c$ and $\sigma^2$ obtained from the 1st stage ABC and produced better MAP tree estimates that are on average closer to the simulation truths than $\MH_{\sf default}$ and {Hclust}.
    
\subsection{Estimation of Treatment Similarities}\label{supp:Two_Single_Alg_Similarity} 
\begin{figure}[!htb]
        \centering
        \includegraphics[width=\linewidth]{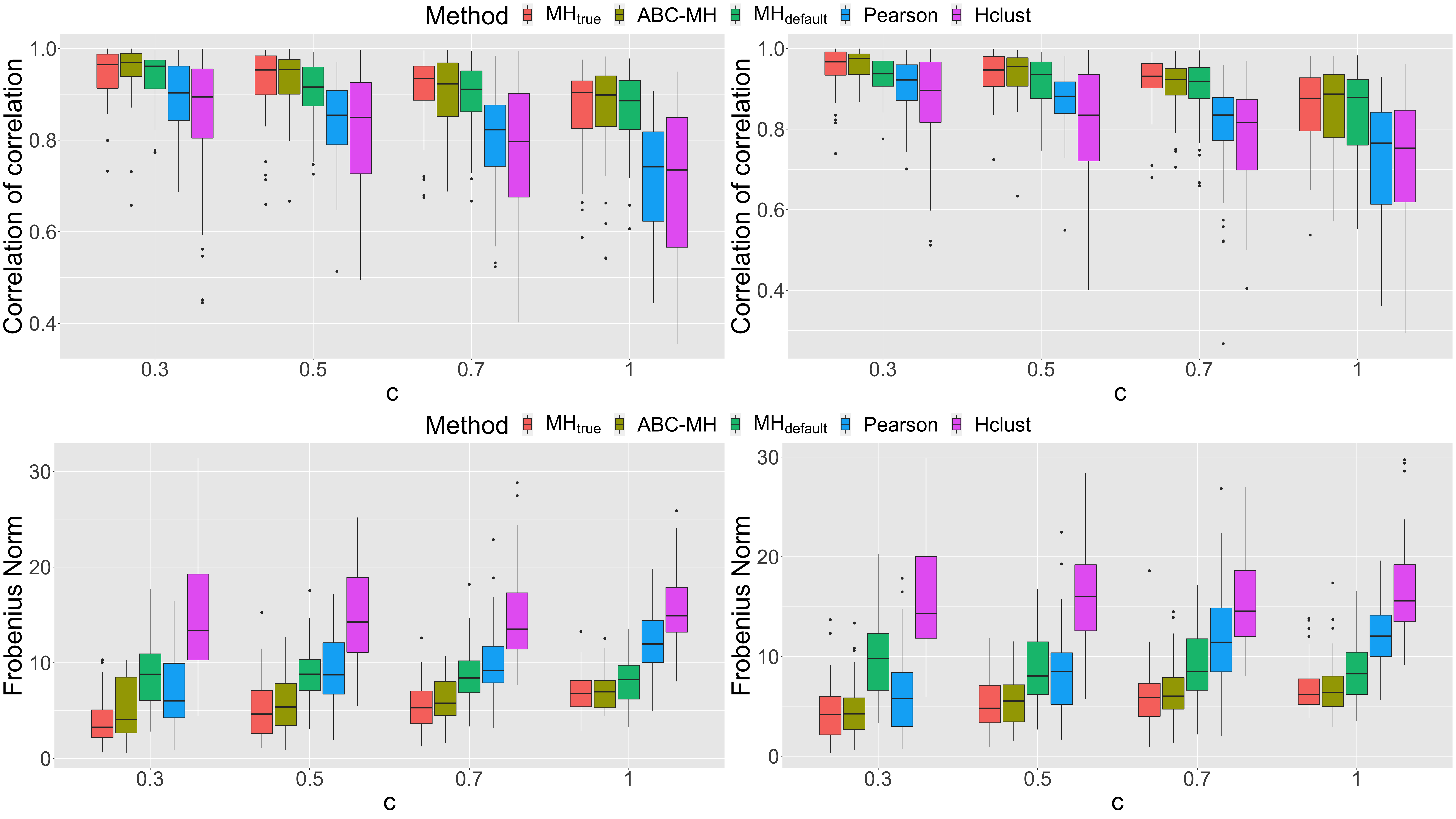}
    \caption{Under different $c$ and $\sigma^2$, two-stage algorithm better estimates the pairwise similarities than classical single-stage MCMC in terms of correlation of correlation (upper panels) and Frobenius norm (lower panels). (Left) $\sigma^2=0.5$; (Right) $\sigma^2=1$. Results are based on 50 replications.}
    \label{sfig:iPCP_ABCMH_JtMCMC}
    \end{figure}

The two-stage algorithm also produces better iPCPs due to decoupling strategy and superior inference for Euclidean parameters in the first stage. Similar to the results for MAP, pairwise iPCPs from ABC-MH better recover the true branching time than $\MH_{\sf default}$, {Hclust} and Pearson correlation and reach similar quality to the iPCPs from $\MH_{\sf true}$ (See Figure \ref{sfig:iPCP_ABCMH_JtMCMC}). Since $\MH_{\sf true}$ requires unrealistic true parameters, $\MH_{\sf true}$ is not attainable. From the simulations above, MAP and iPCPs from ABC-MH outperform $\MH_{\sf default}$ and take care of overall and local tree details, respectively. We apply the ABC-MH to obtain posterior DDT samples for the real data analysis section. 
    
\section{Additional Results for PDX Analysis}\label{supp:PDXAnalysis}
In this section, we provide the pre-processing procedures of NIBR-PDXE and present the results for non-small lung cancer (NSCLC) and pancreatic ductal adenocarcinoma (PDAC) with tables including treatment and pathway information.

\subsection{PDX Data Pre-Processing} \label{supp:preProcess}
We followed pre-processing procedure in \citet{doi:10.1080/01621459.2020.1828091} and imputed the missing data by k-nearest neighbor method. We take the best average response (BAR) as the response and scale the BAR by the standard deviation over all patients, treatments and across five cancers. Since the scaled BAR contains missing values, we impute the missing data by the k-nearest neighbor with $k=10$ and compare all treatments to the untreated group. Specifically, we take $x_{ij}={\rm BAR}_{ij}-{\rm BAR}_{0j}, i=1\ldots I, j=1\ldots J$ as the observed data, where ${\rm BAR}_{0,j}$ is the untreated BAR for patient $j$.


\subsection{Test for Multivariate Normal Assumption} 
\begin{figure}[!htb]
    \centering
    \includegraphics[width=\linewidth]{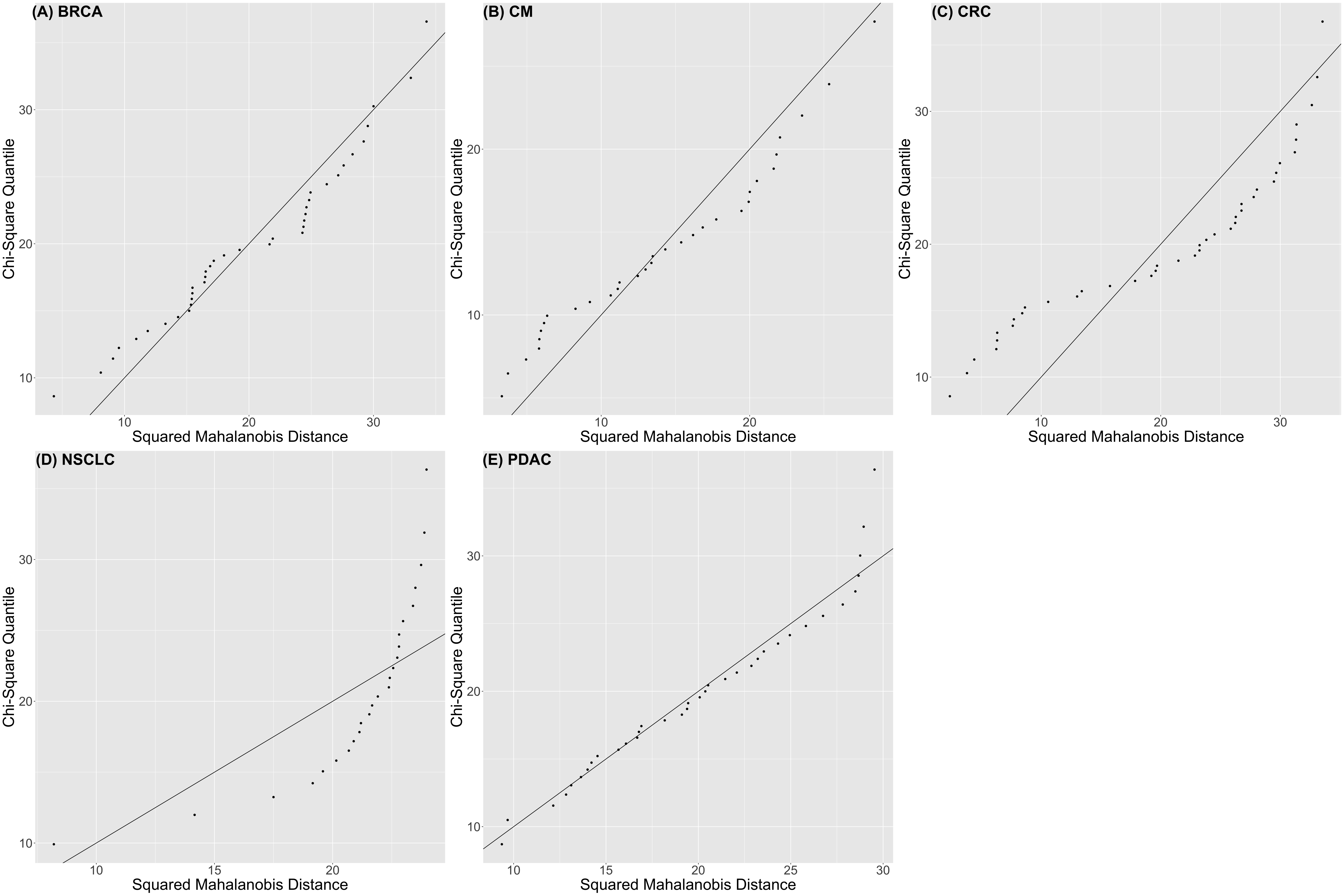}
    \caption{The multivariate normality QQ-plot for (A) breast cancer, (B) melanoma, (C) colorectal cancer, (D) non-small lung cancer and (E) pancreatic adenocarcinoma.}
    \label{sfig:MVN_qq}
\end{figure}

From the Proposition 1 and the Equation (5), the observed PDX data follow a multivariate normal distribution. We plot the QQ-plot to check the normal assumption for each cancer and show the result in Figure \ref{sfig:MVN_qq}. Among fiver cancers, BRCA (panel (A)) and CM (panel (B)) roughly fall on the 45-degree line and PDAC (panel (E)) is slightly away from the 45-degree line with some extreme data. Contrary, CRC (panel (C)) and NSCLC (panel (D)) deviate from the the 45-degree lines indicating departure from normality.

\subsection{\texorpdfstring{\Rx-Trees}{TEXT} for Non-Small Lung Cancer (NSCLC) and Pancreatic Ductal Adenocarcinoma (PDAC)}\label{supp:NSCLC_PDAC}
\begin{figure}[!htb]
        \centering
        \includegraphics[width=\linewidth]{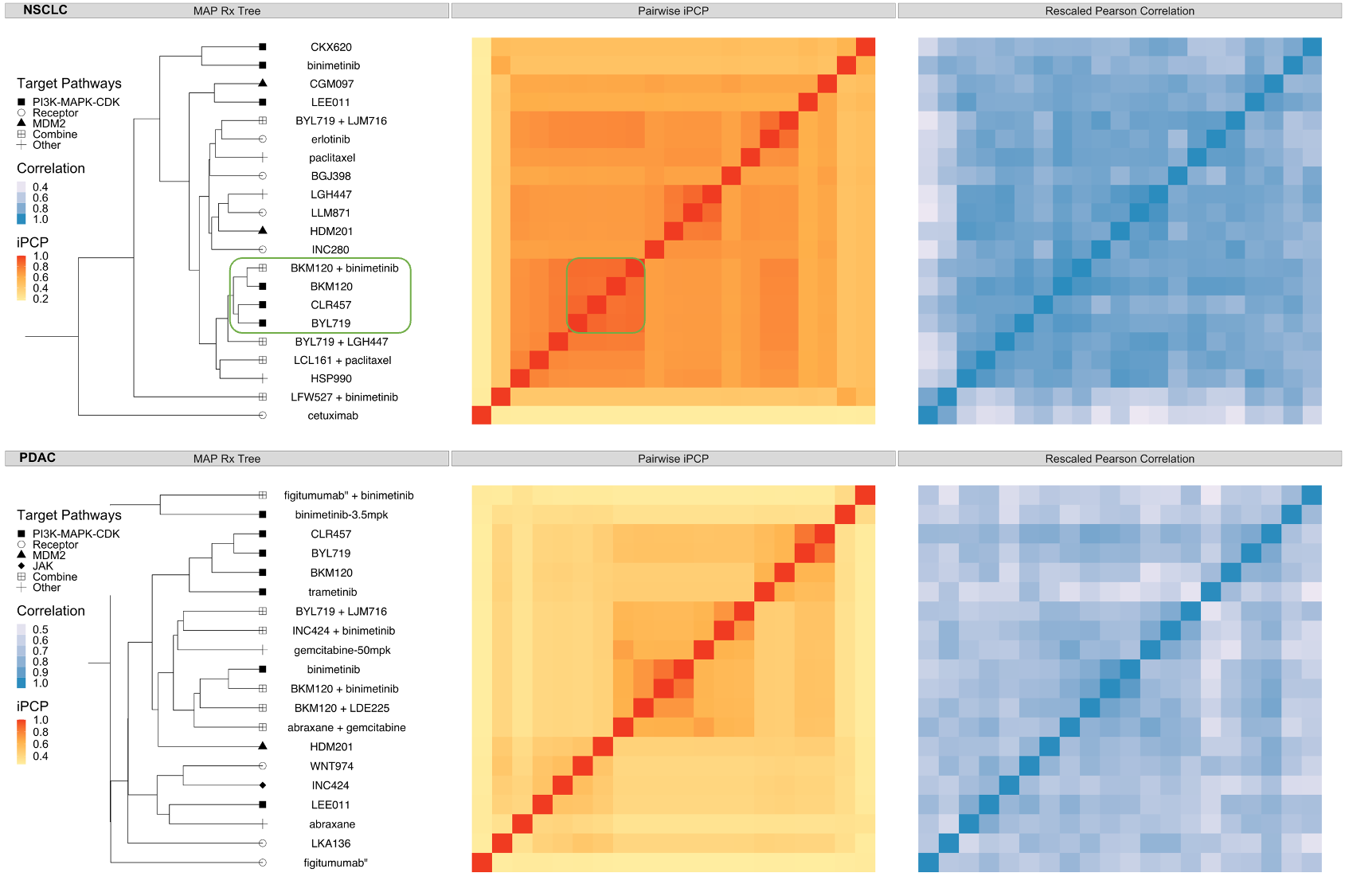}
\caption{The \Rx-tree and iPCP for non-small cell lung cancer (NSCLC, top row) and pancreatic ductal adenocarcinoma (PDAC, lower row). Three panels in each row represent: (left) estimated \Rx-tree (MAP); distinct external target pathway information is shown in distinct shapes for groups of treatments on the leaves; (middle) Estimated pairwise iPCP, i.e., the posterior mean divergence time for pairs of entities on the leaves (see the result paragraph for definition for any subset of entities); (right) Scaled Pearson correlation for each pair of treatments. The Pearson correlation $\rho\in [-1,1]$ was scaled by $\frac{\rho+1}{2}$ to fall into $[0,1]$. Note that the MAP visualizes the hierarchy amongst treatments; the iPCP is not calculated based on the MAP, but based on posterior tree samples (see definition in Main Paper Section 3.2)}
\label{sfig:ggtree_supp}
\end{figure}

We applied the \Rx-tree on the rest two cancers in the data: non-small lung cancer (NSCLC) and pancreatic ductal adenocarcinoma (PDAC). Similar to the Figure 5 in the Main Paper, \Rx-tree, pairwise iPCP and (scaled) Pearson correlation are shown in the left, middle and right panels in Figure \ref{sfig:ggtree_supp}, respectively. Again, we observe that the \Rx-tree and the pairwise iPCP matrix show the similar clustering patterns. For example, three PI3K inhibitors (BKM120, BYL719 and CLR457) and a combination therapy (BKM120 + binimetibin) in NSCLC form a tight subtree and are labeled by a box in the \Rx-tree of Figure \ref{sfig:ggtree_supp} and a block with higher values of iPCP among therapies above also shows up in the corresponding iPCP matrix. The \Rx-tree roughly clusters monotherapies targeting oncogenic process (PI3K-MAPK-CDK, MDM2 and JAK) and agrees with the biology mechanism. For example, three PI3K inhibitors (BKM120, BYL719 and CLR457) belong to a tighter subtree in both cancers. Following the same idea as the Main Paper, we further quantify the treatment similarity through iPCP. However, compared to three cancers (BRCA, CRC and CM) in the Main Paper, different problems of model fitting or interpretation lie in NSCLC and PDAC: NSCLC deviates from the normal assumption of Equation (4) (Figure \ref{sfig:MVN_qq}) and PDAC shows lower iPCP (average iPCP of PDAC: 0.4119 $<$ BRCA: 0.6734, CRC: 0.5653, CM: 0.7535, NSCLC: 0.5817). For concerns raised above, we only verify the model through the monotherapies with known biology for each cancer.

\noindent \underline{Non-small cell lung cancer.}
Our model suggests high iPCP values for treatments share the same target. For example, our model shows a high iPCP among three PI3K (BKM120, BYL719 and CLR457) inhibitors: (BKM120, BYL719): 0.8402, (BKM120, CLR457): 0.8321, (BYL719, CLR457): 0.8710. For treatments with different targets, our model also exhibits a high iPCP values. For example, the monotherapy HSP990 that inhibits the heat shock protein 90 (HSP90) shows a high iPCP with PI3K inhibitors ((BKM120, HSP990): 0.7108, (BYL719, HSP990): 0.7114, (CLR457,HSP990): 0.7109). Since the inhibiting of HSP90 also suppresses PI3K \citep{pmid28619753}, it is not surprising to ses a high iPCP between PI3K and HSP90 inhibitors.
	
\noindent \underline{Pancreatic ductal adenocarcinoma.}
For PDAC, our model overall suggests a lower iPCP (average iPCP of PDAC: 0.4119). Out of $91$ pairs of monotherapies, only BYL719 and CLR457 share a higher iPCP (0.8415). The higher iPCP can be explained by the common target PI3K of BYL719 and CLR457.

\subsection{R Shiny Application}\label{supp:RShiny}
\begin{figure}[!htb]
    \centering
    \includegraphics[width=\linewidth]{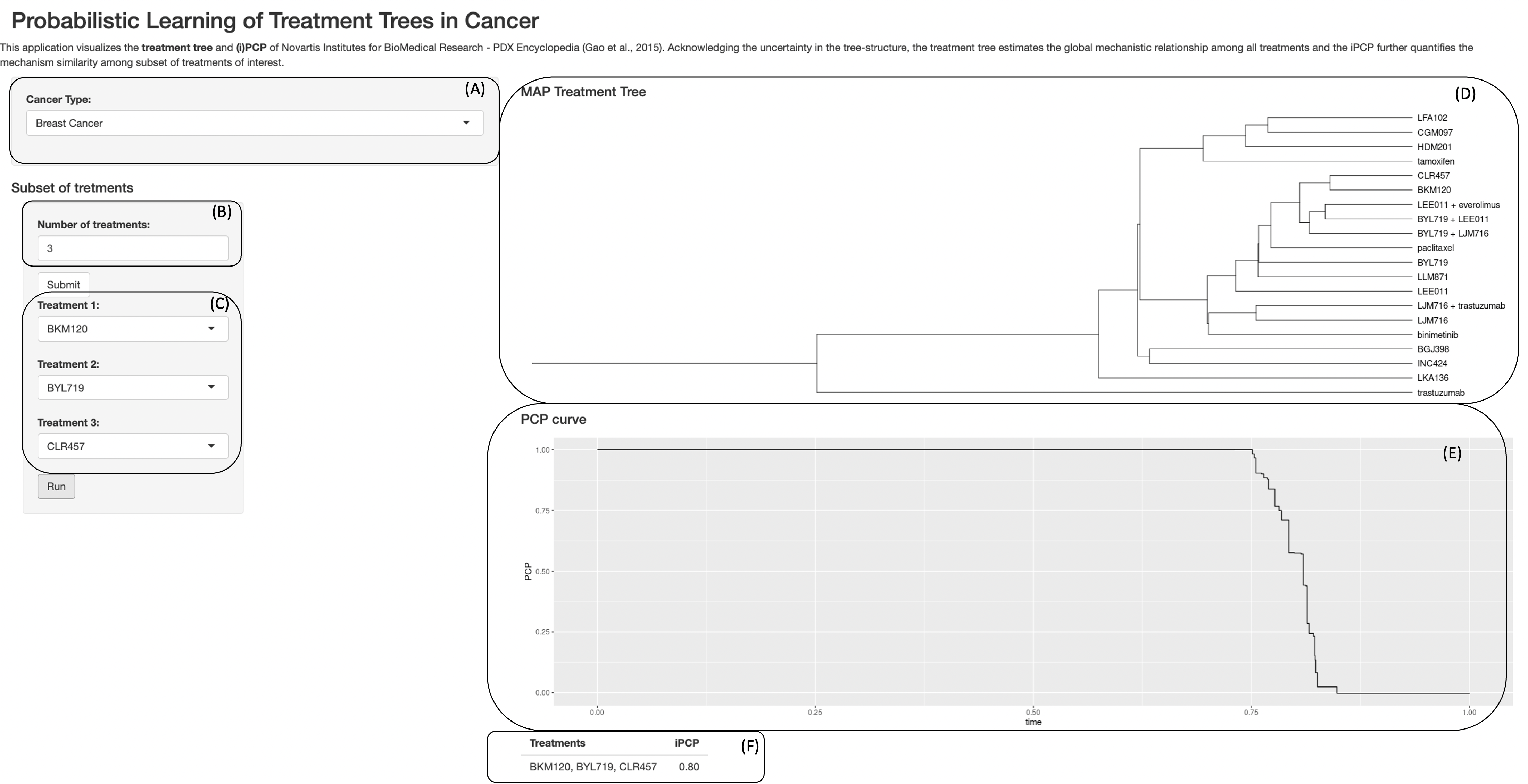}
\caption{R Shiny app screenshot for illustrating model inputs and outputs for analyzing PDX data (20 treatments for breast cancer); the PCP curve and iPCP value are computed for a subset of three selected treatments.}
\label{fig:R_shiny1}
\end{figure}

We illustrate the input and outputs of the proposed method via a R Shiny application hosted on the web (Figure \ref{fig:R_shiny1}). The visualizations are based on completed posterior computations for illustrative purposes. A user needs to specify the following inputs:
\begin{enumerate}[label=(\Alph*)]
\itemsep 0em
    \item Cancer type to choose the subset of data for analysis
    \item Number of treatments of interest in the subset $\cA$ to evaluate synergy via iPCP
    \item Names of the treatments in the subset $\cA$
\end{enumerate}
Given the inputs above, the Shiny app visualizes the outputs:
\begin{enumerate}[resume, label=(\Alph*)]
\itemsep 0em
    \item \textit{maximum a posteriori} treatment tree for all the available treatments
    \item $\textrm{PCP}_{\cA}(t)$ curve for the subset of treatments, $\cA$
    \item $\textrm{iPCP}_\cA$ value calculated from the corresponding $\textrm{PCP}_{\cA}(t)$
\end{enumerate}

\begin{table}[!htb]
    \caption{Full CPUs series used for computations.}
    \label{stab:CPU}
    \centering
    \setlength\extrarowheight{-7pt}
    \begin{tabular}{ll}
    \toprule
    \multirow[c]{2}{*}{Intel Xeon X series} & X5660@2.80GHz\\
    & X5680@3.33GHz\\
    \midrule
    \multirow[c]{7}{*}{Intel Xeon E series} & E5-24400@2.40GHz\\
    & E5-24700@2.30GHz\\
    & E5-24500@2.10GHz\\
    & E5-2650v3@2.30GHz\\
    & E5-2650v4@2.20GHz\\
    & E5-2690v4@2.60GHz\\
    & E5-2690v4@2.60GHz\\
    \bottomrule
    \end{tabular}
\end{table}

\begin{table}[!htb]
    \caption{Pathways full names and the corresponding abbreviations.}
    \label{stab:tgtName}
    \centering
    \setlength\extrarowheight{-7pt}
    \begin{tabular}{ll}
    \toprule
    Abbreviation & Target Name\\
    \midrule
    PI3K & Phosphoinositide 3-kinases\\
    \midrule
    CDK & Cyclin-dependent kinases\\
    \midrule
    MAPK & Mitogen-activated protein kinases\\
    \midrule
    JAK & Janus kinase\\
    \midrule
    MDM2 & Murine double minute 2\\
    \midrule
    BRAF & Serine/threonine-protein kinase B-Raf\\
    \midrule
    MTOR & Mechanistic target of rapamycin \\
    \midrule
    EGFR/ERBB & Epidermal growth factor receptor\\
    \midrule
    SMO & Smoothened \\
    \midrule
    TNKS & Tankyrase \\
    \midrule
    PIM & Proto-oncogene serine/threonine-protein kinase Pim-1 \\
    \midrule
    BIRC2 & Baculoviral IAP repeat-containing protein 2 \\
    \midrule
    IGF1R & Insulin-like growth factor 1 receptor \\
    \bottomrule
    \end{tabular}
\end{table}

\begin{table}[!htb]
    \caption{Monotherapy names with targets. Different target groups are labeled differently in the Figure 5 and Figure \ref{sfig:ggtree_supp}.}
    \label{stab:trtName}
    \centering
    \setlength\extrarowheight{-7pt}
    \begin{tabular}{lllll}
    \toprule
    Treatment name & Other names & Trade name & Target & Target Group \\
    \midrule
    5FU & Fluorouracil &Adrucil & chemotherapy & Other \\ 
    \midrule
    abraxane & nab-paclitaxel &abraxane & Tubulin & Other \\ 
    \midrule
    BGJ398 & Infigratinib & & FGFR & Receptor \\ 
    \midrule
    binimetinib & MEK162 &Mektovi & MAPK & PI3K-MAPK-CDK \\ 
    \midrule
    BKM120 & Buparlisib & & PI3K & PI3K-MAPK-CDK \\ 
    \midrule
    BYL719 & Alpelisib &Piqray & PI3K & PI3K-MAPK-CDK \\ 
    \midrule
    cetuximab &  &Erbitux & EGFR & Receptor \\ 
    \midrule
    CGM097 &  & & MDM2 & MDM2 \\ 
    \midrule
    CKX620 &  & & MAPK & PI3K-MAPK-CDK \\ 
    \midrule
    CLR457 &  & & PI3K & PI3K-MAPK-CDK \\ 
    \midrule
    dacarbazine &  &DTIC-Dome & chemotherapy & Other \\ 
    \midrule
    encorafenib & LGX818 &Braftovi & BRAF & BRAF \\ 
    \midrule
    erlotinib & Erlotinib hydrochloride &Tarceva & EGFR & Receptor \\ 
    \midrule
    figitumumab & CP-751871 & & IGF1R & Receptor \\ 
    \midrule
    gemcitabine &  &Gemzar & chemotherapy & Other \\ 
    \midrule
    HDM201 & Siremadlin & & MDM2 & MDM2 \\ 
    \midrule
    HSP990 &  & & HSP90 & Other \\ 
    \midrule
    INC280 & Capmatinib &Tabrecta & MET & Receptor \\ 
    \midrule
    INC424 & Ruxolitinib &Jakafi and Jakavi & JAK & JAK \\ 
    \midrule
    LDE225 & Sonidegib &Odomzo & SMO & Receptor \\ 
    \midrule
    LDK378 & Ceritinib &Zykadia & ALK & Receptor \\ 
    \midrule
    LEE011 & Ribociclib &Kisqal & CDK & PI3K-MAPK-CDK \\ 
    \midrule
    LFA102 &  & & PRLR & Receptor \\ 
    \midrule
    LGH447 &  & & PIM & Other \\ 
    \midrule
    LGW813 &  & & IAP & Other \\ 
    \midrule
    LJC049 &  & & TNKS & Other \\ 
    \midrule
    LJM716 & Elgemtumab & & ERBB3 & Receptor \\ 
    \midrule
    LKA136 &  & & NTRK & Receptor \\ 
    \midrule
    LLM871 &  & & FGFR2/4 & Receptor \\ 
    \midrule
    paclitaxel &  &Taxol & Tubulin & Other \\ 
    \midrule
    tamoxifen &  &Nolvadex & ESR1 & Receptor \\ 
    \midrule
    TAS266 &  & & DR5 & Receptor \\ 
    \midrule
    trametinib & GSK1120212 &Mekinist & MAPK & PI3K-MAPK-CDK \\ 
    \midrule
    trastuzumab &  &Herceptin & ERBB2 & Receptor \\ 
    \midrule
    WNT974 &  & & PORCN & Receptor \\ 
    \bottomrule
    \end{tabular}
\end{table}

\begin{table}[!htb]
    \caption{Combination therapy full names with known targets.}
    \label{stab:cmbTrt}
    \centering
    \setlength\extrarowheight{-7pt}
    \begin{tabular}{lll}
    \toprule
    Combination Therapies & Known Target Pathways & Cancer\\
    \midrule
    abraxane+gemcitabine & Tubulin+chemotherapy & PDAC \\
    \midrule
    BKM120+binimetinib & PI3K+MAPK & NSCLC,PDAC \\ 
    \midrule
    BKM120+encorafenib & PI3K+BRAF & CM \\ 
    \midrule
    BKM120+LDE225 & PI3K+SMO & PDAC \\ 
    \midrule
    BKM120+LJC049 & PI3K+TNKS & CRC \\ 
    \midrule
    BYL719+binimetinib & PI3K+MAPK & CRC \\ 
    \midrule
    BYL719+cetuximab & PI3K+EGFR & CRC \\ 
    \midrule
    BYL719+cetuximab+encorafenib & PI3K+EGFR+BRAF & CRC \\ 
    \midrule
    BYL719+encorafenib & PI3K+BRAF & CRC \\ 
    \midrule
    BYL719+LEE011 & PI3K+CDK & BRCA \\ 
    \midrule
    BYL719+LGH447 & PI3K+PIM & NSCLC \\ 
    \midrule
    BYL719+LJM716 & PI3K+ERBB3 & BRCA,CRC,NSCLC,PDAC \\ 
    \midrule
    cetuximab+encorafenib & EGFR+BRAF & CRC \\ 
    \midrule
    encorafenib+binimetinib & BRAF+MAPK & CM \\ 
    \midrule
    figitumumab+binimetinib & IGF1R+MAPK & PDAC \\ 
    \midrule
    INC424+binimetinib & JAK+MAPK & PDAC \\ 
    \midrule
    LCL161+paclitaxel & BIRC2+Tubulin & NSCLC \\ 
    \midrule
    LEE011+encorafenib & CDK+BRAF & CM \\ 
    \midrule
    LEE011+everolimus & CDK+MTOR & BRCA \\ 
    \midrule
    LFW527+binimetinib & IGF1R+MAPK & NSCLC \\ 
    \midrule
    LJM716+trastuzumab & ERBB3+ERBB2 & BRCA \\ 
    \bottomrule
    \end{tabular}
\end{table}

\clearpage
\bibliographystyle{unsrtnat}
\bibliography{Appen_DDT}